\documentclass{jfm-generic}
\usepackage{floatrow}
\usepackage{graphicx}
\usepackage{epstopdf,subfig,epsfig}
\usepackage{amsmath,amssymb,amsfonts,mathtools,euscript,changebar,multicol,amscd,comment}
\usepackage{color,epstopdf,url}
\usepackage{bm}
\usepackage[colorlinks]{hyperref}
\usepackage{times}
\usepackage{xcolor}
\usepackage{soul}
\usepackage[titletoc]{appendix}
\usepackage{enumitem}

\usepackage{tikz}
\newcommand*\circled[1]{\tikz[baseline=(char.base)]{
            \node[shape=circle,draw,inner sep=2pt] (char) {#1};}}

\hypersetup{
    colorlinks=true,       
    linkcolor=blue,        
    citecolor=blue,        
    filecolor=black,       
    urlcolor=black,        
    bookmarks=false,
    pdffitwindow=true,
    pdfpagelayout=SinglePage,
    plainpages=false
}

\renewcommand{\i}{\mathrm{i}}

\renewcommand{\d}{\mathrm{d}}

\newcommand{\const}{\mathrm{const}}

\DeclareMathOperator{\Ci}{Ci}

\DeclareMathOperator{\sech}{sech}
\DeclareMathOperator{\atan}{atan}

\newcommand{\marginlabel}[1]
{\mbox{}\marginpar{\center\hspace{0pt}\bf\tiny\color{white} #1}}

\shorttitle{Transverse instability of concentric soliton waves}
\shortauthor{R. Krechetnikov}

\title{Transverse instability of concentric soliton waves}

\author{R. Krechetnikov\aff{1}
  \corresp{\email{krechet@ualberta.ca}}}

\affiliation{\aff{1}Mathematical and Statistical Sciences, University of Alberta, Edmonton, AB, T6G 2G1, Canada}

\begin{document}

\setlist[description]{font=\normalfont\itshape}

\maketitle

\begin{abstract}
Should it be a pebble hitting water surface or an explosion taking place underwater, concentric surface waves inevitably propagate. Except for possibly early times of the impact, finite amplitude concentric water waves emerge from a balance between dispersion or nonlinearity resulting in solitary waves. While stability of plane solitary waves on deep and shallow water has been extensively studied, there are no analogous analyses for concentric solitary waves. On shallow water, the equation governing soliton formation -- the nearly concentric Korteweg-de Vries -- has been deduced before without surface tension, so we extend the derivation onto the surface tension case. On deep water, the envelope equation is traditionally thought to be the nonlinear Schr\"{o}dinger type originally derived in the Cartesian coordinates. However, with a systematic derivation in cylindrical coordinates suitable for studying concentric waves we demonstrate that the appropriate envelope equation must be amended with an inverse-square potential, thus leading to a Gross-Pitaevskii equation instead.

Properties of both models for deep and shallow water cases are studied in detail, including conservation laws and the base states corresponding to axisymmetric solitary waves. Stability analyses of the latter lead to singular eigenvalue problems, which dictate the use of analytical tools. We identify the conditions resulting in the transverse instability of the concentric solitons revealing crucial differences from their plane counterparts. Of particular interest here are the effects of surface tension and cylindrical geometry on the occurrence of transverse instability.
\end{abstract}

\tableofcontents

\clearpage

\section{Introduction} \label{sec:introduction}

The dissonance between the \citet{Airy:1845} linear water wave theory and \citet{Russell:1844} observations of a solitary wave, resolved by \citet{Boussinesq:1871,Boussinesq:1872,Boussinesq:1877} and \citet{Rayleigh:1876}, led to the idea of balancing dispersion and nonlinearity, now known as the \citet{Ursell:1953} criterion, leading to the classical Korteweg-de Vries (KdV) and nonlinear Schr\"{o}dinger (NLS) equations, both possessing soliton solutions. The transverse instability of solitary waves in the shallow and deep water regimes has been explored for a long time since the classical works of \citet{Kadomtsev:1970} and \citet{Zakharov:1974}, respectively, but all the subsequent studies have been mostly limited to plane KdV and NLS solitons, i.e. the effects of curvature of the solitary wave as well as of its amplitude decay due to cylindrical geometry have not been systematically addressed; for review of the vast literature on the topic see \citet{Kivshar:2000,Yang:2010}. In fact, despite the classical setting and ongoing interest \citep{Peregrine:1983,Grimshaw:2007,Vitanov:2013}, this set of problems proved to be understudied as not only instability to transverse perturbations has not been explored, but also the equations governing the solitary wave dynamics in the deep-water limit have not been derived, while in the shallow-water limit the underlying asymptotic assumptions and the structure of the equation and its axisymmetric solutions in the presence of surface tension have not been fully understood. It is the goal of the present study to fill in this gap and, applying the Ursell criterion, to deduce appropriate weakly nonlinear models generating concentric solitary waves in the presence of surface tension and to analyze their transverse instability.

As models for studying the transverse instability of the finite-amplitude concentric carrier waves, we will consider the deep water case resulting in the nearly concentric NLS-type (ncNLS) and the shallow water case resulting in the nearly concentric Korteweg-de Vries (ncKdV) equations, respectively; the term ``near concentric'' is used to distinguish from the ``concentric'' limits cNLS and cKdV having no azimuthal dependence. The former (NLS-type), to the author's knowledge, has not been derived before. Since the systematic derivation is technically involved, to make it more transparent in \S \ref{subsec:ncNLS} we guide the reader through the key steps in the derivation of a NLS-type equation in cylindrical geometry including the effects of surface tension, while highlighting the differences in the derivation from the plane case. As for the latter (ncKdV), it has been previously derived by \citet{Johnson:1980} in a certain asymptotic regime without surface tension effects; therefore, in \S \ref{subsec:ncKdV} we revisit the derivation not only to include these effects, but also to highlight the physics necessary for understanding the transverse instability implications, and to identify an asymptotic regime relevant for our purposes, under which the ncKdV arises.

Since both models correspond to the limit of a finite-amplitude narrow wavepacket evolution leading to a solitary wave solution, we are interested in here, to set the stage we contrast it with the case when the initial axisymmetric free surface deflection $\eta_{0}(r)$ dependent on the radial coordinate $r$ only is infinitesimally small and contains a wide range of wavenumbers $k$, e.g. in the Hankel-space $\widehat{\eta}_{0}(k)=1$, which corresponds to a localized initial free surface deflection in the physical space $\eta_{0}(r) = \left\{2 \pi b \ \text{for} \ 0 \le r \le r_{0}, \ \text{and} \ 0 \ \text{for} \ r > r_{0}\right\}$; here $\pi b \, r_{0}^{2} = 1$ with $r_{0} \rightarrow 0$ and $b \rightarrow \infty$. For example, in the deep water case of pure gravity-driven waves the stationary phase analysis \citep{Koshlyakov:1964} identifies the `stationary' wavenumber $k = \frac{g t^{2}}{4 r^{2}}$ proportional to the gravity acceleration $g$ and gives for the free surface evolution:
\begin{align}
\label{free-surface:stationary-phase}
\eta(t,r) \sim \frac{g t^{2}}{r^{3}} \cos{k r} = \frac{g t^{2}}{r^{3}} \cos{\frac{g t^{2}}{4 r}}
\end{align}
indicating that at a fixed time $t$ the waves become of longer wavelength and smaller amplitude with increasing $r$, while for fixed $r$ the amplitude of the wave increases and the wavelength shortens; equation \eqref{free-surface:stationary-phase} was calculated by \citet{Lamb:1904}, though some elements of this analysis were known to \citet{Poisson:1816}. Of course, in reality such idealized Dirac delta-function signals $\eta_{0}(r)$ do not exist: replacing it with a smoothed (delta-sequence) deflection $\eta_{0}(r) \sim e^{-\frac{1}{2} \delta^{2} r^{2}}$ of characteristic width $\delta$ gives $\widehat{\eta}_{0}(k) \sim \frac{1}{\delta^{2}} e^{-\frac{k^{2}}{2 \delta^{2}}}$ thus regularizing the solution at short distances and long times. One may also arrive at \eqref{free-surface:stationary-phase} informally \citep{Kadomtsev:1982}: namely, given that the initial perturbation consists of all wavenumbers, a wavepacket centered around $k$ propagates with the group velocity $\frac{\d \omega}{\d k} = \frac{1}{2} \left(\frac{g}{k}\right)^{1/2}$ and hence in time $t$ will arrive at the point $r = \frac{1}{2} \left(\frac{g}{k}\right)^{1/2} t$, i.e. at a given time $t$ we find the wavenumber $k$ of the wave arriving at the point $r$, identified above with the stationary phase method.

For the comparison with the subsequent results in the present study, let us remind the reader the key conclusions of the earlier cited classical works on stability of 1D solitons. \textit{First}, the 1D case of a general 2D NLS with focusing nonlinearity in the Cartesian coordinates (written here in the adopted in the present work scaling with $\xi$ standing for the longitudinal direction, in which the 1D-soliton propagates, and $Y$ for the transverse direction):
\begin{align}
\label{eqn:NLS-1D}
\i \psi_{\tau} + \psi_{\xi\xi} + \alpha \psi_{YY} + |\psi|^{2} \psi = 0,
\end{align}
with $\alpha = 0$ and $\psi$ being the slow envelope amplitude of a traveling wave, admits the solution in the standing wave (Stokes) form $\psi = e^{\i \omega \tau} v(\xi)$ leading to
\begin{align}
\label{base-state:1D-NLS}
- \omega \, v + v^{\prime\prime} + v^{3} = 0,
\end{align}
which can be supplied with boundary conditions (BCs) $v^{\prime}(0)=0$, $v(\infty)=0$. Multiplying \eqref{base-state:1D-NLS} with $v^{\prime}$, and integrating w.r.t. $\xi$, we get
\begin{align}
\label{base-state:1D-NLS:reduced}
- \omega \, v^{2} + v^{\prime 2} + \frac{1}{2} v^{4} = 0 \ \Rightarrow \ w^{\prime 2} = \omega \, w^{2} - \frac{1}{2},
\end{align}
where we took into account the BCs and switched to $w=v^{-1}$. The solution of \eqref{base-state:1D-NLS:reduced} is a localized in the $\xi$-space soliton, $v = (2 \, \omega)^{1/2} \sech{\left(\omega^{1/2} \xi\right)}$. Other solutions can be generated from the fact that \eqref{eqn:NLS-1D} is amenable to translational symmetry $(\tau,\xi) \rightarrow (\tau^{\prime},\xi^{\prime} = \xi - u \tau)$ such that $\psi(\tau,\xi) \rightarrow e^{\i \frac{u}{2} \left(\xi^{\prime} + \frac{u}{2} \tau^{\prime}\right)} \psi(\tau^{\prime},\xi^{\prime})$. As shown by \citet{Zakharov:1968} (see also \citet{Grimshaw:2007}), plane waves solutions of 1D NLS are modulationally unstable in the focusing case \eqref{eqn:NLS-1D}. However, spectrally the solitons are neutrally stable, i.e. all eigenvalues are located on the imaginary axis; this fact is also consistent with the \citet{Vakhitov:1973} criterion \citep{Kuznetsov:1986,Yang:2010} based on the slope of the power curve $P(\mu) = \int{U^{2}(\xi;\mu) \, \d \xi}$ for the 1D solitary wave $\psi(\tau,\xi) = U(\xi) e^{\i \mu \tau}$, where $\mu$ is the propagation constant -- however spectral stability does not imply even linear, not to mention nonlinear, stability \citep{Krechetnikov:2007}. Later, \citet{Zakharov:1974} also established 1D-NLS soliton instability to transverse $Y$-modulations regardless of the sign of the transverse dispersion coefficient $\alpha$ in \eqref{eqn:NLS-1D}; $\alpha=+1$ corresponds to the elliptic and $-1$ to the hyperbolic case, respectively.

\textit{Second}, the nearly plane KdV equation (npKdV) deduced by \citet{Kadomtsev:1970}
\begin{align}
\label{eqn:KP}
\left(2 \, \eta_{\tau} + 3 \, \eta \, \eta_{\xi} + \frac{1}{3} \eta_{\xi\xi\xi}\right)_{\xi} - \beta \, \eta_{YY} = 0,
\end{align}
in absence of $Y$-dependence possesses not only a self-similar solution $\eta(\tau,\xi) = \tau^{-2/3} F(\zeta), \ \zeta = \tau^{-1/3} \xi$, but also the 1D soliton
\begin{align}
\label{soliton:1D:KdV}
\eta(\tau,\xi) = A \, f(\widetilde{\xi}), \ \widetilde{\xi} = \sqrt{A} \left(\xi - A \, \tau\right),
\end{align}
governed by
\begin{align}
\label{eqn:self-similar:1D-KdV}
\frac{1}{3} f^{\prime\prime\prime} - 2 \, f^{\prime} + 3 f f^{\prime} = 0.
\end{align}
Equation \eqref{eqn:self-similar:1D-KdV} can be integrated once to $\frac{1}{3} f^{\prime\prime} - 2 \, f + \frac{3}{2} f^{2} = 0$, assuming that the solution $f$ and its derivatives decay at infinity, and then its order can be reduced even further via $f^{\prime} = g(f)$ and integrated to yield the usual localized $f(\widetilde{\xi}) = \sech^{2}{\left(\frac{3}{2}\widetilde{\xi}\right)^{1/2}}$ soliton, qualitatively anticipated by Boussinesq and \citet{Rayleigh:1876} before the work of \cite{Korteweg:1895}. As first shown by \citet{Kadomtsev:1970} based on \eqref{eqn:KP}, this plane soliton exhibits transverse instability in the medium with positive dispersion ($\beta>0$) in the corresponding dispersion law $\omega(k;\beta)$, meaning that the phase velocity of linear waves increases with the wavenumber $k$, while for negative dispersion ($\beta<0$) it is spectrally stable. As the structure \eqref{soliton:1D:KdV} of the 1D-KdV soliton suggests, its speed $A$ relative to the frame of reference traveling with the phase speed of the carrier linear wave $c_{0} = \omega/k$, where $\omega^{2} = k^{2} h \left(g + \sigma k^{2}/\rho\right)$, depends on the soliton amplitude $A$, namely the larger the amplitude the faster the soliton travels. As we know from the transverse stability analysis of such a soliton \citep{Alexander:1997}, there exists the most amplified (preferred) transverse wavelength, which also depends on the soliton amplitude.

The latter property is not an issue in the plane (1D) case as the soliton amplitude does not change with time in non-dissipative media. However, as soon as we try to translate this knowledge of 1D soliton behavior onto the cylindrical case, we meet with two immediate complications both resulting from intrinsic time-dependence of the base state. To start with, the cylindrical soliton is being stretched in the transverse direction as it travels outwards and hence, according to the stability theory on time-dependent spatial domains \citep{Knobloch:2014,Knobloch:2015,Krechetnikov:2017,Ghadiri:2019}, an Eckhaus instability must insert new wavelengths (cells). However, as is obvious from the energy conservation, the soliton amplitude must decrease as it propagates outwards, which means that if one applies the intuition developed in the plane case than the wavelength of instability must change as well. Also, due to the lack of Galilean symmetry of ncKdV, only a self-similar solution of the form $\eta(\tau,\xi) = \tau^{-2/3} F(\zeta), \ \zeta = \tau^{-1/3} \xi$ exists in the cylindrical case -- its speed dependence upon its amplitude is obscured compared to \eqref{soliton:1D:KdV}; however, one may still adopt approximately the qualitative 1D picture to the cylindrical case as it was done in numerical studies of \citet{Maxon:1974b,Ko:1979}\footnote{While starting with $\sech^{2}$-soliton shape as an IC approximately follows this quasi-1D picture, it is clear that due to amplitude decrease with the radial distance the dynamics will eventually exit the KdV regime and switch to the NLS one as suggested by the fact that small amplitude solutions of KdV are governed by NLS \citep{Dias:2005}, in which case the soliton assumes $\sech$-form.}. As a result, the mechanism of self-focusing existing in the plane soliton case, i.e. when the soliton amplitude change leads to a variation in its speed and hence self-focusing and instability \citep{Askaryan:1962,Kadomtsev:1982}, must be modified in the cylindrical case. Moreover, the soliton stretching in the transverse direction should counteract to any other possible self-focusing mechanisms leading to transverse instability. Hence, the question arises if cylindrical solitons can experience a transverse instability.

Besides that, there is yet another crucial difference between plane and cylindrical geometries -- the single-soliton solutions in the latter case \citep{Maxon:1974b} no longer have exponential decay both in front and behind the soliton, but instead possess a slowly decaying oscillatory tail, i.e. there exists no localized soliton in the cylindrical case which makes the theory more difficult \citep{Freeman:1980}; this motivated one to name the corresponding solutions as `nonlocal' solitons \citep{Boyd:1988}, though the governing equations are local and the semantics of the term ``soliton'' is a subject of recurrent contemplation \citep{Infeld:2000}. While this fact of oscillatory tails in solitons is well-known in the context of KdV \citep{Ablowitz:1977,Johnson:1980}, it is less so for the NLS case. To illustrate this point, note that in the case of a \textit{radial NLS}, i.e. when $\psi_{\xi\xi} \pm \alpha \psi_{YY} \rightarrow \partial_{r}^{2} + \frac{1}{r} \partial_{r}$ in \eqref{eqn:NLS-1D}, one can still reduce \eqref{eqn:NLS-1D} to an equation of the type \eqref{base-state:1D-NLS:reduced}. Indeed, looking for a solution in the form $\psi = e^{\i \omega \tau} v(r)$, multiplying \eqref{eqn:NLS-1D} by $v_{r}$, and integrating w.r.t. the cylindrical measure $r \d r$, instead of \eqref{base-state:1D-NLS:reduced} we get $- \omega \, v^{2} - v^{\prime 2} + \frac{1}{2} v^{4} = 0$, provided the term arising from integration by parts vanishes, $\left.r v^{\prime 2}\right|_{0}^{\infty} = 0$. As a result, instead of \eqref{base-state:1D-NLS:reduced} we obtain
\begin{align}
\label{base-state:radial-NLS:reduced}
w^{\prime 2} = - \omega \, w^{2} + \frac{1}{2},
\end{align}
where the difference in signs from \eqref{base-state:1D-NLS:reduced} is notable. The resulting general solution is either constant everywhere, $v(r) = \pm (2 \, \omega)^{1/2}$ -- the extreme case of nonlocalized soliton -- or $v(r) = (2 \, \omega)^{1/2} \sec{\left(\omega^{1/2} r + \varphi\right)}$ with arbitrate phase $\varphi$; the latter solution does not satisfy the condition $\left.r v^{\prime 2}\right|_{r=\infty} = 0$ necessary to arrive at \eqref{base-state:radial-NLS:reduced} and is singular periodically. This demonstrates the lack of localization in cylindrical geometry characteristic to the plane 1D case. One implication of that is the fact that an attempt to apply the Vakhitov-Kolokolov stability approach for plane (1D) solitons mentioned above in the cylindrical case fails not only because the power curve $P(\mu)$ diverges, but also because $P(\mu)$ does not depend on the propagation constant as follows from a simple scaling argument.

With this introduction to a range of general questions, the outline of the manuscript is as follows. Following the derivations of the governing equations in the deep (\S \ref{subsec:ncNLS}) and shallow (\S \ref{subsec:ncKdV}) water limits, we will discuss the origin and implications of the potential term in the NLS model (\S \ref{subsec:heuristic-analysis}). In \S\S \ref{subsec:ncNLS:BS} and \ref{subsec:soliton:ncKdV} we will construct the ground state solitary waves for deep and shallow water, respectively. Since the envelope equation derived in the deep water case -- the Gross-Pitaevskii (GP) equation with a potential -- is new, its properties and key base state solutions will be studied in detail, including with the help of dynamical systems tools in order to get a better insight into their structure. Stability of these solutions will be studied in \S \ref{subsec:spectral-analysis:ncNLS},\ref{subsec:Lagrange-Dirichlet} and \S\S \ref{subsec:KdV:stability-preliminary},\ref{subsec:KP:ncKdV:analysis}. In the case of GP equation the stability analysis will be done from both spectral (\S \ref{subsec:spectral-analysis:ncNLS}) and nonlinear Hamiltonian (\S \ref{subsec:Lagrange-Dirichlet}) perspectives, while in the case of ncKdV the general considerations in \S \ref{subsec:KdV:stability-preliminary} will be followed in \S \ref{subsec:KP:ncKdV:analysis} with the derivation of the linear amplitude equation governing instability in the spirit of \cite{Kadomtsev:1970} along with its analysis. Finally, while conservation laws will be constructed and discussed for both GP (\S \ref{subsec:GP:conservation-laws}) and ncKdV (\S \ref{subsec:soliton:ncKdV}) equations, in the former case the condition for self-focusing and singularity formation will be identified in analogy to that of the standard NLS equation.

\section{Waves on deep water} \label{sec:deep-water}

\subsection{Derivation of the envelope equation} \label{subsec:ncNLS}

Let us first consider concentric water waves on deep water in the inviscid potential approximation, for which it is natural to adopt a cylindrical system of coordinates. The corresponding non-dimensional system for the velocity potential $\phi$ and interfacial deflection $\eta$ from quiescent state coupled through kinematic and dynamic boundary conditions (BCs) reads
\begin{subequations}
\label{system:deep-water:non-dimensional:cylindrical}
\begin{align}
\label{bulk:Laplace:non-dimensional:cylindrical}
z \le \varepsilon \, \eta(t,x)&: \quad
\left\{\begin{array}{c} \Delta \phi \equiv \frac{1}{r} \frac{\partial}{\partial r}\left(r \frac{\partial \phi}{\partial r}\right) + \frac{1}{r^{2}} \frac{\partial^{2} \phi}{\partial \theta^{2}} + \frac{\partial^{2} \phi}{\partial z^{2}} = 0, \\
\nabla \phi \rightarrow 0, \ z \rightarrow - \infty,
\end{array}\right. \\
\label{interface:kinematic:non-dimensional:cylindrical} z = \varepsilon \, \eta(t,x)&: \quad
\phi_{z} = \eta_{t} + \varepsilon \, \nabla_{\perp} \phi \cdot \nabla_{\perp} \eta, \\
\label{interface:dynamic:non-dimensional:cylindrical} z = \varepsilon \, \eta(t,x)&: \quad
\phi_{t} + \eta + \frac{\varepsilon}{2} \left|\nabla \phi\right|^{2} + We \, \nabla \cdot \mathbf{n} = 0,
\end{align}
\end{subequations}
where $\nabla_{\perp} = (\partial_{r}, r^{-1} \partial_{\theta})$ and the scaled interfacial curvature
\begin{align}
\label{eqn:curvature:cylindrical}
\nabla \cdot \mathbf{n} &= - \frac{\eta_{rr} \left(1 + \varepsilon^{2} \frac{\eta_{\theta}^{2}}{r^{2}}\right) + (1 + \varepsilon^{2} \eta_{r}^{2}) \frac{1}{r} \left(\frac{\eta_{\theta\theta}}{r} + \eta_{r}\right) + 2 \varepsilon^{2} \frac{\eta_{r}\eta_{\theta}}{r^{2}} \left(\frac{\eta_{\theta}}{r} - \eta_{\theta r}\right)}{\left(1 + \varepsilon^{2} \eta_{r}^{2} + \varepsilon^{2} \eta_{\theta}^{2}/r^{2}\right)^{3/2}}.
\end{align}
Above, the Weber number $We = \sigma k_{0}^{2} / (\rho \, g)$ measures the effect of surface tension relative the wave intertia (driven by gravity) and $\varepsilon = a k_{0}$ is the wave amplitude (wave steepness) scaled w.r.t. the wavenumber $k_{0}$ of the carrier wave. The non-dimensionalization that led to \eqref{system:deep-water:non-dimensional:cylindrical} is dictated by the following considerations. Since our interest is to analyze the evolution of a narrow wavepacket centered around a wavenumber $k_{0}$, the latter sets the natural lengthscale for non-dimensionalization:
\begin{align}
\label{eqn:non-dimensionalization}
(r,z) \rightarrow k_{0}^{-1} (r,z), \ t \rightarrow \omega_{0}^{-1} t, \ \eta \rightarrow a \, \eta, \ \phi \rightarrow a \, \omega_{0} \, k_{0}^{-1} \, \phi,
\end{align}
where $\omega_{0} = \omega(k_{0})$ is dictated by the deep water dispersion relation $\omega^{2} = g \, k$ for pure gravity-driven waves, $a$ is the wave amplitude, and the scaling for $\phi$ follows from balancing the fluid velocity at the interface with that of the interface itself, $\phi_{z} \sim \eta_{t}$.

The scaled wave amplitude $\varepsilon$ is treated here as small since we are interested in the balance of nonlinear and dispersive effects, which happens at small solution amplitudes only. Because of the latter, we expand the kinematic and dynamic BCs (\ref{interface:kinematic:non-dimensional:cylindrical},\ref{interface:dynamic:non-dimensional:cylindrical}) in Taylor series around $z=0$, $f(z=\varepsilon\eta) = \left.f(0) + f^{\prime}(0) z + f^{\prime\prime}(0) \frac{z^{2}}{2} + \ldots\right|_{z=\varepsilon\eta}$ thereby making the spatial domain to be the perfect half-space, as well as look for solutions in the series
\begin{align}
\label{expansion:NLS}
\phi = \phi_{0} + \varepsilon \, \phi_{1} + \varepsilon^{2} \, \phi_{2} + \ldots, \
\eta = \eta_{0} + \varepsilon \, \eta_{1} + \varepsilon^{2} \, \eta_{2} + \ldots.
\end{align}
However, solving problem \eqref{system:deep-water:non-dimensional:cylindrical} with such a regular perturbation approach is known to lead to secular divergencies, which necessitates the introduction of multiple scales, cf. \cite{Hakim:1998} and \S \ref{subsec:heuristic-analysis}:
\begin{align}
\label{scales:multiple:NLS}
(t,x,z) \rightarrow (t,T = \varepsilon \, t, \tau = \varepsilon^{2} \, t; x, R = \varepsilon \, r; z, Z = \varepsilon \, z)
\end{align}
with the corresponding transformation of derivatives, i.e. $\partial_{t} \rightarrow \partial_{t} + \varepsilon \, \partial_{T} + \varepsilon^{2} \, \partial_{\tau}$, $\partial_{r} \rightarrow \partial_{r} + \varepsilon \, \partial_{R}$, and $\partial_{z} \rightarrow \partial_{z} + \varepsilon \, \partial_{Z}$.

The NLS proves to appear at the radii $r \sim \varepsilon^{-1}$, so we would have to consider the balance at the lengthscale $R = \varepsilon \, r$. As a result, at the leading order we get the system
\begin{equation}
\label{ncNLS:O-0}
\mathcal{O}(\epsilon^{0}): \begin{cases}
\phi_{0 zz} + \phi_{0 rr} = 0 & z < 0, \\
|\nabla\phi_{0}| < \infty & z \rightarrow -\infty, \\
\phi_{0 z} - \eta_{0 t} = 0 & z = 0, \\
\phi_{0 t} + \eta_{0} = 0 & z = 0,
\end{cases}
\end{equation}
shown here for $We = 0$ as our first goal is to illustrate the derivation in the simplest possible case and then to point out the differences in the derivation when surface tension effects are included. The solution of \eqref{ncNLS:O-0} is
\begin{subequations}
\label{sln:ncNLS:O-0}
\begin{align}
\phi_{0} &= \psi_{0}(T,\tau,R,Z,\theta) \, e^{\i (\widehat{k}_{0} r - \widehat{\omega}_{0} t) + \widehat{k}_{0} z} + \mathrm{c.c.}, \\
\eta_{0} &= \i \, \widehat{\omega}_{0} \, \psi_{0}(T,\tau,R,0,\theta) \, e^{\i (\widehat{k}_{0} r - \widehat{\omega}_{0} t)} + \mathrm{c.c.},
\end{align}
\end{subequations}
where $\widehat{k}_{0}$ and $\widehat{\omega}_{0} = \widehat{k}_{0}^{1/2}$ are equal to one due to our choice of non-dimensionalization \eqref{eqn:non-dimensionalization}, but are kept here explicitly for now, which will be useful when we discuss the case $We \neq 0$ since the dispersion relation $\omega(k)$ will be different. Notably, at $\mathcal{O}(\epsilon^{0})$ the problem is identical to the plane (1D) case. At the next order, however, we start observing some differences
\begin{equation}
\label{ncNLS:O-1}
\mathcal{O}(\epsilon^{1}): \begin{cases}
\phi_{1 zz} + \phi_{1 rr} = - 2 \left(\phi_{0 zZ} + \phi_{0 rR}\right) - \frac{\phi_{0 r}}{R} & z < 0, \\
|\nabla\phi_{1}| < \infty & z \rightarrow -\infty, \\
\phi_{1 z} + \phi_{1 tt} = - 2 \phi_{0 t T} - \phi_{0 Z} - \eta_{0 t} \phi_{0 z t} - \phi_{0 z} \phi_{0 z t} & \\
\qquad - \eta_{0} \left(\phi_{0 z tt} + \phi_{0 zz}\right) + \eta_{0 r} \phi_{0 r} - \phi_{0 r} \phi_{0 r t} & z = 0, \\
\phi_{1 t} + \eta_{1} = - \phi_{0 T} - \frac{1}{2} \left(\phi_{0 z}^{2} + \phi_{0 r}^{2}\right) - \eta_{0} \phi_{0 z t} & z = 0,
\end{cases}
\end{equation}
where instead of the kinematic condition we provided a \textit{combined} one constructed by adding the dynamic condition \eqref{interface:dynamic:non-dimensional:cylindrical}, differentiated with respect to time $t$, to the kinematic condition \eqref{interface:kinematic:non-dimensional:cylindrical}, and subsequently applying the multiple-scales expansion outlined earlier; the use of the combined boundary condition makes it easier to identify the resonances compared to dealing with the system of kinematic and dynamic conditions. The entire right-hand side of the Poisson equation in \eqref{ncNLS:O-1} leads to secular terms containing exponents $e^{\pm \i (\widehat{k}_{0} r - \widehat{\omega}_{0} t)}$, the factors of which vanish provided that the no-resonance condition holds
\begin{align}
\label{conditions:no-resonance:Laplace-1:cylindrical}
\psi_{0 Z}(T,\tau,R,Z,\theta) = - \i \, \left(\psi_{0 R}(T,\tau,R,Z,\theta) + \frac{\psi_{0}(T,\tau,R,Z,\theta)}{2R}\right),
\end{align}
along with the complex conjugate of this expression; both render the Poisson equation in \eqref{ncNLS:O-1} to be homogeneous. For future simplifications, differential consequences of \eqref{conditions:no-resonance:Laplace-1:cylindrical} will be needed:
\begin{align}
\psi_{0 ZZ}(T,\tau,R,Z,\theta) = \frac{\psi_{0}(T,\tau,R,Z,\theta)}{4 R^{2}} - \frac{\psi_{0 R}(T,\tau,R,Z,\theta)}{R} - \psi_{0 RR}(T,\tau,R,Z,\theta).
\end{align}

Similarly, the right-hand side of the combined boundary condition in \eqref{ncNLS:O-1}, after simplification with \eqref{conditions:no-resonance:Laplace-1:cylindrical} evaluated at $Z=0$, leads to the following conditions necessary for avoiding secularities:
\begin{align}
\label{conditions:no-resonance:combinedBC-1:cylindrical}
\psi_{0 T}(T,\tau,R,0,\theta) + \frac{\psi_{0 R}(T,\tau,R,0,\theta)}{2 \widehat{k}_{0}^{1/2}} + \frac{\psi_{0}(T,\tau,R,0,\theta)}{4 R \widehat{k}_{0}^{1/2}} = 0
\end{align}
along with its complex conjugate and the differential consequence
\begin{align}
\psi_{0 TT}(T,\tau,R,0,\theta) = - \frac{\psi_{0}(T,\tau,R,0,\theta)}{16 R^{2} \widehat{k}_{0}} + \frac{\psi_{0 R}(T,\tau,R,0,\theta)}{4 R \widehat{k}_{0}} + \frac{\psi_{0 RR}(T,\tau,R,0,\theta)}{4 \widehat{k}_{0}}.
\end{align}
Integration of \eqref{conditions:no-resonance:combinedBC-1:cylindrical} with initial conditions (ICs) $\psi_{0}(0)=\psi_{0}\left(0\right)$ at $T=0$ and $R(0)$ using the method of characteristics gives $\psi_{0} = (R(0)/R) \, \psi_{0}(0)$, $R = R(0) + T$ and shows that the first two terms in \eqref{conditions:no-resonance:combinedBC-1:cylindrical} represent advection, i.e. the envelope traveling at the group velocity $\widehat{\omega}_{0}^{\prime}(\widehat{k}_{0}) = \frac{1}{2} \widehat{k}_{0}^{-1/2}$, and the last one -- dilution affecting the amplitude of the wavepacket, i.e. decreasing it with $R$ as $\psi_{0} \sim R^{-1}$ on the timescale $T$; as we will see, on the timescale $\tau$ the amplitude varies as $\psi_{0} \sim R^{-1/2}$. The condition \eqref{conditions:no-resonance:combinedBC-1:cylindrical} nullifies the inhomogeneous terms in the combined BC and results in the following solution for $\phi_{1}$:
\begin{align}
\label{sln:ncNLS:O-1}
\phi_{1} &= \psi_{1}(T,\tau,R,Z,\theta) \, e^{\i (\widehat{k}_{0} r - \widehat{\omega}_{0} t) + \widehat{k}_{0} z} + \mathrm{c.c.},
\end{align}
while $\eta_{1}$ is found straightforwardly from the dynamic condition in \eqref{ncNLS:O-1}. Finally, the Laplace equation at the order required for balancing the nonlinearity and dispersion reads
\begin{equation}
\label{NLS:O-2}
\mathcal{O}(\epsilon^{2}):
\begin{cases}
\phi_{2 zz} + \phi_{2 rr} = - 2 \left(\phi_{1 zZ} + \phi_{1 rR}\right) - \left(\phi_{0 ZZ} + \phi_{0 RR}\right) & \\
\qquad\qquad\qquad- \frac{\phi_{1 r}}{R} - \frac{\phi_{0 R}}{R} - \frac{\phi_{0 \theta\theta}}{R} & z < 0, \\
|\nabla\phi_{2}| < \infty & z \rightarrow -\infty,
\end{cases}
\end{equation}
which brings about the no-resonance conditions
\begin{multline}
\label{conditions:no-resonance:Laplace-2:cylindrical}
\psi_{1 Z}(T,\tau,R,Z,\theta) = - \frac{\psi_{0}(T,\tau,R,Z,\theta) + 4 \, \psi_{0 \theta\theta}(T,\tau,R,Z,\theta)}{8 R^{2} \widehat{k}_{0}} \\
- \i \, \frac{\psi_{1}(T,\tau,R,Z,\theta) - 2 R \psi_{1 R}(T,\tau,R,Z,\theta)}{2 R}.
\end{multline}
The corresponding combined boundary condition at $\mathcal{O}(\epsilon^{2})$ (not shown due to excessive number of terms), simplified with the conditions (\ref{conditions:no-resonance:Laplace-1:cylindrical},\ref{conditions:no-resonance:combinedBC-1:cylindrical},\ref{conditions:no-resonance:Laplace-2:cylindrical})\footnote{The perturbation $\psi_{1}$ must obey the same condition \eqref{conditions:no-resonance:combinedBC-1:cylindrical} as $\psi_{0}$ since, for coherence of the envelope, both perturbations $\psi_{0}$ and $\psi_{1}$ must travel at the same group velocity.} and their differential consequences, leads to the no-resonance condition in the form of ncNLS amended with an inverse-square potential:
\begin{align}
\label{ncNLS}
- 2 \, \i \, \widehat{k}_{0}^{1/2} \psi_{0 \tau} + \frac{1}{4 \, \widehat{k}_{0}} \left[\psi_{0 RR} + \frac{\psi_{0 R}}{R} - \frac{3}{4} \frac{\psi_{0}}{R^{2}}\right] - \frac{1}{2 R^{2} \widehat{k}_{0}} \psi_{0 \theta\theta} + 4 \, \widehat{k}_{0}^{4} |\psi_{0}|^{2} \psi_{0} = 0,
\end{align}
which in the limit $R \rightarrow \infty$, obviously, reduces to the 1D NLS if the dependence on the transverse coordinate is neglected or to the 2D NLS derived by \citet{Zakharov:1968} in the Cartesian coordinates if one lets $R \, \theta \rightarrow y$ (and $r \rightarrow x$).

Inclusion of surface tension ($We > 0$) brings about several key differences and the associated algebraic complications. First, the dynamic condition in \eqref{ncNLS:O-0} is amended with the leading order curvature terms $\nabla \cdot \mathbf{n} \approx - \eta_{rr} \left(1 - \frac{3}{2} \varepsilon^{2} \eta_{r}^{2}\right) + \mathcal{O}(\varepsilon^{3})$ contributing to the resulting envelope equation, so that the frequency $\widehat{\omega}_{0}$ in \eqref{sln:ncNLS:O-0} modifies to $\widehat{\omega}_{0}^{2} = \widehat{k}_{0} \left(1 + We \, \widehat{k}_{0}^{2}\right)$, where $\widehat{k}_{0}=1$ as in the case $We = 0$. Naturally, the derivation of the combined boundary condition requires not only substitution of $\eta_{t}$ from the kinematic condition \eqref{interface:kinematic:non-dimensional:cylindrical}, but also calculating $\eta_{t}$, $\eta_{rt}$, $\eta_{rrt}$, $\eta_{\theta t}$, $\eta_{\theta\theta t}$ and $\eta_{r \theta t}$ from the kinematic condition \eqref{interface:kinematic:non-dimensional:cylindrical} to substitute them in the differentiated dynamic condition \eqref{interface:dynamic:non-dimensional:cylindrical}. While the condition \eqref{conditions:no-resonance:Laplace-1:cylindrical} and its differential consequences are not affected by surface tension, equation \eqref{conditions:no-resonance:combinedBC-1:cylindrical} now reads
\begin{align}
\label{condition:no-resonance:ST}
\psi_{0 T}(T,\tau,R,0,\theta) = - \left(1 + 3 We \, \widehat{k}_{0}^{2}\right) \left[\frac{\psi_{0 R}(T,\tau,R,0,\theta)}{2 \, \widehat{\omega}_{0}} + \frac{\psi_{0}(T,\tau,R,0,\theta)}{4 R \widehat{\omega}_{0}}\right],
\end{align}
and still retains the meaning that the envelope of the wavepacket (and its complex conjugate $\psi_{0}^{*}$) propagates at the group velocity; differential consequences \eqref{condition:no-resonance:ST} are computed similar to the no-surface tension case above. Next, as opposed to \eqref{sln:ncNLS:O-1} the solution for $\phi_{1}$ now contains the inhomogeneous part leading to
\begin{multline}
\phi_{1} = \psi_{1}(T,\tau,R,Z,\theta) \, e^{\i (\widehat{k}_{0} r - \widehat{\omega}_{0} t) + \widehat{k}_{0} z} \\
+ \frac{3 \, \i \, We \, \widehat{k}_{0}^{4} \, \psi_{0}^{2}(T,\tau,R,Z,\theta)}{\left(1 - 2 \, We \, \widehat{k}_{0}^{2}\right) \widehat{\omega}_{0}} \, e^{2 \, \i (\widehat{k}_{0} r - \widehat{\omega}_{0} t) + 2 \, \widehat{k}_{0} z} + \mathrm{c.c.}.
\end{multline}
The no-resonance condition \eqref{conditions:no-resonance:Laplace-2:cylindrical} arising at $\mathcal{O}(\varepsilon^{2})$ stays intact. As a result, the envelope equation in the presence of surface tension now generalizes from \eqref{ncNLS} to
\begin{multline}
\label{ncNLS-ST}
- 2 \, \i \, \widehat{\omega}_{0} \psi_{0 \tau} + \frac{1 - 6 \, We \, \widehat{k}_{0}^{2} - 3 \, We^{2} \, \widehat{k}_{0}^{4}}{4 \, \widehat{\omega}_{0}^{2}} \left(\psi_{0 RR} + \frac{\psi_{0 R}}{R}\right) - \frac{3 + We \, \widehat{k}_{0}^{2} \left(2 + 3 \, We \, \widehat{k}_{0}^{2}\right)}{16 \, R^{2} \, \widehat{\omega}_{0}^{2}} \psi_{0} \\
- \frac{1 + 3 \, We \, \widehat{k}_{0}^{2}}{2 \, R^{2} \, \widehat{k}_{0}} \psi_{0 \theta\theta} + \frac{\widehat{k}_{0}^{5} \left(8 + We \, \widehat{k}_{0}^{2} + 2 \, We^{2} \, \widehat{k}_{0}^{4}\right)}{2 \, \widehat{\omega}_{0}^{2} (1 - 2 \, We \, \widehat{k}_{0}^{2})} |\psi_{0}|^{2} \psi_{0} = 0,
\end{multline}
which in the limit $R \rightarrow \infty$ reduces to the nearly plane NLS \citep{Kawahara:1975,Djordjevic:1977,Ablowitz:1979}. Adopting the notation for the coefficients in the NLS from the latter reference, \eqref{ncNLS-ST} can be compactly rewritten as
\begin{align}
\label{ncNLS-ST:abstract}
\i \, \psi_{\tau} + \lambda_{\infty} \Delta_{R} \psi_{R} + \lambda_{\infty}^{\prime} \frac{\psi}{R^{2}} + \frac{\mu_{\infty}}{R^{2}} \psi_{\theta\theta} = \chi_{\infty} |\psi|^{2} \psi,
\end{align}
where we dropped index $0$ and introduced the notation for the radial Laplacian $\Delta_{R} = \partial_{R}^{2} + \frac{1}{R}\partial_{R}$; the coefficients in \eqref{ncNLS-ST:abstract} are
\begin{subequations}
\label{coefficients:GP}
\begin{align}
\lambda_{\infty} &= - \frac{1 - 6 \, We \, \widehat{k}_{0}^{2} - 3 \, We^{2} \, \widehat{k}_{0}^{4}}{8 \, \widehat{\omega}_{0}^{3}} \mathop{\longrightarrow}_{We \rightarrow 0} - \frac{1}{8}, \\
\lambda_{\infty}^{\prime} &= \frac{3 + We \, \widehat{k}_{0}^{2} \left(2 + 3 \, We \, \widehat{k}_{0}^{2}\right)}{32 \, \widehat{\omega}_{0}^{3}} \mathop{\longrightarrow}_{We \rightarrow 0} \frac{3}{32}, \\
\mu_{\infty} &= \frac{1 + 3 \, We \, \widehat{k}_{0}^{2}}{4 \, \widehat{k}_{0} \, \widehat{\omega}_{0}} \mathop{\longrightarrow}_{We \rightarrow 0} \frac{1}{4}, \\
\chi_{\infty} &= \frac{\widehat{k}_{0}^{5} \left(8 + We \, \widehat{k}_{0}^{2} + 2 \, We^{2} \, \widehat{k}_{0}^{4}\right)}{4 \, \widehat{\omega}_{0}^{3} (1 - 2 \, We \, \widehat{k}_{0}^{2})} \mathop{\longrightarrow}_{We \rightarrow 0} 2,
\end{align}
\end{subequations}
where and in what follows we put $\widehat{k}_{0} = 1$ based on the non-dimensionalization \eqref{eqn:non-dimensionalization}. Once surface tension effects are introduced, $\chi_{\infty}$ changes sign from positive to negative at $We = \frac{1}{2}$, while $\lambda_{\infty}$ changes sign from negative to positive at $We = \frac{2}{\sqrt{3}} - 1$. The latter implies that the type of equation \eqref{ncNLS-ST:abstract} changes from hyperbolic to elliptic in accordance with the classification of its Cartesian counterpart \eqref{eqn:NLS-1D}. This will have certain consequences for the stability of solutions of \eqref{ncNLS-ST:abstract}, as we will see in \S \ref{subsec:spectral-analysis:ncNLS}.

\subsection{On the origin of the potential and its implications} \label{subsec:heuristic-analysis}

Since the work of \citet{Zakharov:1968}, where 2D NLS was derived, it has been tacitly assumed the Laplacian $\Delta$ stays intact when applies the NLS to axisymmetric case \citep{Zakharov:1976b,Jones:1988}; however, the principle of covariance (coordinate-independence) is applicable only to the fundamental physical laws such as Euler's equations of fluid motion, not amplitude equations deduced from them under concrete asymptotic assumptions despite their `universal' character.

To understand the origin of the potential term $V(R) \sim \frac{1}{R^{2}}$ in (\ref{ncNLS},\ref{ncNLS-ST:abstract}), let us perform a heuristic derivation of the linear part of the envelope equation in the case of pure gravity-driven waves. To bring in more physical intuition, let us consider the linear part of \eqref{system:deep-water:non-dimensional:cylindrical} back in dimensional variables:
\begin{subequations}
\label{system:deep-water:dimensional:cylindrical}
\begin{align}
\label{bulk:Laplace:dimensional:cylindrical}
z \le \varepsilon \, \eta(t,x)&: \quad
\left\{\begin{array}{c} \Delta \phi \equiv \frac{1}{r} \frac{\partial}{\partial r}\left(r \frac{\partial \phi}{\partial r}\right) + \frac{1}{r^{2}} \frac{\partial^{2} \phi}{\partial \theta^{2}} + \frac{\partial^{2} \phi}{\partial z^{2}} = 0, \\
\nabla \phi \rightarrow 0, \ z \rightarrow - \infty,
\end{array}\right. \\
\label{interface:kinematic:dimensional:cylindrical} z = \varepsilon \, \eta(t,x)&: \quad
\phi_{z} = \eta_{t}, \\
\label{interface:dynamic:non-dimensional:cylindrical} z = \varepsilon \, \eta(t,x)&: \quad
\phi_{t} + g \eta = 0,
\end{align}
\end{subequations}
the straightforward analysis of which in the axisymmetric case leads to the following form of the solution for the free surface elevation:
\begin{align}
\eta(t,r) = \int_{0}^{\infty}{\widehat{\eta}_{0}(k) J_{0}(k r) e^{-\i \omega(k) t} \, k \d k} + \mathrm{c.c.},
\end{align}
where $\widehat{\eta}_{0}(k)$ is the Hankel transform of the initial free surface deflection $\eta_{0}(r)$. The asymptotic expansion of this expression away from the origin, $k r \gg 1$, and in the form of a narrow wavepacket $|\delta k| = |k - k_{0}| \ll k_{0}$ near some fixed wavenumber $k_{0}$ yields
\begin{multline}
\label{free-surface:deep-water:asymptotics}
\eta(t,r) = e^{\i (k_{0} r - \omega_{0} t)} \frac{\varepsilon^{1/2}}{R^{1/2}} \int_{-\infty}^{\infty} \widehat{\eta}_{0}(k_{0},\kappa) \bigg[e^{\i \left(\kappa R - \omega^{\prime}(k_{0}) T - \frac{\omega^{\prime\prime}(k_{0})}{2} \kappa^{2} \tau - \frac{\pi}{4}\right)} \\
+ \mathcal{O}\left(\frac{\varepsilon}{R}\right)\bigg] \, \kappa^{1/2} \d \kappa + \mathrm{c.c.},
\end{multline}
that is $\eta(t,r)$ is a traveling wave $e^{\i (k_{0} r - \omega_{0} t)}$ modulated with an envelope function
\begin{align}
\label{envelope:NLS:heuristic}
\psi_{0}(T,\tau,R) \sim \int_{-\infty}^{\infty}{\widehat{\eta}_{0}(k_{0},\kappa) e^{\i \left(\kappa R - \omega^{\prime}(k_{0}) T - \frac{\omega^{\prime\prime}(k_{0})}{2} \kappa^{2} \tau - \frac{\pi}{4}\right)} \, \kappa^{1/2} \d \kappa}
\end{align}
evolving on slow time $T = \varepsilon t$, $\tau = \varepsilon^{2} t$ and spatial $R = \varepsilon r$ scales, which naturally appear in this narrow wavepacket approximation $\kappa = \delta k/\varepsilon$. Taking the derivatives of \eqref{envelope:NLS:heuristic}, we get the following factors for the integrand in \eqref{envelope:NLS:heuristic}:
\begin{align}
\begin{split}
\psi_{0T} \sim - \frac{\i \kappa \omega^{\prime}(k_{0})}{R^{1/2}}, \ &\psi_{0\tau} \sim - \frac{\i \kappa^{2} \omega^{\prime\prime}(k_{0})}{2 R^{1/2}}, \\
\psi_{0R} \sim - \frac{1}{2} R^{-3/2} + \frac{\i \kappa}{R^{1/2}}, \ \psi_{0RR} &\sim \frac{3}{2} R^{-5/2} - \i \kappa R^{-3/2} - \kappa^{2} R^{-1/2},
\end{split}
\end{align}
where we omitted the sign of integration for brevity, and immediately find that
\begin{align}
\psi_{0T} + \frac{\omega^{\prime}(k_{0})}{2 R} \psi_{0} + \omega^{\prime}(k_{0}) \psi_{0R} = 0,
\end{align}
which is the no-resonance condition \eqref{conditions:no-resonance:combinedBC-1:cylindrical} identified above in the course of the formal analysis, as well as
\begin{align}
\label{eqn:SE:cylindrical}
\psi_{0\tau} - \frac{\i \, \omega^{\prime\prime}(k_{0})}{2} \left(\psi_{0RR} + \frac{1}{R} \psi_{0R} - \frac{1}{4 R^{2}} \psi_{0}\right) = 0,
\end{align}
which is almost the same as the linear part of \eqref{ncNLS} except for the coefficient in front of the potential, i.e. $-3/4$ vs $-1/4$. Notably, with the transformation $\psi_{0} = R^{-1/2} \widetilde{\psi}_{0}(\tau,R)$ the above equation reduces to the 1D Schrodinger equation
\begin{align}
\widetilde{\psi}_{0\tau} - \frac{\i\omega^{\prime\prime}(k_{0})}{2} \widetilde{\psi}_{0RR} = 0,
\end{align}
i.e. the effect the potential $- \frac{1}{4 R^{2}} \psi_{0}$ plays in \eqref{eqn:SE:cylindrical} is to modify the amplitude of the wave as it travels either to or from the origin; this, in turn, explains the appearance of the potential in our system -- without it the wave would travel as a ``free particle'' with unmodified amplitude.

A salient feature of the above heuristic derivation was the assumption that the wavepacket changes its width in the same fashion as in the 1D case. This is evident from the approximation \eqref{free-surface:deep-water:asymptotics}, which is valid only in the limit $k r \rightarrow \infty$. However, as the behavior of the Bessel function $J_{0}(k r)$ entails for large, but finite $k r$, the speed of propagation changes as one gets closer to the origin: this effect leads to the more severe change in the wavepacket width and, in fact, when the corresponding wavelength $\lambda = 2 \pi / k$ becomes shorter than the distance $r$ from the origin, is responsible for the formation of a singularity in the form of a spike jet. Therefore, in order to account for a stronger wavepacket width change, the potential must be modified from that of $- \frac{1}{4 R^{2}} \psi_{0}$, and, as we saw from the formal derivation in \S \ref{subsec:ncNLS}, the potential indeed becomes stronger (through a modified factor), in the sense that it will lead to a stronger singularity of the solution near the origin compared to $\sim R^{-1/2}$ in \eqref{eqn:SE:cylindrical} as we will see in \S \ref{subsec:ncNLS:BS}.

The resulting envelope equations (\ref{ncNLS},\ref{ncNLS-ST:abstract}) arise from a balance between nonlinearity and dispersion of the wavepacket, which occurs only at some distance from the origin as the wave amplitude varies with it -- this is a crucial difference from the translationally invariant case when one can take the limit of small amplitude solutions and be left with the same linear part; in the case of cylindrical waves this is no longer the case, i.e. the linear part of \eqref{ncNLS}, when nonlinearity and dispersion are balanced, does not correspond to \eqref{eqn:SE:cylindrical}, when nonlinearity is absent. Notably, for both potentials $- \frac{1}{4 R^{2}} \psi_{0}$ and $- \frac{3}{4 R^{2}} \psi_{0}$ the wave amplitude drops as $R^{-1/2}$, but the behavior near the origin proves to be different (\S \ref{subsec:ncNLS:BS}). Finally, the technical reason for the appearance of the $- \frac{3}{4 R^{2}} \psi_{0}$ potential instead of $- \frac{1}{4 R^{2}} \psi_{0}$ is due to the first term in the second-order no-resonance condition \eqref{conditions:no-resonance:Laplace-2:cylindrical}, which entangles both $\psi_{1}$ and $\psi_{0}$ -- this effect is absent in the plane (1D and 2D) cases. In any case, the appearance of an inverse-square potential is a generic property of cylindrical envelope wave equations; for example, a derivation of NLS from Maxwell's equations in nonlinear options gives the factor $-1$ at the inverse-square potential. As we saw from \eqref{ncNLS-ST:abstract}, in the case of waves on deep water this factor changes with surface tension as $\lambda_{\infty}^{\prime}/\lambda_{\infty}$.

Our NLS equations (\ref{ncNLS},\ref{ncNLS-ST:abstract}) with the inverse-square potential belong to the Gross-Pitaevskii type \citep{Gross:1961,Pitaevskii:1961}, originally derived to describe the ground state wavefunction of a quantum system composed of a Bose-Einstein condensate in an external potential and nonlinearity is responsible for the interaction between particles. The interested reader may find a mechanistic interpretation of equation \eqref{ncNLS-ST:abstract} in Appendix~\ref{appx:mechanistic-interpretation}. Notably, an inverse-square potential also arises, though not in the context of NLS, in the motion of a charged particle in the field of a stationary electric dipole, in quantum mechanics \citep{Case:1950,Kalf:1975,Reed:1979}, molecular physics \citep{Camblong:2001}, nuclear physics \citep{Beane:2001,Esteve:2002}, black holes \citep{Regge:1957,Zerilli:1970,Moncrief:1974,Strominger:1998,Claus:1998,Azcarraga:1999,Solodukhin:1999,Michelson:2000,Papadopoulos:2000,Bellucci:2002,Carlip:2002},
in wave propagation on conic manifolds \citep{Cheeger:1982}, and in the theory of combustion \citep{Bebernes:1989}.

Since in our case the potential is $V(R) = - \frac{3}{4 R^{2}}$ and the Laplacian are of equal strength, the former cannot be neglected and the GP equation retains the NLS scaling symmetry
\begin{align}
u(\tau,R) \mapsto \lambda u(\lambda^{2} \tau,\lambda R).
\end{align}
Because of that it is known to have some peculiar properties such as no ground state, i.e. there is no lower limit on the allowed energies \citep{Essin:2006} and symmetry breaking anomaly emerges in the process of renormalization \citep{Essin:2006,Camblong:2000,Coon:2002}. The spatial operator in \eqref{ncNLS} or, more generally, in \eqref{ncNLS-ST:abstract}:
\begin{align}
L = \Delta_{R} + \frac{\lambda_{\infty}^{\prime}}{\lambda_{\infty}} \frac{1}{R^{2}},
\end{align}
has as eigenfunctions $L \phi_{\lambda} = \lambda \phi_{\lambda}$ either modified Bessel function of real order $I_{\nu}\left(\lambda^{1/2} R\right)$, $K_{\nu}\left(\lambda^{1/2} R\right)$ with $\nu^{2} = \lambda_{\infty}^{\prime} / \lambda_{\infty} > 0$, which are unbounded at infinity and origin, respectively, or of imaginary order $I_{\i\nu}\left(\lambda^{1/2} R\right)$, $K_{\i\nu}\left(\lambda^{1/2} R\right)$ with $- \nu^{2} = \lambda_{\infty}^{\prime} / \lambda_{\infty} < 0$, which have highly oscillatory behavior with the period decreasing near the origin. As we will see in \S \ref{subsec:ncNLS:BS}, these observations will have certain implications for the structure of solutions of (\ref{ncNLS},\ref{ncNLS-ST:abstract}), which could be regular and singular.

Without the potential $V(R)$, the corresponding standard NLS is of critical type since the dimension of the problem is $d=2$, while the order of the nonlinearity $|\psi_{0}|^{2n}$ is $n=1$, so that $n \, d = 2$. This borderline case separates the subcritical NLS with $n \, d < 2$ when all solutions exist globally from the supercritical NLS with $n \, d > 0$, where singular solutions exist \citep{Fibich:2015}. Finally, while the standard defocusing NLS has a purely ``dispersive'' character, i.e. no solitary waves of the type
\begin{align}
\label{wave:Stokes}
\psi_{0}(\tau,R) = e^{\i \mu \tau} u(R)
\end{align}
exist and focusing NLS does have ground states \eqref{wave:Stokes} that are unstable leading to a finite-time blow-up, both focusing and defocusing GP have solutions of the form \eqref{wave:Stokes} as we will see in \S \ref{subsec:ncNLS:BS}. Singular solutions of GP equation are as valuable as the widely studied finite-time singularities peculiar to NLS \citep{Glassey:1977} -- such singularities are indicative of a localized behavior in the original unreduced physical system such as the Euler equations, from which \eqref{ncNLS-ST:abstract} is deduced.

Finally, as follows from the derivation in \S \ref{subsec:ncNLS}, the deduced GP equations (\ref{ncNLS},\ref{ncNLS-ST:abstract}) are valid only at asymptotically large distances $R = \mathcal{O}(1)$ from the origin. Hence, while the deduced Gross-Pitaevskii equation captures the singularity at the origin, which  is naturally expected at the origin as in the spike solutions \citep{McAllister::2022}, due to limitations its applicability in that region, one should not seek quantitative accuracy in describing the details of the corresponding singularities. Also, the symmetry of \eqref{ncNLS-ST:abstract} does not preclude from a possibility of ring-type singularities at a finite distance from the origin, where the GP equation is applicable, which will be shown in \S \ref{subsec:ncNLS:BS}. In this context it is worth pointing out that the extensive and controversial research on the rate at which the singularity is approached starting with \citet{Kelley:1965,Zakharov:1976b} (see also overview in \citet{Rypdal:1986} and \citet{Sulem:1999}) is flawed not only because it was unjustifiably assumed that the Laplacian in the 2D NLS deduced in the Cartesian coordinates stays intact when the NLS is applied to an axisymmetric case, but also because the NLS and GP equations in the axisymmetric case are applicable only at sufficiently large distances from the origin. The inapplicability of the NLS model near the blow-up where focusing levels are high (sometimes claimed \citep{Fibich:2015} necessary to be $\gg 10^{48}$ for the self-similar asymptotic rates to be valid) is also obvious as NLS was deduced only for sufficiently small, but finite, amplitudes allowing a balance with the dispersion effects, and the assumptions behind its derivation are no longer valid when the amplitude of the solution becomes incommensurate with the narrow wave-packet assumption.

\subsection{Conservation laws, variance, and finite-time singularity} \label{subsec:GP:conservation-laws}

To analyze the conservation laws of the GP equation \eqref{ncNLS-ST:abstract}, from physical considerations we supply the initial-value problem (IVP) for this equation with the BCs:
\begin{align}
\label{BCs:cNLS}
R=0: \ \psi_{R} = 0; \ R \rightarrow \infty: \ \psi \rightarrow 0.
\end{align}
Multiplying \eqref{ncNLS-ST:abstract} with $\overline{\psi} = \psi^{r} - \i \psi^{i}$,
\begin{align}
\label{eqn:cNLS-times-psiconj}
\i \, \overline{\psi} \, \psi_{\tau} + \lambda_{\infty} \overline{\psi} \, \Delta_{R} \psi + \frac{\lambda_{\infty}^{\prime}}{R^{2}} |\psi|^{2} + \frac{\mu_{\infty}}{R^{2}} \overline{\psi} \psi_{\theta\theta} - \chi_{\infty} \, |\psi|^{4} = 0,
\end{align}
and taking the imaginary part, we get
\begin{align}
\label{eqn:cNLS:psi-squared}
\frac{\d}{\d \tau}|\psi|^{2} + \lambda_{\infty} \left(\psi^{r} \Delta_{R} \psi^{i} - \psi^{i} \Delta_{R} \psi^{r}\right) + \frac{\mu_{\infty}}{R^{2}} \left(\psi^{r} \psi^{i}_{\theta\theta} - \psi^{i} \psi^{r}_{\theta\theta}\right) = 0,
\end{align}
where we took into account that $\overline{\psi} \, \psi_{\tau} = \psi^{r} \psi^{r}_{\tau} + \psi^{i} \psi^{i}_{\tau} + \i \left(\psi^{r} \psi^{i}_{\tau} - \psi^{i} \psi^{r}_{\tau}\right) = \frac{\d}{\d \tau}|\psi|^{2} + \i \left(\psi^{r} \psi^{i}_{\tau} - \psi^{i} \psi^{r}_{\tau}\right)$, $\overline{\psi} \, \Delta_{R} \psi = \psi^{r} \Delta_{R} \psi^{r} + \psi^{i} \Delta_{R} \psi^{i} + \i \left(\psi^{r} \Delta_{R} \psi^{i} - \psi^{i} \Delta_{R} \psi^{r}\right)$ and similar equalities for $\overline{\psi} \, \psi_{\theta\theta}$. Next, since the integral of the second term in \eqref{eqn:cNLS:psi-squared}:
\begin{align}
\label{eqn:cNSL:mass-derivation:1}
\begin{split}
&\int_{0}^{\infty}{\left[\psi^{r} \left(\psi^{i}_{RR} + \frac{1}{R}\psi^{i}_{R}\right) - \psi^{i} \left(\psi^{r}_{RR} + \frac{1}{R}\psi^{r}_{R}\right)\right] R \, \d R} \\
&= R \left[\psi^{r} \psi^{i}_{R} - \psi^{i} \psi^{r}_{R}\right]_{0}^{\infty} -
\int_{0}^{\infty}{\left[\psi^{i}_{R} \left(\psi^{r} R\right)_{R} - \psi^{r}_{R} \left(\psi^{i} R\right)_{R}\right] \, \d R} \\
&+ \int_{0}^{\infty}{\left[\psi^{r} \psi^{i}_{R} - \psi^{i} \psi^{r}_{R}\right] \, \d R} = 0
\end{split}
\end{align}
vanishes in view of the BCs \eqref{BCs:cNLS} as well as the integral of the last term  in \eqref{eqn:cNLS:psi-squared}:
\begin{align}
\int_{0}^{2\pi}{\left[\psi^{r} \psi^{i}_{\theta\theta} - \psi^{i} \psi^{r}_{\theta\theta}\right] \, \d \theta} = \left[\psi^{r} \psi^{i}_{\theta} - \psi^{i} \psi^{r}_{\theta}\right]_{0}^{2\pi} -
\int_{0}^{2\pi}{\left[\psi^{r}_{\theta} \psi^{i}_{\theta} - \psi^{i}_{\theta} \psi^{r}_{\theta}\right] \, \d \theta}  = 0
\end{align}
due to periodicity in $\theta$, equation \eqref{eqn:cNLS:psi-squared} leads to the conservation of the number of particles (in analogy to quantum mechanics):
\begin{align}
\label{conservation-mass:cNLS}
\frac{\d \mathcal{N}}{\d \tau} \equiv \frac{\d}{\d \tau}\int{|\psi|^{2} \, \d \nu} = 0,
\end{align}
which is the consequence of the invariance of \eqref{ncNLS-ST:abstract} under the phase-shift; the integration over the cylindrical measure $\d \nu$ is defined as
\begin{align}
\label{measure:cylindrical}
\int{\circ \, \d \nu} = \int_{0}^{2\pi}{\circ \, \d\theta}\int_{0}^{\infty}{x \d x}.
\end{align}

Similarly, multiplying \eqref{ncNLS-ST:abstract} with $\overline{\psi}_{\tau}$,
\begin{align}
\i \, |\psi_{\tau}|^{2} + \lambda_{\infty} \overline{\psi}_{\tau} \, \Delta_{R} \psi + \frac{\lambda_{\infty}^{\prime}}{R^{2}} \overline{\psi}_{\tau} \psi + \frac{\mu_{\infty}}{R^{2}} \overline{\psi}_{\tau} \psi_{\theta\theta} - \chi_{\infty} \, |\psi|^{2} \overline{\psi}_{\tau} \psi = 0,
\end{align}
and taking the real part of the resulting expression, we get
\begin{multline}
\label{eqn:cNLS:psi-tau-squared}
\lambda_{\infty} \left(\psi^{r}_{\tau} \Delta_{R} \psi^{r} + \psi^{i}_{\tau} \Delta_{R} \psi^{i}\right) + \frac{\lambda_{\infty}^{\prime}}{R^{2}} \frac{1}{2} \frac{\d}{\d \tau}|\psi|^{2} \\
+ \frac{\mu_{\infty}}{R^{2}} \left(\psi^{r}_{\tau} \psi^{r}_{\theta\theta} + \psi^{i}_{\tau} \psi^{i}_{\theta\theta}\right) - \chi_{\infty} |\psi|^{2} \frac{1}{2} \frac{\d}{\d \tau}|\psi|^{2} = 0,
\end{multline}
where we again took into account that $\psi^{r} \psi^{r}_{\tau} + \psi^{i} \psi^{i}_{\tau} = \frac{1}{2} \frac{\d}{\d \tau}|\psi|^{2} = \frac{1}{2} \frac{\d}{\d \tau}\left(\psi^{r2} + \psi^{2i}\right)$. Next, integrating by parts
\begin{multline}
\label{eqn:cNSL:energy-derivation:1}
\int_{0}^{\infty}{f_{\tau} \Delta_{R} f \, R \, \d R} = \int_{0}^{\infty}{f_{\tau} \left(f_{RR} + \frac{1}{R} f_{R}\right) \, R \, \d R} = \left.f_{R} f_{\tau} R\right|_{0}^{\infty} - \int_{0}^{\infty}{f_{R} \left(f_{\tau} \, R\right)_{R} \, \d R} \\
+ \int_{0}^{\infty}{f_{R} f_{\tau} \, \d R} = \left.f_{R} f_{\tau} R\right|_{0}^{\infty} - \frac{1}{2} \frac{\d}{\d \tau} \int_{0}^{\infty}{R f_{R}^{2} \, \d R},
\end{multline}
and applying this result to $f = \psi^{r}$ and $\psi^{i}$ with the BCs \eqref{BCs:cNLS}, equation \eqref{eqn:cNLS:psi-tau-squared} takes the form of the conservation of the Hamiltonian $\mathcal{H}$:
\begin{align}
\label{conservation-energy:cNLS}
\frac{\d \mathcal{H}}{\d \tau} \equiv \frac{\d}{\d \tau}\int{\left[-\frac{\lambda_{\infty}}{2}|\psi_{R}|^{2} + \frac{\lambda_{\infty}^{\prime}}{2 R^{2}}|\psi|^{2} - \frac{\mu_{\infty}}{2 R^{2}}|\psi_{\theta}|^{2} - \frac{\chi_{\infty}}{4} |\psi|^{4}\right] \, \d \nu} = 0;
\end{align}
here we simplified the last term in \eqref{eqn:cNLS:psi-tau-squared}, $|\psi|^{2} \frac{1}{2} \frac{\d}{\d \tau}|\psi|^{2} = \frac{1}{2} (\psi^{r2}+\psi^{i2}) \frac{\d}{\d \tau} (\psi^{r2}+\psi^{i2}) = \frac{1}{4}\frac{\d}{\d \tau} (\psi^{r2}+\psi^{i2})^{2}$, and also took into account that $\int_{0}^{2\pi}{f_{\tau} f_{\theta\theta} \, \d \theta} = \left.f_{\tau} f_{\theta}\right|_{\theta=0}^{2\pi} - \int_{0}^{2\pi}{f_{\tau\theta} f_{\theta} \, \d \theta} = - \frac{1}{2} \frac{\d}{\d\tau}\int_{0}^{2\pi}{f_{\theta}^{2} \, \d \theta}$ when integrating the third term in \eqref{eqn:cNLS:psi-tau-squared}, $\psi^{r}_{\tau} \psi^{r}_{\theta\theta} + \psi^{i}_{\tau} \psi^{i}_{\theta\theta}$. Hence, the Hamiltonian reads
\begin{align}
\label{H:ncNLS:original}
\mathcal{H} = \int{\left[-\frac{\lambda_{\infty}}{2}|\psi_{R}|^{2} + \frac{\lambda_{\infty}^{\prime}}{2 R^{2}}|\psi|^{2} - \frac{\mu_{\infty}}{2 R^{2}}|\psi_{\theta}|^{2} - \frac{\chi_{\infty}}{4} |\psi|^{4}\right] \, \d \nu}.
\end{align}

Finally, given the above expression for the Hamiltonian, it can be shown (cf. Appendix~\ref{appx:variance}) that the evolution of the variance $\mathcal{V}(\tau) = \int{R^{2} |\psi|^{2} \, \d \nu}$, also known as the wave power (a variant of the power curve introduced by \citet{Vakhitov:1973}), obeys
\begin{align}
\label{variance:derivative:second:final}
\frac{1}{4\lambda_{\infty}} \frac{\d^{2} \mathcal{V}}{\d \tau^{2}} = - 4 \, \mathcal{H} + 2 \pi \lambda_{\infty}^{\prime} |\psi(\tau,0)|^{2},
\end{align}
integrating which yields
\begin{align}
\mathcal{V}(\tau) = - 8 \mathcal{H} \tau^{2} + 8 \pi \lambda_{\infty}^{\prime} \int_{0}^{\tau}{\d \tau^{\prime}\int_{0}^{\tau^{\prime}}{|\psi(\tau^{\prime\prime},0)|^{2} \d \tau^{\prime\prime}}} + \mathcal{V}^{\prime}(0) \tau + \mathcal{V}(0).
\end{align}
Should $\lambda_{\infty}^{\prime}=0$ as in the case of the standard NLS, then, if the initial conditions are such that $\mathcal{H} > 0$, i.e. $\mathcal{V}^{\prime\prime}(0) = - 16 \mathcal{H} < 0$, from the solution of the quadratic equation
\begin{align}
\mathcal{V}^{\prime\prime}(0) \frac{\tau_{*}^{2}}{2} + \mathcal{V}^{\prime}(0) \tau_{*} + \mathcal{V}(0) = 0 \ \Rightarrow \ \tau_{*} = \frac{- \mathcal{V}^{\prime}(0) + \sqrt{\mathcal{V}^{\prime 2}(0) - 2 \mathcal{V}(0) \mathcal{V}^{\prime\prime}(0)}}{\mathcal{V}^{\prime\prime}(0)},
\end{align}
where necessarily $\mathcal{V}(0)>0$ and $\mathcal{V}^{\prime}(0)<0$, it follows that there exists a finite time $\tau_{*} > 0$ such that $\mathcal{V} \rightarrow 0$ in contradiction to its definition, which shows that it has to be positive. The $H^{1}$-solution must therefore develop a singularity no later than the time $\tau_{*}$, $|\psi| \rightarrow \infty$, $|\psi_{R}| \rightarrow \infty$ at $R \rightarrow 0$. This means that the solution gets out of the $H^{1}$-space, so that the condition of $\mathcal{V}$ being positive (when $\psi \in H^{1}$) does not need to be satisfied any longer. The analogous behavior is known for the standard NLS equations \citep{Glassey:1977}. However, the presence of the potential leads to an extra term $2 \pi \lambda_{\infty}^{\prime} |\psi(\tau,0)|^{2}$ in \eqref{variance:derivative:second:final}: if $\lambda_{\infty}^{\prime} < 0$ then, since the integral of $|\psi(\tau,0)|^{2}$ is positive-definite, the finite-time singularity still takes the place, while for $\lambda_{\infty}^{\prime} > 0$ the situation may potentially change and prevent the singularity from formation altogether, i.e. if the growth of the second term in \eqref{variance:derivative:second:final} with time is faster than $8 \mathcal{H} \tau^{2}$.

Note that in the above analysis, in particular in equations (\ref{eqn:cNSL:mass-derivation:1},\ref{eqn:cNSL:energy-derivation:1},\ref{eqn:cNSL:variance-derivation:1},\ref{eqn:cNSL:variance-derivation:2},\ref{eqn:cNSL:variance-derivation:3}),  we used the BC \eqref{BCs:cNLS} $\psi_{R}=0$ at $R=0$ and also naturally assumed that at $R=0$ the solution itself is non-singular so that the corresponding terms at $R=0$ vanish in equations (\ref{variance:derivative:second-prelim},\ref{eqn:cNSL:variance-derivation:4}). In all these equations we also assumed sufficiently fast decay of the solution as $R \rightarrow \infty$, which should be valid at least initially if the IC is chosen as a compact/localized perturbation of finite energy; however, at some time the solution at infinity may not decay fast enough to enable neglecting the boundary terms in the above referenced equations. As we will see in \ref{subsec:ncNLS:BS}, there is a class of solutions of the Stokes-type \eqref{wave:Stokes}, which indeed decay only as $R^{-1/2}$ at infinity, though with an oscillatory coefficient.

\subsection{Base states} \label{subsec:ncNLS:BS}

It is known that a truly solitary wave occurs only if the phase speed of the carrier wave coincides with the group velocity of the envelope which happens at a certain wavenumber \citep{Grimshaw:2007}, though, of course, even in the classical case of the KdV soliton \eqref{soliton:1D:KdV} its does not happen as it travels with amplitude-dependent speed relative to the carrier wave. While it may happen in the case of the GP equation \eqref{ncNLS-ST:abstract} at very large distances from the origin, $R \rightarrow \infty$, where its solution behaves as
\begin{align}
\label{wave:traveling:infinity}
\psi_{0}(\tau,R) \sim \frac{A_{0}}{\sqrt{R}} e^{\i (\mu \tau - k R)},
\end{align}
it does not happen everywhere in the cylindrical geometry we consider here, which is easy to see by appending \eqref{wave:traveling:infinity} with next order terms accounting for large, but finite, distances $R$:
\begin{align}
\label{wave:traveling:asymptotics}
\psi_{0}(\tau,R) \sim A(R) e^{\i (\mu t - \varphi(R))},
\end{align}
where
\begin{subequations}
\begin{align}
\varphi(R) &= R \left[k - \frac{A_{0}^{2} \, \chi_{\infty}}{2 \, k \, \lambda_{\infty}} \, \frac{\ln{R}}{R} + \mathcal{O}\left(\frac{1}{R^{2}}\right)\right], \\
A(R) &= \frac{A_{0}}{\sqrt{R}} \left[1 + \frac{A_{0}^{2} \, \chi_{\infty}}{4 \, k^{2} \lambda_{\infty}}\frac{1}{R} + \mathcal{O}\left(\frac{1}{R^{2}}\right)\right].
\end{align}
\end{subequations}
Therefore, as we can see from the expression for the phase $\varphi(R)$, the group velocity of the envelope is changing with the distance from the origin $R$, while the phase speed of the (linear) carrier wave does not. This implies that one cannot identify a single wavenumber $k$ at which those two speeds would match for all $R$.

Therefore, in this section we will focus on axisymmetric standing-wave ground states of \eqref{ncNLS-ST:abstract}, which are sought in the form \eqref{wave:Stokes} also known as a solitary wave (ground state or breather) in the context of NLS. Substituting \eqref{wave:Stokes} in \eqref{ncNLS-ST:abstract} we get
\begin{align}
\label{ncNLS-ST:ground-state}
- \mu \, u + \lambda_{\infty} \left(u_{RR} + \frac{1}{R} u_{R}\right) + \lambda_{\infty}^{\prime} \frac{u}{R^{2}} = \chi_{\infty} |u|^{2} u,
\end{align}
where we keep the modulus sign for the convenience of subsequent calculations, though all the base states we consider are real.
Equation \eqref{ncNLS-ST:ground-state} belongs to a semilinear elliptic type, which has been widely studied \citep{Berestycki:1983,Jones:1986,McLeod:1990,Bartsch:1993,Derrick:1997} and known to possess an infinite number of solutions. However, semilinear elliptic equations with singular and, in particular, inverse-square potentials are considerably less explored \citep{Lin:2019}.

\begin{figure}
	\centering
    \includegraphics[width=2.5in]{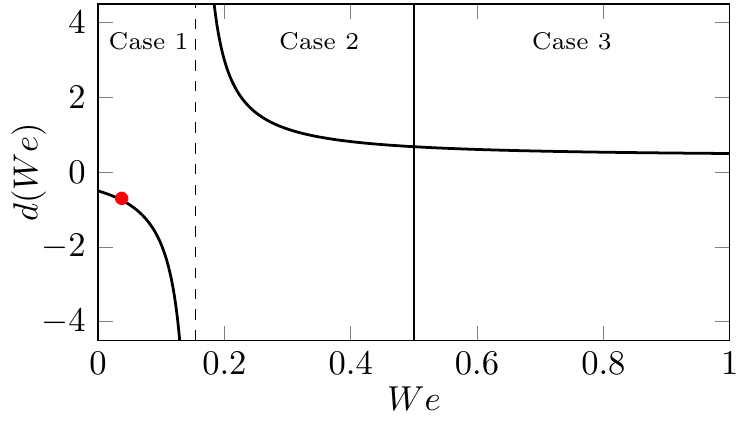}
	\caption{\label{fig:d-parameter} On the variation of the parameter $d$ with the Weber number; the point $d=-\frac{3}{4}$ corresponds to special asymptotics \eqref{asymptotics:special}.}
\end{figure}
Next applying the transformation $u(R) = R^{-1/2} U(R)$, which eliminates the first derivative w.r.t. $R$ in \eqref{ncNLS-ST:abstract} thereby removing the $R^{-1/2}$-factor in the asymptotics $R \rightarrow \infty$, we obtain
\begin{align}
\label{ncNLS-ST:abstract:U}
\lambda_{\infty} R^{2} U^{\prime\prime} + \left[\left(\frac{1}{4}\lambda_{\infty} + \lambda_{\infty}^{\prime}\right) - \mu R^{2}\right] U - \chi_{\infty} \, R \, |U|^{2} U = 0.
\end{align}
In order to bring it to a form convenient for analysis, let us scale variables according to $R = \alpha x$, $U = \beta y$, thus furnishing
\begin{align}
\label{ncNLS-ST:abstract:y:general}
x^{2} y^{\prime\prime} + \left[d - \frac{\mu \alpha^{2}}{\lambda_{\infty}} x^{2}\right] y - \frac{\chi_{\infty}}{\lambda_{\infty}} \alpha \beta^{2} \, x \, |y|^{2} y = 0, \ d \equiv \left(\frac{1}{4} + \frac{\lambda_{\infty}^{\prime}}{\lambda_{\infty}}\right).
\end{align}
Because of the change of signs of $\lambda_{\infty}$ and $\chi_{\infty}$, there are three ranges of Weber numbers to consider, cf. figure~\ref{fig:d-parameter}:
\begin{description}
  \item[Case 1, $0 \le We < \frac{2}{\sqrt{3}} - 1$: \ ] in which case $\lambda_{\infty}<0$, $\chi_{\infty}>0$, and $d<0$, so that we define $\alpha$ and $\beta$ via $\frac{\mu \alpha^{2}}{\lambda_{\infty}} = - 1$, $\frac{\chi_{\infty}}{\lambda_{\infty}} \alpha \beta^{2} = -1$ thus yielding $\alpha = \left(-\lambda_{\infty}/\mu\right)^{1/2}$ and $\beta = \left(-\lambda_{\infty}/\chi_{\infty}\right)^{1/2} \left(-\mu/\lambda_{\infty}\right)^{1/4}$ and reducing \eqref{ncNLS-ST:abstract:y:general} to
      \begin{align}
      \label{ncNLS-ST:abstract:y:case-1}
      y^{\prime\prime} + \left[\frac{d}{x^{2}} + 1\right] y + \frac{1}{x} \, |y|^{2} y = 0.
      \end{align}

  \item[Case 2, $\frac{2}{\sqrt{3}} - 1  < We < \frac{1}{2}$: \ ] in which case $\lambda_{\infty}>0$, $\chi_{\infty}>0$, and $d>0$, so that we define $\alpha$ and $\beta$ via $\frac{\mu \alpha^{2}}{\lambda_{\infty}} = 1$, $\frac{\chi_{\infty}}{\lambda_{\infty}} \alpha \beta^{2} = 1$ thus yielding $\alpha = \left(\lambda_{\infty}/\mu\right)^{1/2}$ and $\beta = \left(\lambda_{\infty}/\chi_{\infty}\right)^{1/2} \left(\mu/\lambda_{\infty}\right)^{1/4}$ and reducing \eqref{ncNLS-ST:abstract:y:general} to
      \begin{align}
      \label{ncNLS-ST:abstract:y:case-2}
      y^{\prime\prime} + \left[\frac{d}{x^{2}} - 1\right] y - \frac{1}{x} \, |y|^{2} y = 0.
      \end{align}

  \item[Case 3, $\frac{1}{2} < We$: \ ] in which case $\lambda_{\infty}>0$, $\chi_{\infty}<0$, and $d>0$, so that we define $\alpha$ and $\beta$ via $\frac{\mu \alpha^{2}}{\lambda_{\infty}} = 1$, $\frac{\chi_{\infty}}{\lambda_{\infty}} \alpha \beta^{2} = -1$ thus yielding $\alpha = \left(\lambda_{\infty}/\mu\right)^{1/2}$ and $\beta = \left(-\lambda_{\infty}/\chi_{\infty}\right)^{1/2} \left(\mu/\lambda_{\infty}\right)^{1/4}$ and reducing \eqref{ncNLS-ST:abstract:y:general} to
      \begin{align}
      \label{ncNLS-ST:abstract:y:case-3}
      y^{\prime\prime} + \left[\frac{d}{x^{2}} - 1\right] y + \frac{1}{x} \, |y|^{2} y = 0.
      \end{align}
\end{description}

\begin{figure}
	\setlength{\labelsep}{-3.0mm}
	\centering
    \sidesubfloat[]{\includegraphics[width=2.5in]{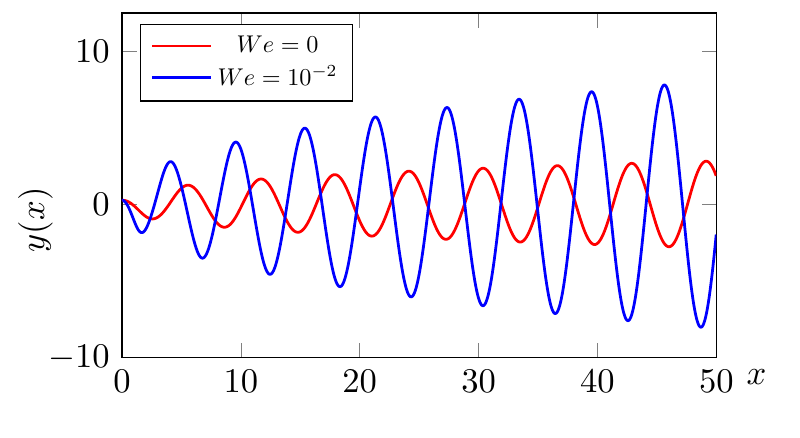}\label{fig:plot-y-We-low}}
	\sidesubfloat[]{\includegraphics[width=2.5in]{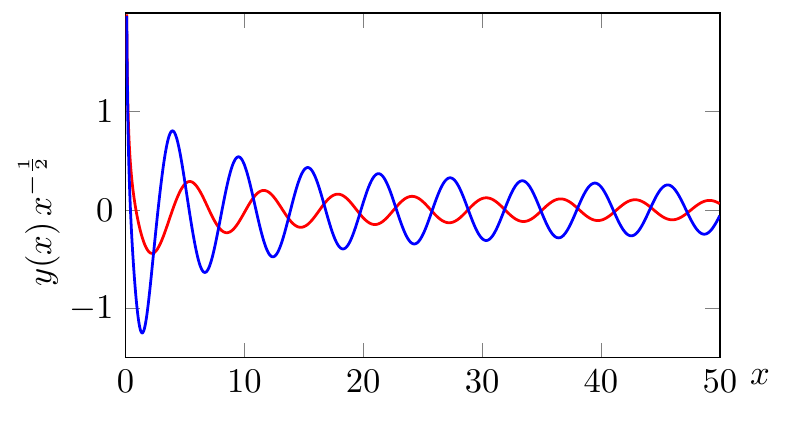}\label{fig:plot-u-We-low}} \\
    \sidesubfloat[]{\includegraphics[width=2.5in]{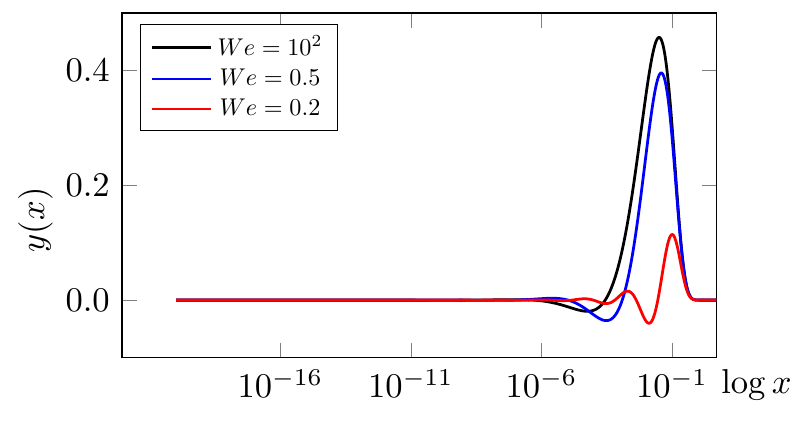}\label{fig:plot-y-We-high}}
	\sidesubfloat[]{\includegraphics[width=2.5in]{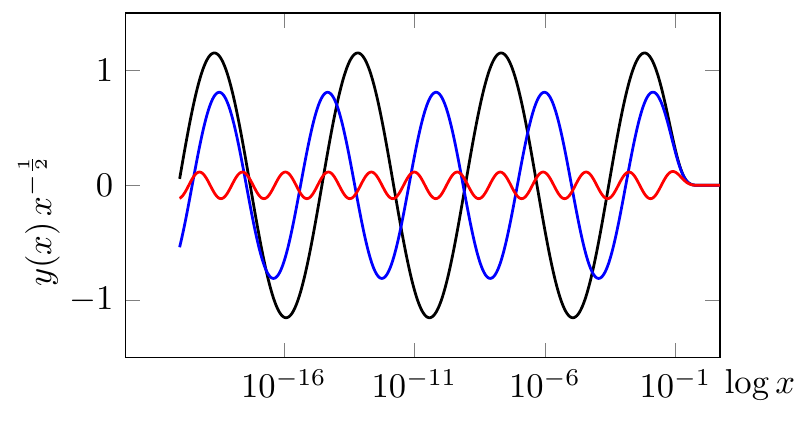}\label{fig:plot-u-We-high}}
\caption{(a) Solutions to \eqref{ncNLS-ST:abstract:y:case-1} and (b) in the unscaled variables for $We < \frac{2}{\sqrt{3}}-1$. (b) Solutions to (\ref{ncNLS-ST:abstract:y:case-2},\ref{ncNLS-ST:abstract:y:case-3}) and (c) in the unscaled variables for $We > \frac{2}{\sqrt{3}}-1$.} \label{fig:plot-GP-solitons}
\end{figure}

The first notable fact about the base states of the Gross-Pitaevskii equation is that, in general, they can be singular at the origin -- this is opposed to the case when the potential $V(R)$ is omitted as was done by \citet{Zakharov:1976b}, for example, which leads to the standing wave-type solutions \eqref{wave:Stokes} regular at the origin and satisfying $u^{\prime}(0)=0$. To get a sense of the structure of the $y$-solution, let us look into the asymptotics near the origin, $x \rightarrow 0$, starting with \textit{case 1}. Expecting a power-law form $y = C x^{\alpha}$, where from now on the notation $C$ is used for a generic constant unless stated otherwise, so that \eqref{ncNLS-ST:abstract:y:case-1} produces:
\begin{align}
\left[\alpha (\alpha-1) + d\right] x^{\alpha-2} + x^{\alpha} + C^{2} x^{3\alpha-1} = 0.
\end{align}
We find that for $d \in [-\frac{1}{2},-\frac{3}{4}]$ the solution is determined by the first two (linear) terms in \eqref{ncNLS-ST:abstract:y:case-1} giving $\alpha = \frac{1 \pm \sqrt{1 - 4 d}}{2} \in \left[\frac{1-\sqrt{3}}{2},-\frac{1}{2}\right]$, where the most singular solution is of interest to us. At $We=0$ the parameter $d=-\frac{1}{2}$ and then decreases with $We$ down to $-\infty$. At $d = - \frac{3}{4}$ the nonlinearity `kicks in' with the power $\alpha = - \frac{1}{2}$ and the solution of \eqref{ncNLS-ST:abstract:y:case-1} has a different asymptotics:
\begin{align}
\label{asymptotics:special}
y(x) \sim \frac{x^{1/2}}{\left(\ln{x}\right)^{1/2}}.
\end{align}
As $d$ varies further in the range $-\infty < d < -\frac{3}{4}$, the power $\alpha$ stays at the same value $\alpha = - \frac{1}{2}$, but the `amplitude' of the solution $C$ in  $y = C x^{\alpha}$ varies with $d$ according to $\alpha (\alpha-1) + d + C^{2} = 0$.

In \textit{cases 2 and 3}, we have $d > 0$ falling in the range $(\frac{1}{2},\infty)$ as the Weber number changes from $\infty$ down to $\frac{2}{\sqrt{3}} - 1$. The power $\alpha = \alpha_{R} + \i \alpha_{I}$ is then complex with $\alpha_{R} = \frac{1}{2}$. Looking for a solution in the form $y = C \, x^{\alpha}$ with $\alpha = \alpha_{R} + \i \alpha_{I}$ yields
\begin{align}
\left[\alpha (\alpha-1) + d\right] - x^{2} \mp |C|^{2} x^{2 \alpha_{R}+1} = 0,
\end{align}
where we took into account that $|x^{\alpha_{R} + \i \alpha_{I}}| = |x^{\alpha_{R}}| \, x^{\i \alpha_{I}} = x^{\alpha_{R}} |e^{\i \alpha_{I} \ln{x}}| = x^{\alpha_{R}}$, i.e. the imaginary part $\alpha_{I}$ does not affect the amplitude because $x^{\i \alpha_{I}} = e^{\i \alpha_{I} \ln{x}}$ and hence $|x^{\i \alpha_{I}}| = 1$ for any $x$. At the leading order the balance occurs due to the first two (linear) terms in \eqref{ncNLS-ST:abstract:y:case-2} and the first two (linear) terms in \eqref{ncNLS-ST:abstract:y:case-3}, respectively, which are the same as in \eqref{ncNLS-ST:abstract:y:case-1} and hence $\alpha_{R} = \frac{1}{2}$, $\alpha_{I} = \pm \frac{\sqrt{4 d - 1}}{2}$. Thus, if $2 \alpha_{R}+1 > 0$, then
\begin{align}
\alpha = \frac{1 \pm \sqrt{1-4d}}{2} \ \Rightarrow \ \alpha_{R} = \frac{1}{2}, \ \alpha_{I} = \pm \frac{1}{2} \sqrt{4 d - 1};
\end{align}
in the considered cases $d \in [\frac{1}{2},\infty)$, which implies $\alpha_{I} \in (-\infty,-\frac{1}{2}] \cup [\frac{1}{2},\infty)$. As a result, the asymptotics of the real solution can be represented as
\begin{align}
\label{asymptotics:y:2-3:origin}
y = C x^{\alpha_{R}} \cos{\left[\alpha_{I} \ln{x} + \varphi(x)\right]},
\end{align}
where $|\varphi(x)| \ll |\ln{x}|$.

\begin{figure}
	\centering
    \includegraphics[width=2.5in]{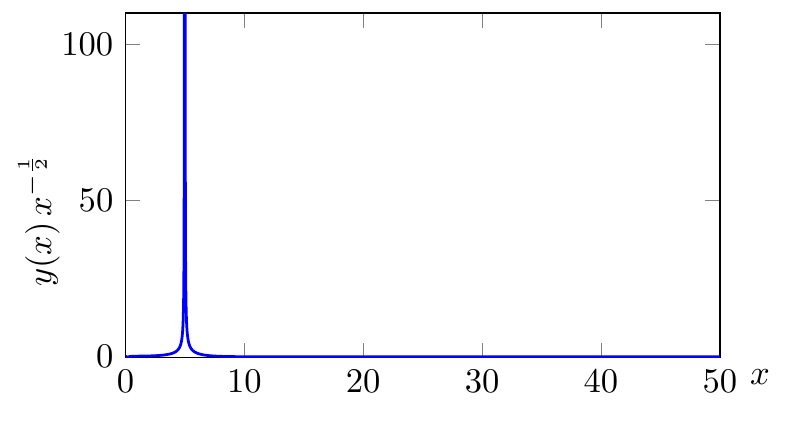}
\caption{Solution to \eqref{ncNLS-ST:abstract:y:case-2} for $We =0.2$ singular along the ring of radius $x_{0}=5$.}\label{fig:plot-u-ring}
\end{figure}

Notably, \textit{case 2} also admits solutions singular along a ring of radius $x_{0} \neq 0$, cf. figure~\ref{fig:plot-u-ring}:
\begin{align}
y(x) \sim \frac{C}{|x-x_{0}|}, \ x_{0} = \frac{1}{2} C^{2}.
\end{align}
It should be noted that the found singular ring ground states are different from the ring-type solitons and the solutions identified in the radial NLS not only because they were constructed without the potential term $V(R)$, but also because they are non-singular (and approximate) dark \citep{Kivshar:1994} and bright \citep{Lomdahl:1980,Afanasjev:1995} ring solitons.

Next, let us determine the asymptotics of solutions at infinity. In \textit{case 1}, we see that as $x \rightarrow \infty$ the leading-order solution is $\cos{x}$ with some corrections to its phase (cf. Appendix~\ref{appx:asymptotics:infinity}):
\begin{align}
\label{asymptotics:BS:cNLS:infinity:case1}
y(x) = C \cos{\left[x + \frac{C^{2}}{4} \ln{x} + \mathcal{O}\left(\frac{1}{x}\right)\right]}.
\end{align}
Because the asymptotics at infinity to this order does not depend on parameter $d$ in this case, numerical integration can be done only starting from the neighborhood of the origin. However, as we saw from the corresponding analysis of the leading order asymptotic terms, the solution is singular with a negative power-law exponent $\alpha_{0}$. Clearly, for numerically accurate solution one needs to improve that asymptotics
\begin{align}
y(x) = x^{\alpha_{0}} \left(C_{0} + C_{1} x^{\alpha_{1}} + \ldots + C_{i} x^{\alpha_{i}} + \ldots\right)
\end{align}
to the order $O(x^{\alpha_{i}})$ with $\alpha_{0}+\alpha_{i}>1$, since the first derivative is needed for numerical integration as well. For values of $We < 0.011$, it proves sufficient to compute the first five terms in the above expansion giving $\alpha_{i} = i (1+2\alpha_{0})$, $i \ge 1$, and the coefficients
\begin{align}
C_{1} &= - \frac{C_{0}^{3}}{\alpha_{1} \left(2 \alpha_{0} - 1 + \alpha_{1}\right)}, \
C_{2} = - \frac{3 C_{1} C_{0}^{2}}{\alpha_{2} \left(2 \alpha_{0} - 1 + \alpha_{2}\right)}, \
C_{3} = - 3 \frac{C_{2} C_{0}^{2} + C_{0} C_{1}^{2}}{\alpha_{3} \left(2 \alpha_{0} - 1 + \alpha_{3}\right)}, \nonumber \\
C_{4} &= - 3 \frac{C_{3} C_{0}^{2} + 2 C_{0} C_{1} C_{2} + C_{1}^{3}}{\alpha_{4} \left(2 \alpha_{0} - 1 + \alpha_{4}\right)}, \
C_{5} = - 3 \frac{C_{4} C_{0}^{2} + 2 C_{0} C_{1} C_{3} + C_{0} C_{2}^{2} + C_{2} C_{1}^{2}}{\alpha_{5} \left(2 \alpha_{0} - 1 + \alpha_{5}\right)}.
\end{align}

Similarly, the asymptotics can be determined in \textit{cases 2 and 3}. The physically meaningful leading-order solution $y(x) = C e^{-x}$ is corrected with a phase $\varphi(x)$, i.e. $y(x) = C e^{-x + \varphi(x)}$. However, as opposed to case 1 in which the phase is found from balance with the nonlinear term, the phase here comes from balance of linear terms; indeed substitution of $y(x) = C e^{-x + \varphi(x)}$ in \eqref{ncNLS-ST:abstract:y:case-2} and \eqref{ncNLS-ST:abstract:y:case-3} gives
\begin{align}
\varphi^{\prime\prime} + \left(-1 + \varphi^{\prime}\right)^{2} + \frac{d}{x^{2}} - 1 \mp \frac{C^{2}}{x} e^{- 2 x + 2 \varphi(x)} = 0,
\end{align}
and hence at the next order the balance is due to $2 \, \varphi^{\prime} = \frac{d}{x^{2}}$, which yields $\varphi(x) = - \frac{d}{2 x} + \varphi(\infty)$ satisfying the underlying assumptions that $|\varphi^{\prime\prime}| \ll |\varphi^{\prime}|$ and $|\varphi^{\prime}|^{2} \ll |\varphi^{\prime}|$. As a result, the corrected asymptotics in both cases 2 and 3 reads
\begin{align}
\label{asymptotics:BS:cNLS:infinity:case2-3}
y(x) = C \exp{\left[- x - \frac{d}{2 x} + \const\right]}.
\end{align}

Despite the singular nature of the ground states in Figs.~\ref{fig:plot-u-We-low},\ref{fig:plot-u-ring}, they are as valuable as the widely studied finite-time singularity peculiar to NLS -- such singularities are indicative of a localized behavior in the original unreduced physical system such as the Euler equations \eqref{system:deep-water:non-dimensional:cylindrical} such as spike waves \citep{McAllister::2022}, from which \eqref{ncNLS-ST:abstract} is deduced. In accordance with physical expectations the identified singular and regular solitons shown in figures~\ref{fig:plot-GP-solitons} and \ref{fig:plot-u-ring} are bright, i.e. localized in space and evanescent at infinity.

\begin{figure}
	\setlength{\labelsep}{-3.0mm}
	\centering
    \sidesubfloat[]{\includegraphics[width=2.5in]{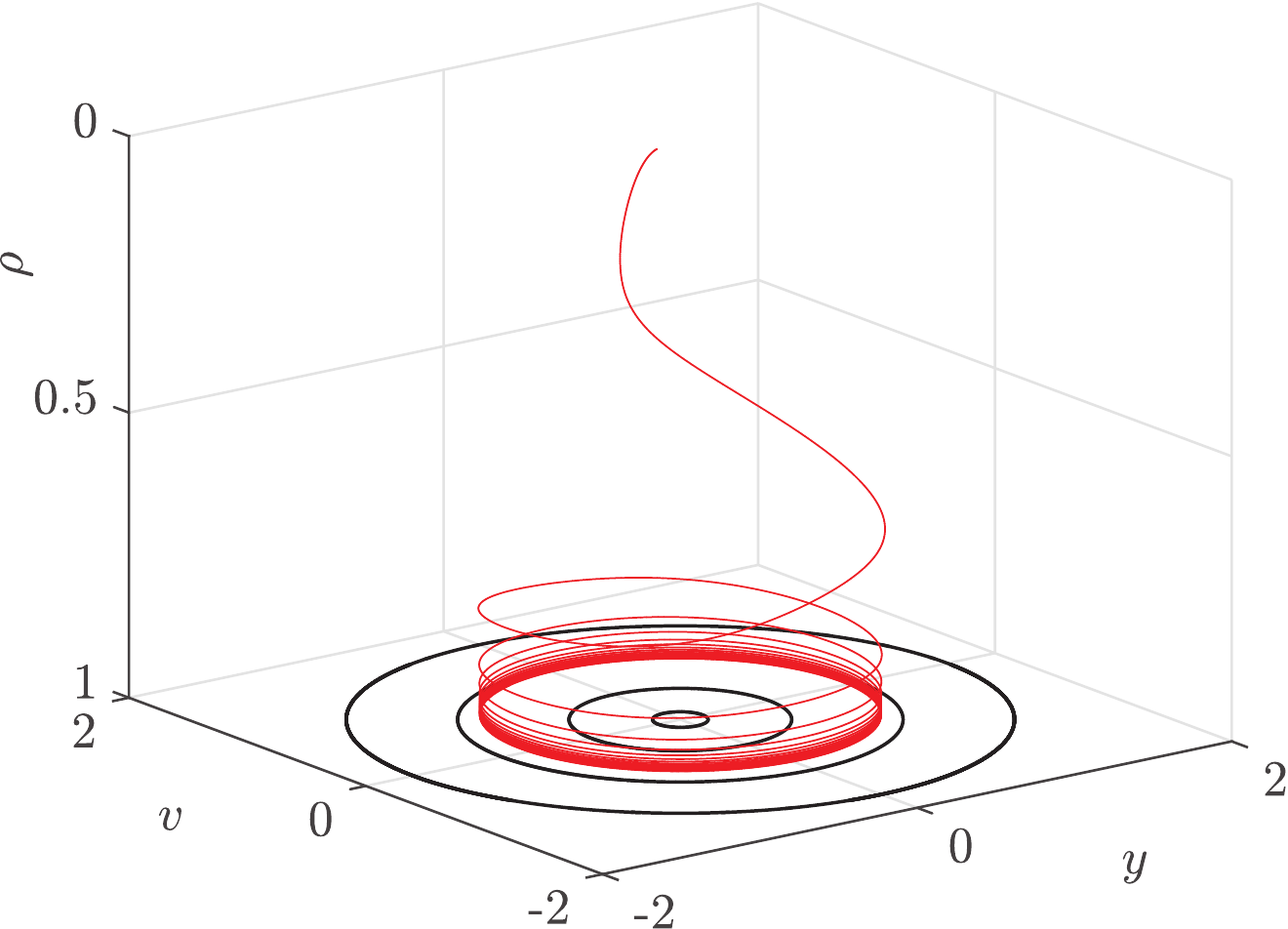}\label{fig:combined1}} \quad
	\sidesubfloat[]{\includegraphics[width=2.5in]{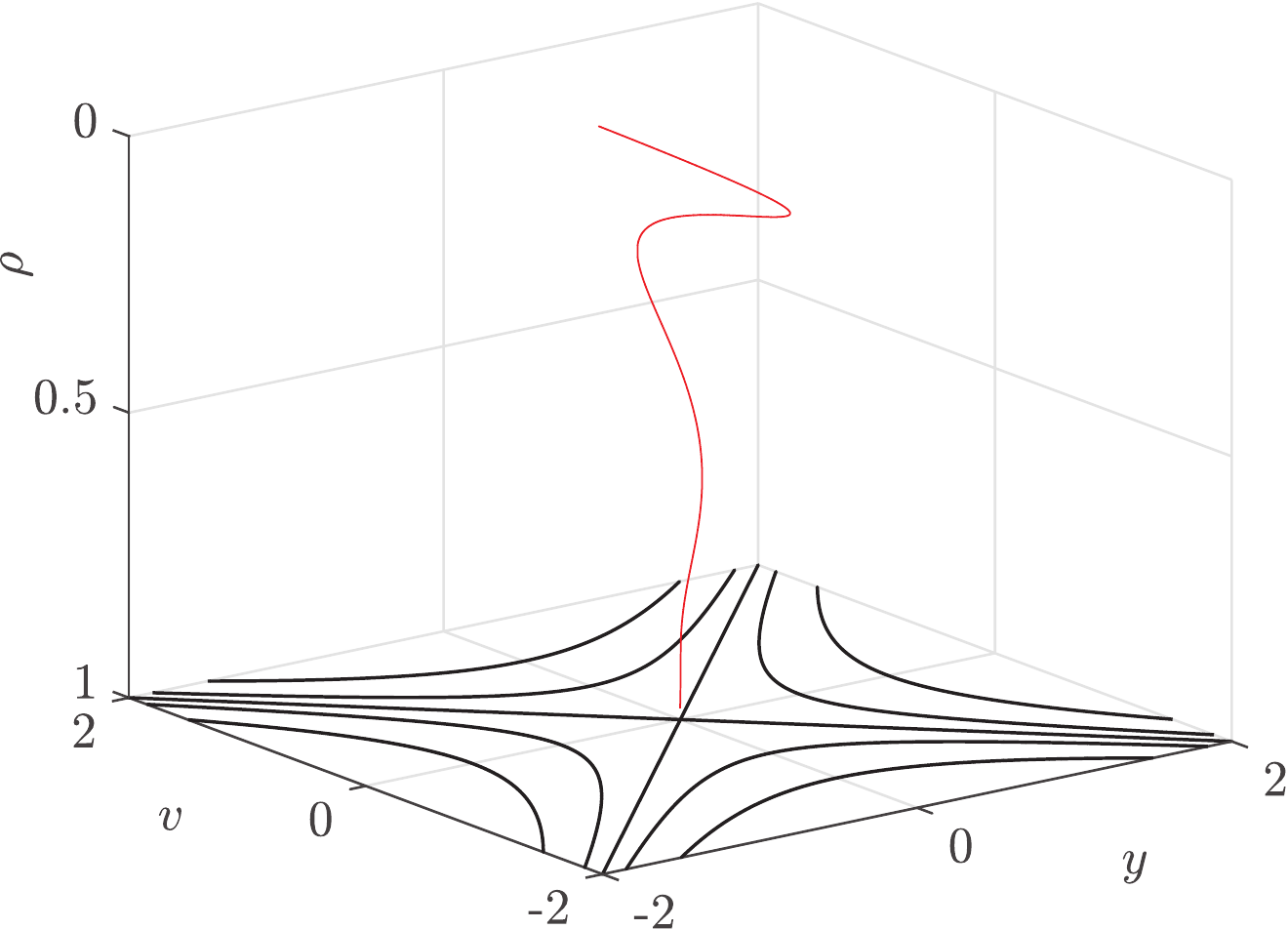}\label{fig:combined2}}
\caption{ A solution trajectory of (a) equation \eqref{ncNLS-ST:abstract:y:case-1} and (b) equation \eqref{ncNLS-ST:abstract:y:case-2}.} \label{fig:combined}
\end{figure}

A convenient way to understand the structure of solution variety of (\ref{ncNLS-ST:abstract:y:case-1}-\ref{ncNLS-ST:abstract:y:case-3}) is through a dynamical systems point of view \citep{Jones:1986,Newton:1993}. The idea is to compactify the problem: the phase space is augmented with a bounded but open dimension and then extended at both ends by gluing in invariant subspaces that carry autonomous dynamics of the limit systems \citep{Wieczorek:2021}. Namely, reducing, for example, (\ref{ncNLS-ST:abstract:y:case-1},\ref{ncNLS-ST:abstract:y:case-2}) to an non-autonomous system of first-order equations:
\begin{subequations}
\label{system:Stokes-wave}
\begin{align}
\dot{y} &= \rho^{2} v, \\
\dot{v} &= - \left[d (1-\rho)^{2} \pm \rho^{2}\right] y \mp \rho (1-\rho) |y|^{2} y, \\
\dot{\rho} &= \rho^{2} (1-\rho)^{2},
\end{align}
\end{subequations}
in which a singularity at the origin is removed by introducing a new independent variable $t = x - \frac{1}{x} + 2 \ln{x} \in (-\infty,+\infty)$ for $x \in [0,\infty)$ and seeing the radial coordinate $x$ via a new dependent variable $\rho = x / (x+1)$; the upper choice of sign corresponds to \eqref{ncNLS-ST:abstract:y:case-1} and the lower one to \eqref{ncNLS-ST:abstract:y:case-2}. From \eqref{system:Stokes-wave} we find that all solutions starting in the invariant plane $\rho=0$ end up being attracted to one of the trajectories in the invariant plane $\rho=1$ shown in figure~\ref{fig:combined}. For example, the solutions of the type in figure~\ref{fig:plot-y-We-low} look like in figure~\ref{fig:combined1} and get attracted to one of the centers. On the way from $\rho=0$ to $\rho=1$ the solution may pierce the $y$-plane many times, which correspond to the number of zeros of a given solution. This dynamical systems approach proved to be fruitful to analyze the number of zeros or existence of a solution with a given number of zeros for the semilinear elliptic equation \eqref{ncNLS-ST:ground-state} without the potential term \citep{Jones:1986}. The dynamical systems view in figure~\ref{fig:combined1} also makes it clear that structurally the solutions must be Lyapunov stable. On the other hand, solutions of the type shown in figure~\ref{fig:plot-y-We-high}, e.g. for $We=0.2$ corresponding to \textit{case 2} represent trajectories approaching a saddle point as one can observe from the phase portrait at $\rho = 1$ in figure~\ref{fig:combined2}. Obviously, unless the boundary condition at infinity, $y,v \rightarrow 0$ as $x \rightarrow \infty$, is enforced, the solution would otherwise be structurally unstable. We will see both scenarios from the subsequent spectral (\S \ref{subsec:spectral-analysis:ncNLS}) and Hamiltonian (\S \ref{subsec:Lagrange-Dirichlet}) stability analyses.

\subsection{Spectral stability of base states} \label{subsec:spectral-analysis:ncNLS}

Superimposing a perturbation on the base state:
\begin{align}
\psi(\tau,R) = u(R) \left[1 + u^{\prime}(\tau,R)\right] e^{\i \left[\mu \tau + \varphi(\tau,R)\right]}
\end{align}
substituting in \eqref{ncNLS-ST:abstract}, and separating real $\Re$ and imaginary $\Im$ parts we get a system
\begin{subequations}
\begin{align}
\Re&: & &- u (1+u^{\prime}) (\mu + \varphi_{\tau}) + \lambda_{\infty} \bigg[u_{RR} (1+u^{\prime}) + 2 u_{R} u^{\prime}_{R} - u (1+u^{\prime}) \varphi_{R}^{2} \nonumber \\
& & &+ u u^{\prime}_{RR} + \frac{1}{R}\left\{u_{R} (1+u^{\prime}) + u u^{\prime}_{R}\right\}\bigg] \\
& & &+ \lambda_{\infty}^{\prime} \frac{1}{R^{2}} u (1+u^{\prime}) + \frac{\mu_{\infty}}{R^{2}} \left[u u^{\prime}_{\theta\theta} - u (1+u^{\prime}) \varphi_{\theta}^{2}\right] - \chi_{\infty} u^{3} (1+u^{\prime})^{3} = 0, \nonumber \\
\Im&: & &u u^{\prime}_{\tau} + \lambda_{\infty} \bigg[2 u_{R} (1+u^{\prime}) \varphi_{R} + 2 u u^{\prime}_{R} \varphi_{R} \nonumber \\
& & &+ u (1+u^{\prime}) \varphi_{RR} + \frac{1}{R} u (1+u^{\prime}) \varphi_{R}\bigg] + \frac{\mu_{\infty}}{R^{2}} \left[2 u u^{\prime}_{\theta} \varphi_{\theta} + u (1+u^{\prime}) \varphi_{\theta\theta}\right] = 0,
\end{align}
\end{subequations}
where $u$ is real as we consider real base states constructed in \S \ref{subsec:ncNLS:BS}\marginlabel{redo for complex $u$, $u^{\prime}$}. Taking into account equation \eqref{ncNLS-ST:abstract:U} for the base state, the linearized system for a perturbation simplifies to
\begin{subequations}
\begin{align}
- \varphi_{\tau} + \lambda_{\infty} \Delta_{R} u^{\prime} + \frac{\mu_{\infty}}{R^{2}} u^{\prime}_{\theta\theta} + 2 \lambda_{\infty} \frac{u_{R}}{u} u^{\prime}_{R} - 2 \chi_{\infty} u^{2} u^{\prime} &= 0, \\
u^{\prime}_{\tau} + \lambda_{\infty} \Delta_{R} \varphi + \frac{\mu_{\infty}}{R^{2}} \varphi_{\theta\theta} + 2 \lambda_{\infty} \frac{u_{R}}{u} \varphi_{R} &= 0.
\end{align}
\end{subequations}
Next, applying the Fourier transform in $\theta$ and looking for eigenmodes, i.e. $\varphi = \widehat{\varphi} \, e^{\lambda \tau} e^{\i k \theta}$ and $u^{\prime} = \widehat{u} \, e^{\lambda \tau} e^{\i k \theta}$, we arrive at
\begin{subequations}
\begin{align}
\lambda \widehat{\varphi} &= L_{R} \widehat{u} - 2 \chi_{\infty} u^{2} \widehat{u}, \\
- \lambda \widehat{u} &= L_{R} \widehat{\varphi},
\end{align}
\end{subequations}
where $L_{R} = \lambda_{\infty} \Delta_{R} - \frac{\mu_{\infty} k^{2}}{R^{2}} + 2 \lambda_{\infty} \frac{u_{R}}{u} \partial_{R}$. To bring these equations to the canonical form convenient for analysis, first let us apply the transformation of the base state, $u(R) = R^{-1/2} U(R)$ introduced earlier (\S \ref{subsec:ncNLS:BS}), which gives $u^{2} = \frac{U^{2}}{R}$ and $\frac{u_{R}}{u} = - \frac{1}{2} \frac{1}{R} + \frac{U_{R}}{U}$. Second, rescaling the variables $R = \alpha x$, $U = \beta y$, we end up with the following canonical systems:
\begin{subequations}
\label{EVP:cNLS}
\begin{align}
\label{EVP:cNLS:case-1-3}
\text{\textit{cases 1 and 3}}&: & &\left\{\begin{array}{c}
\nu \widehat{\varphi} = L_{x} \widehat{u} + 2 \frac{y^{2}}{x} \widehat{u}, \\
- \nu \widehat{u} = L_{x} \widehat{\varphi};
\end{array}\right. \ \begin{pmatrix}
                       \text{\textit{case 1}}: \ \frac{\alpha^{2}}{\lambda_{\infty}} = - \frac{1}{\mu}, \ \lambda_{\infty} < 0 \\
                       \text{\textit{case 2}}: \ \frac{\alpha^{2}}{\lambda_{\infty}} = \frac{1}{\mu}, \ \lambda_{\infty} > 0
                     \end{pmatrix} \\
\label{EVP:cNLS:case-2}
\text{\textit{case 2}}&: & &\left\{\begin{array}{c}
\nu \widehat{\varphi} = L_{x} \widehat{u} - 2 \frac{y^{2}}{x} \widehat{u}, \\
- \nu \widehat{u} = L_{x} \widehat{\varphi};
\end{array}\right. \ \begin{pmatrix}
                       \frac{\alpha^{2}}{\lambda_{\infty}} = \frac{1}{\mu}, \ \lambda_{\infty} > 0
                     \end{pmatrix},
\end{align}
\end{subequations}
where $\nu = \lambda \alpha^{2} / \lambda_{\infty}$ and
\begin{align}
L_{x} = \Delta_{x} - \frac{\mu_{\infty} k^{2}}{\lambda_{\infty} x^{2}} + 2 \left(-\frac{1}{2} \frac{1}{x} + \frac{y_{x}}{y}\right) \frac{\d}{\d x} = \frac{\d^{2}}{\d x^{2}} - \frac{\widetilde{\mu}}{x^{2}} + 2 \frac{y_{x}}{y} \frac{\d}{\d x}, \ \widetilde{\mu} = \frac{\mu_{\infty} k^{2}}{\lambda_{\infty}}.
\end{align}
As for the BCs, it is natural to impose
\begin{subequations}
\label{EVP:BCs:cNLS}
\begin{align}
x=0&: \ \widehat{u}_{x} = \widehat{\varphi}_{x} = 0, \\
x=\infty&: \ \widehat{u}=\widehat{\varphi}_{x}=0.
\end{align}
\end{subequations}
The challenge of the eigenvalue problem (\ref{EVP:cNLS},\ref{EVP:BCs:cNLS}) is its singularity, i.e. some of the coefficients in \eqref{EVP:cNLS} diverge either at infinity (\textit{case 1}) or at the origin (\textit{cases 2 and 3}) as follows from \S \ref{subsec:ncNLS:BS}. Apparently, it is not feasible to solve the eigenvalue problems \eqref{EVP:cNLS} analytically for all $x$, as well as the numerically accurate treatment of the problem is impeded by the singular behavior mentioned above or non-periodic oscillations propagating to $x \rightarrow \infty$, which requires ever-increasing number of modes/nodes for resolution\footnote{Due to the identified oscillatory behavior of the solution at infinity, truncating the semi-infinite domain to a finite one necessarily introduces significant errors; also mapping the semi-infinite to a finite domain simply compresses oscillations near one of the boundaries with ever-increasing frequency of oscillations.}.\marginlabel{find a ref} However, the latter properties, that makes numerical approach difficult, allow us to resort to an asymptotic way of solving (\ref{EVP:cNLS},\ref{EVP:BCs:cNLS}) based on a peculiar behavior of the corresponding linear operators. The key guiding principle is that if we can solve an eigenvalue problem locally, i.e. for some range of $x$, then due to the linear character of the problem at hand, the thereby determined eigenvalues hold globally.

\textit{Case 1}. The eigenvalue problem assumes the form
\begin{subequations}
\label{EVP:no-ST:cNLS}
\begin{align}
\nu \widehat{\varphi} &= L_{x} \widehat{u} + 2 \frac{y^{2}}{x} \widehat{u}, \\
- \nu \widehat{u} &= L_{x} \widehat{\varphi},
\end{align}
\end{subequations}
where $L_{x} = \frac{\d^{2}}{\d x^{2}} - \frac{\widetilde{\mu}}{x^{2}} + 2 \frac{y_{x}}{y} \frac{\d}{\d x}$ and $\nu = - \frac{\lambda}{\mu}$. Since for large $x$
\begin{align}
\frac{y_{x}}{y} \approx - \left(1 + \frac{C^{2}}{4x}\right) \tan{x},
\end{align}
we get the approximate eigenvalue problem
\begin{subequations}
\label{EVP:no-ST:cNLS:infinity}
\begin{align}
\nu \widehat{\varphi} &= L_{x}^{\infty} \widehat{u}, \\
- \nu \widehat{u} &= L_{x}^{\infty} \widehat{\varphi},
\end{align}
\end{subequations}
where $L_{x}^{\infty} = \frac{\d^{2}}{\d x^{2}} - 2 \tan{x} \frac{\d}{\d x}$. Applying operator $L_{x}^{\infty}$ to the second of equations \eqref{EVP:no-ST:cNLS:infinity} produces an equation for $\widehat{\varphi}$:
\begin{align}
\label{EVP:no-ST:cNLS:infinity:leading-order:original}
L_{x}^{\infty 2} \widehat{\varphi} = - \nu^{2} \widehat{\varphi}.
\end{align}
Let us first treat the simpler problem
\begin{align}
\label{EVP:no-ST:cNLS:infinity:leading-order}
L_{x}^{\infty} \widehat{\varphi} = \left[\frac{\d^{2}}{\d x^{2}} - 2 \tan{x} \frac{\d}{\d x}\right] \widehat{\varphi} = \widetilde{\nu} \, \widehat{\varphi} \ \text{on} \ x \in \left[-\frac{\pi}{2},\frac{\pi}{2}\right],
\end{align}
which will be justified by the constructed solution satisfying \eqref{EVP:no-ST:cNLS:infinity:leading-order:original}; here $\widetilde{\nu}^{2} = - \nu^{2}$, i.e. $\widetilde{\nu} = \pm \i \nu$. Multiplication by the integrating factor $I(x) = \cos^{2}{x}$ gives a self-adjoint Sturm-Liouville problem
\begin{align}
\label{SL-problem:case-1}
\frac{\d}{\d x}\left[\cos^{2}{x} \frac{\d \widehat{\varphi}}{\d x}\right] = \widetilde{\nu} \, \cos^{2}{x} \, \widehat{\varphi} \ \text{on} \ x \in \left[-\frac{\pi}{2},\frac{\pi}{2}\right].
\end{align}
With the change of variables $z = \tan{x}$ equation \eqref{SL-problem:case-1} can be reduced to
\begin{align}
\left(1 + z^{2}\right)^{2} \widehat{\varphi}_{zz} = \widetilde{\nu} \, \widehat{\varphi} \ \text{on} \ z \in \left(-\infty,\infty\right).
\end{align}
The requirement for its solution to be bounded leads to quantization
\begin{align}
\widehat{\varphi}(z) = \sqrt{1 + z^{2}} \left\{C_{1} \cos{(\alpha \atan{z})} + C_{2} \sin{(\alpha \atan{z})}\right\} \ \text{for} \ 1 - \widetilde{\nu} = \alpha^{2} > 0,
\end{align}
or, in the original variables,
\begin{align}
\widehat{\varphi}_{0}(x) = \frac{\cos{\sqrt{1-\widetilde{\nu}} \, x}}{\cos{x}},
\end{align}
where one must put $\sqrt{1-\widetilde{\nu}} = 1 + 2 n$, $n \in \Bbb{Z}$ for the solution to be bounded. As a result, $\widetilde{\nu} = 1 - (1 + 2 n)^{2}$, $n \in \Bbb{Z}$. The original eigenvalue $\lambda$ is then
\begin{align}
\label{EV:cNLS:no-ST:leading}
\lambda = - \mu \nu = \pm \i \mu \widetilde{\nu} = \pm \i \mu \left[1 - (1 + 2 n)^{2}\right], \ n \in \Bbb{Z},
\end{align}
i.e. one has spectral stability. To see the effect of higher-order terms in $L_{x}$ including those due to the transverse perturbations with wavenumber $k$, we represent the operator as
\begin{align}
L_{x} = L_{x}^{\infty} + L_{x}^{\prime} \ \text{with} \ L_{x}^{\prime} = - \frac{C^{2}}{x} \tan{x} \frac{\d}{\d x} - \frac{\widetilde{\mu}}{x^{2}}.
\end{align}
From \eqref{EVP:no-ST:cNLS} we deduce a stand-alone equation for $\widehat{\varphi}$:
\begin{align}
- \nu^{2} \widehat{\varphi} = L_{x}^{2} \widehat{\varphi} + 2 \frac{y^{2}}{x} L_{x} \widehat{\varphi}.
\end{align}
Linearizing around the zero eigenvalue $\widetilde{\nu}=0$, i.e. $\nu_{0}=0$ as well, and the corresponding eigensolution $\widehat{\varphi}_{0} = 1$, we find for the eigenvalue $\nu^{\prime}=\nu-\nu_{0}$ and the eigenfunction $\widehat{\varphi}^{\prime}$ perturbations:
\begin{align}
\label{Fredholm:ncNLS:case-1}
L_{x}^{\infty 2} \widehat{\varphi}^{\prime} = - \nu^{\prime 2} \widehat{\varphi}_{0} - L_{x}^{\infty} \left(L_{x}^{\prime} \widehat{\varphi}_{0}\right) - 2 \frac{y^{2}}{x} L_{x}^{\prime} \widehat{\varphi}_{0}.
\end{align}
While the operator $L_{x}^{\infty 2}$ is not self-adjoint, we know that its solution corresponding to zero eigenvalue is $\widehat{\varphi}^{\prime} = \widehat{\varphi}_{0}$, so we may apply the Fredholm alternative using the same integrating factor $I(x) = \cos^{2}{x}$, which allows us to determine the eigenvalue deviation from zero:
\begin{align}
\nu^{\prime 2} = \widetilde{\mu} \frac{\int_{-\frac{\pi}{2}}^{\frac{\pi}{2}}{L_{x}^{\infty}\left(\frac{1}{x^{2}}\right) I(x) \, \d x}}{\int_{-\frac{\pi}{2}}^{\frac{\pi}{2}}{I(x) \, \d x}}, \ \text{where} \ L_{x}^{\infty}\left(\frac{1}{x^{2}}\right) = \frac{6}{x^{4}} + \frac{4}{x^{3}} \tan{x};
\end{align}
we also took into account that the last term in \eqref{Fredholm:ncNLS:case-1} does not contribute as it is odd in $x$. Since integrands in both integrals are positive-definite, then $\nu^{\prime 2} < 0$ since $\widetilde{\mu}<0$. Hence, corrections to \eqref{EV:cNLS:no-ST:leading} are purely imaginary and spectral stability is retained. Note that while the Fredholm alternative is global in nature, i.e. requires the knowledge of eigenfunction for all $x$, due to periodicity of the solution at infinity, the Fredholm alternative can be applied `locally' over the period of the solution in this asymptotic limit.

\textit{Cases 2-3}. The corresponding equations (\ref{EVP:cNLS:case-2},\ref{EVP:cNLS:case-1-3}) for perturbations:
\begin{subequations}
\label{EVP:cNLS:cases-2-3}
\begin{align}
\label{EVP:a:cNLS:cases-2-3}
\nu \widehat{\varphi} &= L_{x} \widehat{u} \mp 2 \frac{y^{2}}{x} \widehat{u}, \\
\label{EVP:b:cNLS:cases-2-3}
- \nu \widehat{u} &= L_{x} \widehat{\varphi},
\end{align}
\end{subequations}
can be rewritten in the new variable $z = \alpha_{I} \ln{x}$. Splitting the operator into the main and perturbation parts $L_{x} = L_{x}^{\infty} + L_{x}^{\prime}$, where $L_{x}^{\infty} = \frac{\d^{2}}{\d x^{2}} + 2 \frac{y_{x}}{y} \frac{\d}{\d x}$ and $L_{x}^{\prime} =  - \frac{\widetilde{\mu}}{x^{2}}$, yields
\begin{align}
\label{operator:EVP:cNLS:cases-2-3}
L_{x}^{\infty} = \alpha_{I}^{2} e^{-2z/\alpha_{I}} \, \left[\frac{\d^{2}}{\d z^{2}} - 2 \tan{z} \frac{\d}{\d z}\right], \ L_{x}^{\prime} = \alpha_{I}^{2} e^{-2z/\alpha_{I}} \, \left[- \frac{\widetilde{\mu}}{\alpha_{I}^{2}}\right],
\end{align}
where we assumed that $\widetilde{\mu}$ is small, i.e. corresponding to the short wavenumber limit. Since $z \rightarrow - \infty$ as $x \rightarrow 0$, the last term in equation \eqref{EVP:a:cNLS:cases-2-3}:
\begin{align}
\frac{y^{2}}{x} \approx C^{2} \cos^{2}{z} = \mathcal{O}(1)
\end{align}
can be considered as a perturbation; here we used the asymptotics \eqref{asymptotics:y:2-3:origin}. Hence, at the leading order, \eqref{EVP:cNLS:cases-2-3} reduces to
\begin{align}
- \nu^{2} \widehat{\varphi} = L_{x}^{\infty 2} \widehat{\varphi},
\end{align}
or, taking $\widetilde{\nu}^{2} = - \nu^{2}$, to a simpler problem
\begin{align}
L_{x}^{\infty} \widehat{\varphi} = \widetilde{\nu} \widehat{\varphi},
\end{align}
which similar to \textit{case 1} allows us to justify that $\nu = 0$ is an eigenvalue. Hence, we may drop the factor $\alpha_{I}^{2} e^{-2z/\alpha_{I}}$ in the operator \eqref{operator:EVP:cNLS:cases-2-3} and consider the problem on the periodic interval $z \in \left[-\frac{\pi}{2},\frac{\pi}{2}\right]$.

Next, treating $\nu^{\prime}$ as a perturbation around the zero eigenvalue, from \eqref{EVP:cNLS:cases-2-3} we find $\nu^{\prime 2} \widehat{\varphi} = - L_{x}^{2} \widehat{\varphi} \pm \frac{2 y^{2}}{x} L_{x} \widehat{\varphi}$ and hence for the perturbation
\begin{align}
\nu^{\prime 2} \widehat{\varphi}_{0} = - L_{x}^{\infty 2} \widehat{\varphi}^{\prime} - \left(L_{x}^{\infty} L_{x}^{\prime} + L_{x}^{\prime} L_{x}^{\infty}\right) \widehat{\varphi}_{0} \pm \frac{2 y^{2}}{x} \left(L_{x}^{\infty} + L_{x}^{\prime}\right) \widehat{\varphi}_{0}.
\end{align}
Since $\widehat{\varphi}_{0} = \const$, then the second term on the right gives only $L_{x}^{\infty} L_{x}^{\prime} = - 4 \, \widetilde{\mu} \, x^{-4} \left[1 + \alpha_{I} \tan{z}\right]$, which is of higher order compared to $\frac{2 y^{2}}{x} L_{x}^{\prime} = -2 \widetilde{\mu} C^{2} \frac{\cos^{2}{z}}{x^{2}}$. Thus, to the leading order we get
\begin{align}
L_{x}^{\infty 2} \widehat{\varphi}^{\prime} = - \nu^{\prime 2} \widehat{\varphi}_{0} \mp 2 \widetilde{\mu} C^{2} \frac{\cos^{2}{z}}{x^{2}} \widehat{\varphi}_{0}.
\end{align}
From the Fredholm solvability condition it then follows:
\begin{align}
\int_{-\pi/2}^{\pi/2}{I(z) \left[\mp 2 \, \widetilde{\mu} \, C^{2} \, e^{- 2 z/\alpha_{I}} \, \cos^{2}{z} - \nu^{\prime 2}\right] \d z} = 0,
\end{align}
where $I(z) = \cos^{2}{z}$, that is in \textit{case 2} we have spectral stability, while in \textit{case 3} spectral instability. The above stability analysis conclusions will be compared with the Lagrange-Dirichlet approach in \S \ref{subsec:Lagrange-Dirichlet}.

\subsection{Lagrange-Dirichlet stability analysis} \label{subsec:Lagrange-Dirichlet}

While the above spectral analysis provides certain insights into stability, strictly speaking only spectral instability implies linear (and hence nonlinear) instability, while spectral stability does not even imply linear stability \citep{Krechetnikov:2007}. A good visual understanding of the solution stability picture is provided by figure~\ref{fig:combined}, which shows, in particular, that if we infinitesimally perturb the trajectory in figure~\ref{fig:combined1}, it should stay Lyapunov stable by displacing it to a nearby center orbit, while the trajectory in figure~\ref{fig:combined1} is structurally unstable as any small perturbation will drive it away from the saddle point. With these considerations in mind, let us look at the stability picture from the Hamiltonian finite-amplitude viewpoint starting with equation \eqref{ncNLS-ST:abstract}. Applying the scaling
\begin{align}
\label{scalings:Lagrange-Dirichlet}
R = \alpha \, x, \ \psi = \alpha^{-1/2} \beta \, y, \ \tau = \gamma \, t,
\end{align}
with factors appropriate for \textit{case 1} as per \S \ref{subsec:ncNLS:BS}, i.e.
\begin{align}
\alpha = \left(-\lambda_{\infty}/\mu\right)^{1/2}, \ \beta = \left(-\lambda_{\infty}/\chi_{\infty}\right)^{1/2} \left(-\mu/\lambda_{\infty}\right)^{1/4}, \ \gamma = \mu^{-1},
\end{align}
we arrive at
\begin{align}
\i y_{t} - \left(-\frac{\gamma \lambda_{\infty}}{\alpha^{2}}\right) \Delta_{x} y + \frac{\gamma \lambda_{\infty}^{\prime}}{\alpha^{2}} \frac{y}{x^{2}} + \frac{\gamma \mu_{\infty}}{\alpha^{2}} \frac{y_{\theta\theta}}{x^{2}} = \frac{\gamma \beta^{2} \chi_{\infty}}{\alpha} |y|^{2} y,
\end{align}
where
\begin{align}
-\frac{\gamma \lambda_{\infty}}{\alpha^{2}} = 1, \ \frac{\gamma \lambda_{\infty}^{\prime}}{\alpha^{2}} = - \frac{\lambda_{\infty}^{\prime}}{\lambda_{\infty}} = \frac{1}{4} - d, \ \frac{\gamma \mu_{\infty}}{\alpha^{2}} = - \frac{\mu_{\infty}}{\lambda_{\infty}}, \
\frac{\gamma \beta^{2} \chi_{\infty}}{\alpha} = 1,
\end{align}
so at the end we get a two-parameter equation:
\begin{align}\label{ncNLS:scaled:case-1}
\i y_{t} - \Delta_{x} y - \left(d - \frac{1}{4}\right) \frac{y}{x^{2}} - \frac{\mu_{\infty}}{\lambda_{\infty}} \frac{y_{\theta\theta}}{x^{2}} = |y|^{2} y, \ \text{where} \ d = \frac{1}{4} - \frac{\lambda_{\infty}^{\prime}}{\lambda_{\infty}}.
\end{align}
To bring it to a Hamiltonian form, let $y = u + \i v$, which gives a system for the real and imaginary parts:
\begin{subequations}
\label{system:ncNLS}
\begin{align}
- v_{t} &= \Delta_{x} u + \left(d - \frac{1}{4}\right) \frac{u}{x^{2}} + \frac{\mu_{\infty}}{\lambda_{\infty}} \frac{u_{\theta\theta}}{x^{2}} + \left(u^{2}+v^{2}\right) u, \\
u_{t} &= \Delta_{x} v + \left(d - \frac{1}{4}\right) \frac{v}{x^{2}} + \frac{\mu_{\infty}}{\lambda_{\infty}} \frac{v_{\theta\theta}}{x^{2}} + \left(u^{2}+v^{2}\right) v,
\end{align}
\end{subequations}
respectively. The canonical Hamiltonian form of this system is
\begin{align}
\label{system:Hamiltonian:GP}
J U_{t} = \frac{\delta \mathrm{H}}{\delta U}, \ \text{where} \ J = \begin{pmatrix}
                                                            0 & -1 \\
                                                            1 & 0
                                                          \end{pmatrix}, \
                                                      U = \begin{pmatrix}
                                                            u \\
                                                            v
                                                          \end{pmatrix},
\end{align}
and the Hamiltonian, being a scaled variant of \eqref{H:ncNLS:original}, reads
\begin{align}
\label{H:ncNLS}
\mathrm{H} = - \frac{1}{2} \int{|U_{x}|^{2} \, \d \nu} + \frac{1}{2}\left(d-\frac{1}{4}\right) \int{\frac{|U|^{2}}{x^{2}}\d \nu}
- \frac{1}{2}\frac{\mu_{\infty}}{\lambda_{\infty}} \int{\frac{|U_{\theta}^{2}|}{x^{2}}\d \nu} + \frac{1}{4} \int{|U|^{4}\d \nu},
\end{align}
where the cylindrical measure \eqref{measure:cylindrical} in these scaled variables becomes $\d \nu = \d\theta \, x \d x$.

Assuming that after integration by parts all boundary terms do not contribute (in the azimuthal $\theta$-variable this follows from the periodicity of the solution and its derivatives, while in the radial $x$-variable the boundary terms disappear due to the solution being symmetric, i.e. $U_{x}=0$ at $x=0$ as per \eqref{BCs:cNLS}, or due to vanishing variation $\delta U$ at $x \rightarrow 0$; for $x \rightarrow \infty$ the decay could be due to considering an IVP with compact ICs or also due to vanishing variation $\delta U$), we find for the first variation:
\begin{multline}
\delta \mathrm{H} = \int{\frac{1}{x}\frac{\partial}{\partial x}\left(x\frac{\partial U}{\partial x}\right) \cdot \delta U\d\nu} + \left(d-\frac{1}{4}\right) \int{\frac{U \cdot \delta U}{x^{2}}\d\nu} \\
+ \frac{\mu_{\infty}}{\lambda_{\infty}} \int{\frac{U_{\theta\theta}}{x^{2}} \cdot \delta U \d\nu} + \int{|U|^{2} U \cdot \delta U \d\nu},
\end{multline}
where all the terms are arranged in the same order as in \eqref{H:ncNLS}; also dot denotes scalar product, e.g. $U \cdot \delta U = u \delta u + v \delta v$. Obviously, the base state $y = e^{\i t} \mathcal{Y}(x)$, the stability of which we are studying, is not a fixed point of $\frac{\delta \mathrm{H}}{\delta U}$, but rather that of a Hamiltonian constrained by the conservation of particle number \eqref{conservation-mass:cNLS}, which in rescaled variables reads $\mathrm{N} = \int{|y|^{2} \, \d \nu} = \int{|U|^{2} \, \d \nu} = \const$, so that $\mathcal{Y}(x)$ satisfies both
\begin{align}
\label{eqn:Y}
\mathcal{Y} + \Delta_{x} \mathcal{Y} + \left(d - \frac{1}{4}\right) \frac{\mathcal{Y}}{x^{2}} + \mathcal{Y}^{3} = 0 \ \text{and} \ \frac{\delta \mathrm{H}}{\delta U} + \lambda \frac{\delta \mathrm{N}}{\delta U} = 0,
\end{align}
where $\lambda=\frac{1}{2}$. Notably, while \eqref{eqn:Y} is expectedly Hamiltonian as it is derived from \eqref{system:Hamiltonian:GP}, the Hamiltonian for \eqref{eqn:Y} is non-local, which follows from multiplying \eqref{eqn:Y} with $x \, \mathcal{Y}_{x}$ and integrating w.r.t. $x$ resulting in
\begin{align}
\frac{1}{2}\int_{0}^{\infty}{x \frac{\d}{\d x} \mathcal{Y}^{2} \d x} - \frac{1}{2}\int_{0}^{\infty}{x \frac{\d}{\d x} \mathcal{Y}_{x}^{2} \d x} + \left(d - \frac{1}{4}\right) \int_{0}^{\infty}{x \frac{\mathcal{Y} \mathcal{Y}_{x}}{x^{2}} \d x} + \frac{1}{4}\int_{0}^{\infty}{x \frac{\d}{\d x} \mathcal{Y}^{4} \d x} = 0, \nonumber
\end{align}
after integration by parts. Therefore, the Hamiltonian for the reduced Hamiltonian system \eqref{eqn:Y} is
\begin{align}
\mathrm{H}_{V} = \mathrm{H}_{0} - \left(d - \frac{1}{4}\right) \int_{x}^{\infty}{\frac{\mathcal{Y}(x^{\prime}) \mathcal{Y}_{x}(x^{\prime})}{x^{\prime 2}} \, \d x^{\prime}}, \ \mathrm{H}_{0} = \frac{1}{2} \mathcal{Y}^{2} - \frac{1}{2} \mathcal{Y}_{x}^{2} + \frac{1}{4} \mathcal{Y}^{4},
\end{align}
where the lower limit of integration in the last term of $\mathrm{H}_{V}$ can be chosen arbitrarily, though it should be fixed. One way to interpret the nonlocality of $\mathrm{H}_{V}$ is that the trajectory of \eqref{eqn:Y} crosses the level curves of the Hamiltonian $\mathrm{H}_{0}$ of the system without the potential, i.e. locally the energy $\mathrm{H}_{0}$ changes, but the integral quantity $\mathrm{H}_{V}$ is conserved.

Returning to the Hamiltonian $\mathrm{H}$ \eqref{H:ncNLS}, its second variation reads
\begin{multline}
\label{variation:second:case-1}
\delta^{2} \mathrm{H} = \frac{1}{2}\int \Big\{- \left[(\delta u_{x})^{2}+(\delta v_{x})^{2}\right] + \left(d-\frac{1}{4}\right) \frac{(\delta u)^{2} + (\delta v)^{2}}{x^{2}} - \frac{\mu_{\infty}}{\lambda_{\infty}} \frac{(\delta u_{\theta})^{2}+(\delta v_{\theta})^{2}}{x^{2}} \\
+ \left[\left(3 u^{2} + v^{2}\right) (\delta u)^{2} + \left(3 v^{2} + u^{2}\right) (\delta v)^{2} + 4 u v \, \delta u \delta v\right] \Big\} \d \nu.
\end{multline}
Hence, formally, the Hessian density can be written as
\begin{align}\label{Hessian:ncNLS:case-1}
             \resizebox{0.975\hsize}{!}{$\begin{pmatrix}
               \delta u \\
               \delta v \\
               \delta u_{x} \\
               \delta v_{x} \\
               \delta u_{\theta} \\
               \delta v_{\theta}
             \end{pmatrix}^{T}
             \begin{pmatrix}
               \left(d-\frac{1}{4}\right) \frac{1}{x^{2}} + \left(3 u^{2} + v^{2}\right) & 2 u v & 0 & 0 & 0 & 0 \\
               2 u v & \left(d-\frac{1}{4}\right) \frac{1}{x^{2}} + \left(3 v^{2} + u^{2}\right) & 0 & 0 & 0 & 0 \\
               0 & 0 & -1 & 0 & 0 & 0 \\
               0 & 0 & 0 & -1 & 0 & 0 \\
               0 & 0 & 0 & 0 & -\frac{\mu_{\infty}}{\lambda_{\infty}}\frac{1}{x^{2}} & 0 \\
               0 & 0 & 0 & 0 & 0 & -\frac{\mu_{\infty}}{\lambda_{\infty}}\frac{1}{x^{2}}
             \end{pmatrix}
             \begin{pmatrix}
               \delta u \\
               \delta v \\
               \delta u_{x} \\
               \delta v_{x} \\
               \delta u_{\theta} \\
               \delta v_{\theta}
             \end{pmatrix}$}
\end{align}
and alone suggests instability of the base state. However, according to the \citet{Dirac:1964} theory of constrained Hamiltonian systems, we must consider second variation of the constrained Hamiltonian $\delta^{2} \mathrm{H} + \lambda \, \delta^{2} \mathrm{N}$ and only dynamically accessible variations, i.e. tangent to the constraint,
\begin{align}
\label{variations:constraint}
\delta \mathrm{N} = 0, \ \text{i.e.} \ U \cdot \delta U = 0,
\end{align}
along with its differential consequences (consistency conditions), thus reducing the dimension of \eqref{Hessian:ncNLS:case-1} in half. Without detailed calculations, from the structure of \eqref{variation:second:case-1} it is clear that the second variation is sign-indefinite implying instability with the transverse perturbations playing destabilizing role.

Similar calculations for \textit{case 2}, using appropriate expressions for scaling constants in \eqref{scalings:Lagrange-Dirichlet} from \S \ref{subsec:ncNLS:BS}, instead of \eqref{ncNLS:scaled:case-1} yield for the scaled GP equation
\begin{align}\label{ncNLS:scaled:case-2}
\i y_{t} + \Delta_{x} y + \left(d - \frac{1}{4}\right) \frac{y}{x^{2}} + \frac{\mu_{\infty}}{\lambda_{\infty}} \frac{y_{\theta\theta}}{x^{2}} = |y|^{2} y,
\end{align}
and the Hessian density matrix
\begin{align}\label{Hessian:ncNLS:case-2}
             \begin{pmatrix}
               -\left(d-\frac{1}{4}\right) \frac{1}{x^{2}} + \left(3 u^{2} + v^{2}\right) & 2 u v & 0 & 0 & 0 & 0 \\
               2 u v & -\left(d-\frac{1}{4}\right) \frac{1}{x^{2}} + \left(3 v^{2} + u^{2}\right) & 0 & 0 & 0 & 0 \\
               0 & 0 & 1 & 0 & 0 & 0 \\
               0 & 0 & 0 & 1 & 0 & 0 \\
               0 & 0 & 0 & 0 & \frac{\mu_{\infty}}{\lambda_{\infty}}\frac{1}{x^{2}} & 0 \\
               0 & 0 & 0 & 0 & 0 & \frac{\mu_{\infty}}{\lambda_{\infty}}\frac{1}{x^{2}}
             \end{pmatrix},
\end{align}
which, under the same constrained conditions \eqref{variations:constraint}, again implies instability due to sign-indefiniteness of the second variation $\delta^{2} \mathrm{H} + \lambda \, \delta^{2} \mathrm{N}$; notably, the potential now plays a destabilizing role (w.r.t. the longitudinal perturbations) compared to \textit{case 1}, while the transverse perturbations have a stabilizing effect. In \textit{case 3}, however, we get for the scaled GP equation
\begin{align}\label{ncNLS:scaled:case-3}
\i y_{t} + \Delta_{x} y + \left(d - \frac{1}{4}\right) \frac{y}{x^{2}} + \frac{\mu_{\infty}}{\lambda_{\infty}} \frac{y_{\theta\theta}}{x^{2}} = - |y|^{2} y,
\end{align}
and the Hessian density matrix
\begin{align}\label{Hessian:ncNLS:case-3}
             \begin{pmatrix}
               -\left(d-\frac{1}{4}\right) \frac{1}{x^{2}} - \left(3 u^{2} + v^{2}\right) & - 2 u v & 0 & 0 & 0 & 0 \\
               - 2 u v & -\left(d-\frac{1}{4}\right) \frac{1}{x^{2}} - \left(3 v^{2} + u^{2}\right) & 0 & 0 & 0 & 0 \\
               0 & 0 & 1 & 0 & 0 & 0 \\
               0 & 0 & 0 & 1 & 0 & 0 \\
               0 & 0 & 0 & 0 & \frac{\mu_{\infty}}{\lambda_{\infty}}\frac{1}{x^{2}} & 0 \\
               0 & 0 & 0 & 0 & 0 & \frac{\mu_{\infty}}{\lambda_{\infty}}\frac{1}{x^{2}}
             \end{pmatrix},
\end{align}
which under the same constrained conditions \eqref{variations:constraint}, implies instability as nonlinearity now plays the destabilizing role due to the change of sign (from defocusing in \textit{case 2} to focusing in \textit{case 3}). The limits of (\ref{Hessian:ncNLS:case-2},\ref{Hessian:ncNLS:case-3}) for $x \rightarrow \infty$ correspond to defocusing/focusing cases of NLS, respectively. The corresponding Hessian \eqref{Hessian:ncNLS:case-3} thus recovers the known fact that solutions of the focusing 1D NLS are both longitudinally \citep{Zakharov:1968} and transversely \citep{Zakharov:1974} unstable (see further discussion in \S \ref{sec:introduction}), leading to a finite-time singularity when nonlinearity overpowers the dispersive spreading.

In conclusion, we are in the position to compare the above stability results with the spectral approach in \S \ref{subsec:spectral-analysis:ncNLS}. While in \textit{case 3} the conclusions of the Lagrange-Dirichlet method from the Hessian \eqref{Hessian:ncNLS:case-2} are in agreement with the spectral instability results of \S \ref{subsec:spectral-analysis:ncNLS}, in \textit{cases 1} and \textit{2} they appear to be at variance. However, as mentioned in \S \ref{sec:introduction}, spectral stability does not imply even linear stability, not to mention nonlinear (finite-amplitude) stability -- hence, the contradiction is only apparent. Having said that, the above spectral and Hamiltonian stability analyses apply to different conditions: the spectral approach giving spectral stability in \textit{cases 1} and \textit{2}, while instability in \textit{case 3} -- to the base states in the form of standing envelope solitary waves that are potentially singular at the origin as in \textit{case 1}, while the Hamiltonian approach -- to the base states which are smooth including at the origin and decay fast enough at infinity \textit{or} in the case when variations (and hence admissible perturbations) vanish at the origin and infinity. Lastly, it should be noted that since in all three cases the Lagrange-Dirichlet method implies instability, we do not have to deal with infinite-dimensional nature of the problem, which would otherwise impose extra work on establishing stability since positive-definiteness of the constrained Hamiltonian is not a sufficient condition for a local minimum to occur in infinite dimensions \citep{Krechetnikov:2009}.

\section{Waves on shallow water}

\subsection{Nearly concentric KdV with surface tension} \label{subsec:ncKdV}

Let us next consider nearly concentric water waves on shallow water, also in the inviscid potential approximation. Since our interest is to analyze the evolution of an envelope of a wave with wavelength $\ell$, the latter sets the natural lengthscale for non-dimensionalization in the horizontal direction, while the quiescent fluid layer depth $h$ -- in the vertical direction:
\begin{align}
\label{eqn:non-dimensionalization}
(r,z) \rightarrow \left(\ell r, h z\right), \ t \rightarrow \frac{\ell}{c_{0}} t, \ \eta \rightarrow a \, \eta, \ \phi \rightarrow a \, h^{-1} \, c_{0} \, \ell \, \phi,
\end{align}
where the phase speed $c_{0} = (g \, h)^{1/2}$ is dictated by the shallow water dispersion relation $\omega^{2} = k^{2} g \, h$, $a$ is the wave amplitude, and the scaling for $\phi$ follows from balancing the fluid acceleration at the interface with the hydrostatic pressure, $\phi_{t} \sim g \, \eta$. Altogether, this leads to the following non-dimensional system analogous to \eqref{system:deep-water:non-dimensional:cylindrical} in the deep water case
\begin{subequations}
\label{system:shallow-water:non-dimensional:cylindrical}
\begin{align}
\label{bulk:Laplace:non-dimensional}
z \le 1 + \alpha \, \eta(t,x)&: \quad
\left\{\begin{array}{c} \phi_{zz} + \delta^{2} \nabla_{\perp}^{2} \phi = 0, \\
\nabla \phi \rightarrow 0, \ z = 0,
\end{array}\right. \\
\label{interface:kinematic:non-dimensional} z = 1 + \alpha \, \eta(t,x)&: \quad
\phi_{z} = \delta^{2} \left[\eta_{t} + \alpha \, \nabla_{\perp} \phi \cdot \nabla_{\perp} \eta\right], \\
\label{interface:dynamic:non-dimensional} z = 1 + \alpha \, \eta(t,x)&: \quad
\phi_{t} + \eta + \frac{\alpha}{2} \left[\nabla_{\perp} \phi \cdot \nabla_{\perp} \phi + \frac{1}{\delta^{2}} \phi_{z}^{2}\right] + We \, \delta^{2} \nabla \cdot \mathbf{n} = 0,
\end{align}
\end{subequations}
where $\nabla_{\perp}=\left(\partial_{r},\frac{1}{r} \partial_{\theta}\right)$, the leading-order terms in the curvature are $\nabla \cdot \mathbf{n} = \eta_{rr} + \frac{1}{r} \eta_{r} + \frac{1}{r^{2}} \eta_{\theta\theta} + \mathcal{O}(\alpha^{2} \delta^{2})$, the Weber number $We = \sigma / (\rho \, g \, h^{2})$ measures the effect of surface tension relative the wave intertia (driven by gravity), $\delta = h / \ell$ is the shallowness parameter, and $\alpha = a / h$ is the scaled wave amplitude (the wave steepness). The latter is treated as small since we are interested in the balance of nonlinear and dispersive effects, which happens at small solution amplitudes only. As motivated by the study of \citet{Kadomtsev:1970} of transverse instability of plane (1D) solitons described the KdV equation \eqref{eqn:KP}, there is a natural generalization to weak 2D dependence (npKdV), which was initially done in the plane case by the aforementioned authors. In the nearly concentric case, it was argued by \citet{Johnson:1980} that in order to derive a ncKdV one needs the scaling $\tau = \alpha^{6} \delta^{-4} t$, $\xi = \alpha^{2} \delta^{-2} (r-t)$, $\Theta = \delta \alpha^{-2} \theta$, $\Phi = \alpha^{-1} \phi$, and $H = \alpha^{2} \delta^{-3} \eta$ since the balance occurs at large enough distance from the origin (and hence time) so that the wave amplitude is small due to radial spreading. However, one can derive the ncKdV following the same scaling as in the derivation of 1D KdV on the line \citep{Kano:1986}, i.e. choosing $\alpha = \delta^{2}$, because the wave amplitude $\alpha$ has not been fixed yet:
\begin{align}
\xi = r - t, \ \tau = \alpha \, t, \ \Theta = \frac{1}{\alpha^{1/2}} \theta,
\end{align}
where all new variables are $\mathcal{O}(1)$ meaning that $\theta \sim \mathcal{O}(\alpha^{1/2})$ belongs to a narrow sector as opposed to \eqref{ncNLS}, in which the azimuthal coordinate is defined for the entire circle $\theta \in[0,2\pi)$; thus, same as with $\xi$, we may consider $\Theta \in (-\infty,+\infty)$ in the limit $\alpha \rightarrow 0$. Also, if we are looking for large time behavior, $r$ is large too and must be replaced with $r = \left(\alpha \, \xi + \tau\right) / \alpha$; effectively, this means that geometric spreading measured by the ratio of dimensional quantities $\ell/r \ll 1$ is weak, which in the context of made approximations amounts to $r h / \ell \gg 1$ for non-dimensional $r$. The Laplace equation \eqref{bulk:Laplace:non-dimensional} then transforms to
\begin{align}
\phi_{zz} + \alpha \left[\phi_{\xi\xi} + \frac{\alpha}{\alpha \, \xi + \tau} \phi_{\xi} + \frac{\alpha}{\left(\alpha \, \xi + \tau\right)^{2}} \phi_{\Theta\Theta}\right] = 0,
\end{align}
with the solution being
\begin{multline}
\label{sln:ncKdV-2D}
\phi = \widetilde{\phi}_{0}(\tau,\xi,\Theta) + \alpha \left(\widetilde{\phi}_{1}(\tau,\xi,\Theta) - \frac{z^{2}}{2} \widetilde{\phi}_{0 \xi\xi}(\tau,\xi)\right) + \alpha^{2} \bigg[\widetilde{\phi}_{2}(\tau,\xi,\Theta) - \frac{z^{2}}{2} \bigg(\widetilde{\phi}_{1 \xi\xi}(\tau,\xi,\Theta) \\
+ \frac{1}{\tau}\widetilde{\phi}_{0 \xi}(\tau,\xi,\Theta) + \frac{1}{\tau^{2}}\widetilde{\phi}_{0 \Theta\Theta}(\tau,\xi,\Theta)\bigg) + \frac{z^{4}}{24} \widetilde{\phi}_{0 \xi\xi\xi\xi}(\tau,\xi,\Theta)\bigg] + \mathcal{O}(\alpha^{3}).
\end{multline}
The dynamic boundary condition yields
\begin{align}
\eta_{0} - \widetilde{\phi}_{0 \xi} + \alpha \left[\eta_{1} - \widetilde{\phi}_{1 \xi} + \frac{1}{2} \widetilde{\phi}_{0 \xi\xi\xi} + \widetilde{\phi}_{0 \tau} + \frac{1}{2} \widetilde{\phi}_{0 \xi}^{2} - We \, \eta_{0 \xi\xi}\right] + \mathcal{O}(\alpha^{2}) = 0,
\end{align}
while the kinematic one produces
\begin{multline}
- \alpha \left(1 + \alpha \eta_{0}\right) \widetilde{\phi}_{0 \xi\xi} + \alpha^{2} \left[- \widetilde{\phi}_{1 \xi\xi} - \frac{1}{\tau} \widetilde{\phi}_{0 \xi} - \frac{1}{\tau^{2}} \widetilde{\phi}_{0 \Theta\Theta} + \frac{1}{6} \widetilde{\phi}_{0 \xi\xi\xi\xi}\right] = \\
- \alpha \eta_{0 \xi} + \alpha^{2} \left[\eta_{0 \tau} - \eta_{1 \xi} + \widetilde{\phi}_{0 \xi} \eta_{0 \xi}\right] + \mathcal{O}(\alpha^{3}).
\end{multline}
Collecting terms of the same order gives $\eta_{0} = \widetilde{\phi}_{0 \xi}$ as well as the following two equations for the difference $\eta_{1} - \widetilde{\phi}_{1 \xi}$:
\begin{subequations}
\begin{align}
\eta_{1} - \widetilde{\phi}_{1 \xi} &= - \frac{1}{2} \widetilde{\phi}_{0 \xi\xi\xi} - \widetilde{\phi}_{0 \tau} - \frac{1}{2} \widetilde{\phi}_{0 \xi}^{2} + We \, \eta_{0 \xi\xi}, \\
\eta_{1 \xi} - \widetilde{\phi}_{1 \xi\xi} &= \eta_{0 \tau} + \frac{1}{\tau} \widetilde{\phi}_{0 \xi} + \frac{1}{\tau^{2}} \widetilde{\phi}_{0 \Theta\Theta} - \frac{1}{6} \widetilde{\phi}_{0 \xi\xi\xi\xi} + \eta_{0} \widetilde{\phi}_{0 \xi\xi} + \eta_{0 \xi} \widetilde{\phi}_{0 \xi} + We \, \eta_{0 \xi\xi\xi},
\end{align}
\end{subequations}
which after eliminating $\eta_{1} - \widetilde{\phi}_{1 \xi}$ furnish
\begin{align}
2 \, \eta_{0 \tau} + \frac{1}{\tau} \eta_{0} + \left(\frac{1}{3} - We\right) \eta_{0 \xi\xi\xi} + 3 \eta_{0} \eta_{0 \xi} + \frac{1}{\tau^{2}} \widetilde{\phi}_{0 \Theta\Theta} - We \, \eta_{0 \xi\xi\xi} = 0 \ \& \
\widetilde{\phi}_{0 \xi} = \eta_{0},
\end{align}
or ncKdV\marginlabel{discuss limit as $\xi \rightarrow \infty$ and relate to KP eqn}
\begin{align}
\label{eqn:cKP}
\left[2 \eta_{0 \tau} + \frac{1}{\tau} \eta_{0} + \left(\frac{1}{3}-We\right) \eta_{0 \xi\xi\xi} + 3 \eta_{0} \eta_{0 \xi}\right]_{\xi} + \frac{1}{\tau^{2}} \eta_{0 \Theta\Theta} = 0.
\end{align}
Without surface tension, $We=0$, equation \eqref{eqn:cKP} reduces to that derived by \citet{Johnson:1980}. The reason why the effect of surface tension enters by replacing the coefficient $\frac{1}{3}$ in front of $\eta_{0 \xi\xi\xi}$ to $\left(\frac{1}{3} - We\right)$ as in 1D KdV generalization onto the surface tension case \citep{Korteweg:1895,Benjamin:1982,Green:1983} is because the leading order curvature term in \eqref{eqn:curvature:cylindrical} in the considered approximation assumes the same form as in the plane (1D) case:
\begin{align}
\nabla \cdot \mathbf{n} = - \eta_{\xi\xi} + \mathcal{O}(\alpha).
\end{align}

\subsection{Single concentric soliton} \label{subsec:soliton:ncKdV}

The single concentric soliton, transverse instability of which we will be studying, is governed by the $\Theta$-independent variant of \eqref{eqn:cKP}:
\begin{align}
\label{eqn:cKdV}
2 H_{\tau} + \frac{1}{\tau} H + \left(\frac{1}{3}-We\right) H_{\xi\xi\xi} + 3 H H_{\xi} = 0,
\end{align}
which is translationally invariant in the radial coordinate $\xi$ as opposed to its deep water counterpart \eqref{ncNLS-ST}. Equation \eqref{eqn:cKdV} is known as a concentric KdV, which was originally derived by \citet{Maxon:1974b} in the context of ion-acoustic waves in a collisionless plasma, whose numerical simulations showed that solitary waves are characterized by $A \, \lambda^{2} \simeq \const$, where $A$ is the amplitude and $\lambda$ wavelength of the solitary wave. \citet{Cumberbatch:1978} further demonstrated that the amplitude dependence on radial position $r$ scales as $A \propto r^{-2/3}$. In the context of free-surface gravity waves, equation \eqref{eqn:cKdV} was first derived by \citet{Miles:1978b} from the Boussinesq equations, though without surface tension effects and $\tau$ replaced by $r$; hence, the self-similar solution was studied in that work in the $(r,\xi)$-variables. Numerically, cylindrical solitary waves were also explored by \citet{Chwang:1976}, on water of constant depth, but using the Boussinesq-type model. Some solutions to \eqref{eqn:cKdV} were constructed, cf. \citet{Calogero:1978,Johnson:1979}, with the inverse scattering transform.

\begin{figure}
	\setlength{\labelsep}{-3.0mm}
	\centering
    \sidesubfloat[]{\includegraphics[width=2.5in]{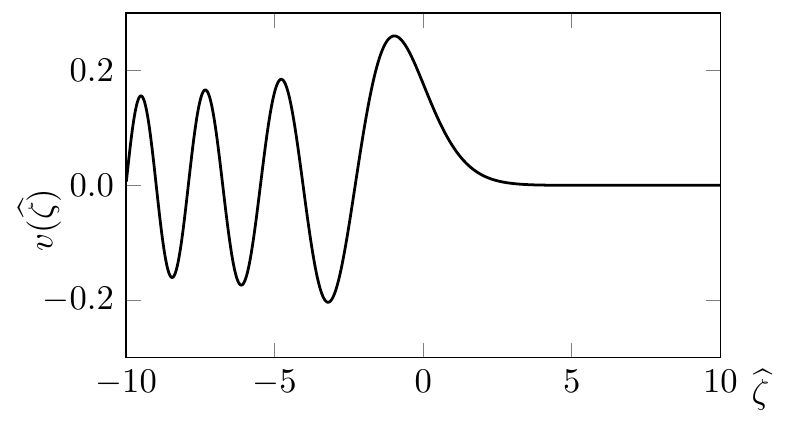}\label{fig:plot-y-a-pos}}
	\sidesubfloat[]{\includegraphics[width=2.5in]{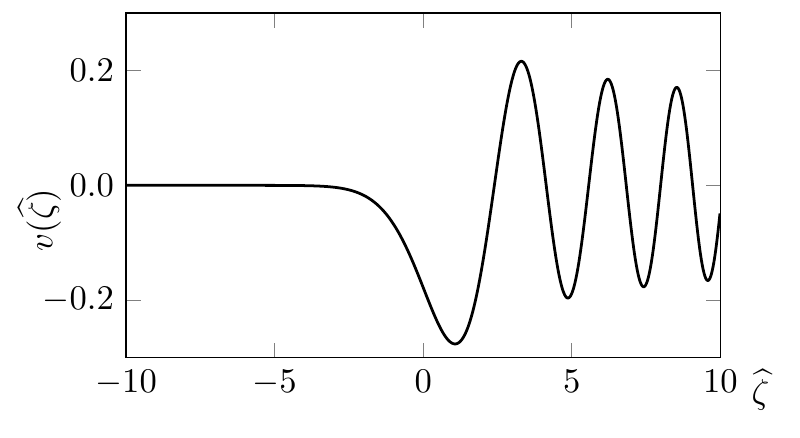}\label{fig:plot-y-a-neg}}
\caption{Solutions to \eqref{eqn:Painleve} for the parameter $a$ taking (a) positive and (b) negative values; for concreteness, we considered $|a|=1$.}
\end{figure}

On the symmetry side, note that the dilatation group of transformations
\begin{align}
\label{group:dilatational}
\tau \rightarrow \gamma^{-3/2} \, \tau^{\prime}, \ \xi \rightarrow \gamma^{-1/2} \, \xi^{\prime}, \ H \rightarrow \gamma \, H^{\prime},
\end{align}
leaves \eqref{eqn:cKdV} invariant. One way to interpret this group is that the scaling constant $\gamma$ falls out when we substitute in \eqref{eqn:cKdV} the solution of the form:
\begin{align}
H \, \gamma^{-1} = f(\tau \, \gamma^{3/2}, \xi \, \gamma^{1/2}).
\end{align}
Clearly, this representation corresponds to the structure of the 1D soliton solution \eqref{soliton:1D:KdV} with $\gamma$ being equivalent to $A$. However, such a solution is not allowed in the cylindrical case due to the lack of Galiliean invariance. Another implication of \eqref{group:dilatational} is the existence of a self-similar solution, which results from the fact that under \eqref{group:dilatational} the following complexes stay invariant:
\begin{align}
H \, \tau^{2/3} = H^{\prime} \, \tau^{\prime 2/3}, \ \xi \, \tau^{-1/3} = \xi^{\prime} \, \tau^{\prime -1/3} \equiv \zeta_{0},
\end{align}
and thus are functionally related via self-similar variables:
\begin{align}
\label{sln:cKdV:self-similar}
H(\tau,\xi) = \tau^{-2/3} F(\zeta), \ \zeta = \tau^{-1/3} \xi,
\end{align}
leading to a single solitary wave solution. While self-similar solutions of equation~\eqref{eqn:cKdV} have been constructed \citep{Johnson:1980} for $We=0$, we are going to explore the general case of $We>0$. The derivatives of \eqref{sln:cKdV:self-similar} are calculated according to
\begin{align}
H_{\tau} = - \frac{2}{3} \tau^{-5/3} F(\zeta) + \tau^{-2/3} F^{\prime}(\zeta) \, \zeta_{\tau}, \ \zeta_{\tau} = - \frac{1}{3} \frac{\zeta}{\tau}, \ H_{\xi} = \tau^{-2/3} F^{\prime}(\zeta) \frac{\zeta}{\xi},
\end{align}
thus leading to an ODE:
\begin{align}
\label{ODE:ncKdV:soliton:original}
 - \frac{1}{3} F - \frac{2 \, \zeta}{3} F^{\prime} + \left(\frac{1}{3} - We\right) F^{\prime\prime\prime} + 3 F F^{\prime} = 0,
\end{align}
where the translational invariance is lost. Multiplying the latter equation by $F$ and integrating once, we get
\begin{align}
\label{ODE:ncKdV:soliton:intermediate}
- \frac{1}{3} \zeta F^{2} + \left(\frac{1}{3} - We\right) \left[F F^{\prime\prime} - \frac{1}{2} F^{\prime 2}\right] + F^{3} = \mathrm{const},
\end{align}
where we used the facts that $\left(\zeta F^{2}\right)^{\prime} = F^{2} + 2 \, \zeta F F^{\prime}$ and $\left(F F^{\prime\prime}\right)^{\prime} = F^{\prime} F^{\prime\prime} + F F^{\prime\prime\prime}$. Further, introducing rescalings $\zeta = 2^{1/3} \widehat{\zeta}$ and $F = 2^{1/3} \widehat{F}/3$ we can simplify \eqref{ODE:ncKdV:soliton:intermediate} to
\begin{align}
\label{ODE:ncKdV:soliton:simplified}
\left(1 - 3 \, We\right) \left[\widehat{F} \widehat{F}^{\prime\prime} - \frac{1}{2} \widehat{F}^{\prime 2}\right] + 2 \left(\widehat{F}^{3} - \widehat{\zeta} \widehat{F}^{2}\right) = \mathrm{const}.
\end{align}
Introducing $\widehat{F} = v^{2}$ and putting the constant in \eqref{ODE:ncKdV:soliton:simplified} to zero (since the solution $\widehat{F}$ decays exponentially to zero at either infinity dictated by the sign of $\left(1 - 3 \, We\right)$), we can reduce \eqref{ODE:ncKdV:soliton:simplified} to the second Painlev\'{e} transcendent \citep{Ince:1944,Miles:1978}:
\begin{align}
\label{eqn:Painleve}
\alpha \, v^{\prime\prime} - \widehat{\zeta} v + v^{3} = 0,
\end{align}
where $\alpha = 1 - 3 \, We$. Naturally, we will require that $v \rightarrow 0$ as $\widehat{\zeta} \rightarrow \pm \infty$, but the rate of decay depends on the direction taken. Also, if one is interested in the solution of \eqref{eqn:Painleve} for negative values of parameter $\alpha$, with the transformation $\alpha \rightarrow -\alpha$, $\widehat{\zeta} \rightarrow -\widehat{\zeta}$, $v \rightarrow -v$, equation \eqref{eqn:Painleve} is transformed to $\alpha \, v^{\prime\prime} - \widehat{\zeta} v - v^{3} = 0$, i.e. only the sign of the nonlinear term changes, which has some noticeable quantitative effect on the form of the solution; however, qualitatively the solution looks similar as one may notice by applying the transformation $\widehat{\zeta} \rightarrow -\widehat{\zeta}$, $v \rightarrow -v$ to figure~\ref{fig:plot-y-a-pos} and comparing with \ref{fig:plot-y-a-neg}. The asymptotics of the solutions to \eqref{eqn:Painleve} is governed by the linearized version of \eqref{eqn:Painleve}, which follows from the fact that $v \rightarrow 0$ as $\widehat{\zeta} \rightarrow \pm \infty$ and hence behaves as the Airy function $\sim \Ai{\left(\widehat{\zeta}\right)}$, e.g. for $\alpha>0$:
\begin{subequations}
\label{asymptotics:Painleve}
\begin{align}
v & \sim C_{+} \frac{e^{-\frac{2}{3} \, \widehat{\zeta}^{3/2}}}{2 \sqrt{\pi} \, \widehat{\zeta}^{1/4}} \ \text{for} \ \widehat{\zeta} \rightarrow \infty; \\
\label{asymptotics:Painleve:minfinity}
v & \sim C_{-} \frac{1}{\sqrt{\pi} \, (-\widehat{\zeta})^{1/4}} \cos{\left[\frac{2}{3} \, (-\widehat{\zeta})^{3/2} - \frac{\pi}{4} + \varphi(\widehat{\zeta})\right]} \ \text{for} \ \widehat{\zeta} \rightarrow -\infty;
\end{align}
\end{subequations}
where $\widetilde{x} = \alpha^{-1/3} x$; for $\alpha<0$ the asymptotics \eqref{asymptotics:Painleve} inverts because with the transformations $\alpha \rightarrow -\alpha$, $\widehat{\zeta} \rightarrow -\widehat{\zeta}$ the linearized part of \eqref{eqn:Painleve} $\alpha \, v^{\prime\prime} - \widehat{\zeta} v = 0$ stays intact. Phase correction $\varphi(\widehat{\zeta})$ to \eqref{asymptotics:Painleve:minfinity} is computed similar to Appendix~\ref{appx:asymptotics:infinity} and yields $\varphi(\widehat{\zeta}) \sim - \frac{3 C_{-}^{2}}{4 \pi} \ln{\left[- \widehat{\zeta}\right]}$, cf. \citep{Ablowitz:1977b,Miles:1978}.

On the conservation law side, previously \citet{Maxon:1974b}, \citet{Cumberbatch:1978}, and \citet{Ko:1979} claimed the existence of the two for \eqref{eqn:cKdV}. The first $\mathrm{I}_{1}$ is found by integrating \eqref{eqn:cKdV} w.r.t. $\xi$ and assuming that the solution and its derivatives up to second order decay at $\xi \pm \infty$, which yields
\begin{align}
\label{conservation:mass:ncKdV}
2 \frac{\d}{\d \tau} \mathrm{I}_{1} + \frac{1}{\tau} \mathrm{I}_{1} = 0, \ \mathrm{I}_{1} = \int_{\Bbb{R}}{H \, \d \xi},
\end{align}
meaning that $\tau^{1/2} \mathrm{I}_{1} = \const$. However, in this derivation the assumption that $H_{\xi\xi} \rightarrow 0$ as $\xi \rightarrow - \infty$ for $\alpha > 0$, cf. figure~\ref{fig:plot-y-a-pos}, and $\xi \rightarrow \infty$ for $\alpha < 0$, cf. figure~\ref{fig:plot-y-a-neg}, is not valid for a self-similar solution \eqref{sln:cKdV:self-similar} unless one considers long enough time limit\marginlabel{figure out the scaling!} or proves that due to fast oscillations the integral of $H_{\xi\xi}$ converges to zero\marginlabel{Riemann–Lebesgue lemma?}. Indeed, as follows from the analysis of equation \eqref{eqn:Painleve}, in the oscillatory tail the solution $H(\tau,\xi)$ behaves as:
\begin{align}
\label{asymptotics:H}
H \sim \xi^{-1/2}, \ H_{\xi} \sim \xi^{0}, \ H_{\xi\xi} \sim \xi^{1/2},
\end{align}
for $\alpha > 0$ and $\xi \rightarrow - \infty$. Similarly, multiplying \eqref{eqn:cKdV} by $H(\tau,\xi)$ and integrating w.r.t. $\xi$ produces the second conservation law
\begin{align}
\label{conservation:energy:ncKdV}
\frac{\d}{\d \tau} \mathrm{I}_{2} + \frac{1}{\tau} \mathrm{I}_{2} = 0, \ \mathrm{I}_{2} = \int_{\Bbb{R}}{H^{2} \, \d \xi},
\end{align}
meaning that $\tau \, \mathrm{I}_{2} = \const$, but the same assumption that $H_{\xi\xi} \rightarrow 0$ as $\xi \rightarrow - \infty$ is invalid. The validity of these conservation laws (\ref{conservation:mass:ncKdV},\ref{conservation:energy:ncKdV}) was asserted only based on the comparison with numerical solutions \citep{Maxon:1974b,Cumberbatch:1978}. The difficulty of comparing with experimental data was discussed by \citet{Stepanyants:1981}, which nevertheless favored the scaling for the amplitude with the radial coordinate $r$ as $\sim r^{-2/3}$ as opposed to $r^{-1/2}$, which one would expect from the above conservation laws (however, the soliton width may change thus affecting the scaling). In the context of cylindrical solitary waves, experiments of \cite{Weidman:1988} in the shallow water regime confirmed that an isolated disturbance evolves into a slowly varying solitary wave with amplitude decaying as $A \propto r^{-2/3}$.

As discussed above, the conservation laws (\ref{conservation:mass:ncKdV},\ref{conservation:energy:ncKdV}) are valid only for localized solutions, which may exist initially or transiently, but not in the long-time limit when the self-similar solutions of the sort \eqref{sln:cKdV:self-similar} establish. The form of both (\ref{conservation:mass:ncKdV},\ref{conservation:energy:ncKdV}) suggests \textit{non-conservative} nature of \eqref{eqn:cKdV}. Indeed, in order to put the latter in a Hamiltonian form, first we would need to transform $H(\tau,\xi) = \tau^{-1/2} u(\tau,\xi)$ to remove the second term in \eqref{eqn:cKdV},
\begin{align}
\label{eqn:cKdV:transformed}
2 u_{\tau} + \left(\frac{1}{3}-We\right) u_{\xi\xi\xi} + 3 \, \tau^{-1/2} u u_{\xi} = 0,
\end{align}
which allows us to put the resulting equation for $u(\tau,\xi)$ in the non-canonical Hamiltonian form:
\begin{align}
\label{form:Hamiltonian:ncKdV}
u_{\tau} = \frac{\partial}{\partial \xi} \frac{\delta \mathcal{H}}{\delta u}, \ \mathcal{H} = \frac{1}{4} \left[\left(\frac{1}{3}-We\right) \int_{-\infty}^{\infty}{u_{\xi}^{2} \, \d\xi} - \tau^{-1/2} \int_{-\infty}^{\infty}{u^{3} \, \d\xi}\right],
\end{align}
i.e. depending upon the sign of $\frac{1}{3}-We$ the Hamiltonian $\mathcal{H}$ changes from focusing to defocusing thus suggesting the corresponding change in stability properties, which we will see in \S \ref{subsec:KP:ncKdV:analysis}. The fact that the Hamiltonian form \eqref{form:Hamiltonian:ncKdV} is non-canonical since the operator $J = \partial_{\xi}$ is non-invertible in general suggests the existence of Casimirs $C_{i}(\tau)$, $i=1,\ldots$. Also, despite the existence of the Hamiltonian $\mathcal{H}$, the non-autonomous character of \eqref{form:Hamiltonian:ncKdV} and the prior transformation from \eqref{eqn:cKdV} to \eqref{eqn:cKdV:transformed} indicates non-conservative nature of the ncKdV in the sense that energy is no longer a constant of motion.

\subsection{Non-existence of a critical transverse wavenumber} \label{subsec:KdV:stability-preliminary}

To analyze the transverse instability of the self-similar solution \eqref{sln:cKdV:self-similar}, we linearize \eqref{eqn:cKP} around the latter, $\eta_{0} = H + \eta^{\prime}$, thus leading to
\begin{align}
\label{eqn:perturbation:ncKdV}
\left[2 \, \eta_{\tau}^{\prime} + \frac{1}{\tau} \eta^{\prime} + \frac{\alpha}{3} \eta_{\xi\xi\xi}^{\prime} + 3 \left(H \eta^{\prime}\right)_{\xi}\right]_{\xi} + \frac{1}{\tau^{2}} \eta_{\Theta\Theta}^{\prime} = 0.
\end{align}
Since the base state $H(\tau,\xi)$ is time-dependent, to make proper interpretation of the stability analysis the perturbation $\eta^{\prime}$ must be scaled in the same fashion as \eqref{sln:cKdV:self-similar}:
\begin{align}
\eta^{\prime} = \tau^{-2/3} h(\tau,\xi),
\end{align}
as well as the independent variables must be transformed according to
\begin{align}
\label{variables:self-similar:ncKdV}
(\tau,\xi,\Theta) \rightarrow (\widehat{\tau} = \ln{\tau},\zeta = \tau^{-1/3} \xi, \widetilde{\Theta} = \tau^{1/3} \Theta),
\end{align}
thus requiring the transformation of derivatives according to
\begin{align}
\partial_{\tau} = \frac{1}{\tau} \partial_{\widehat{\tau}} - \frac{1}{3} \frac{\zeta}{\tau} \partial_{\zeta} + \frac{1}{3} \frac{\widetilde{\Theta}}{\tau} \partial_{\widetilde{\Theta}}, \ \partial_{\xi} = \frac{\zeta}{\xi} \partial_{\zeta} = \frac{1}{\tau^{1/3}} \partial_{\zeta}, \ \partial_{\Theta} = \tau^{1/3} \, \partial_{\widetilde{\Theta}}.
\end{align}
The resulting equation for $h(\widehat{\tau},\zeta,\widetilde{\Theta})$ reads:
\begin{align}
\left[2 \, h_{\widehat{\tau}} - \frac{2}{3} \left(\zeta h_{\zeta} - \widetilde{\Theta} h_{\widetilde{\Theta}}\right) - \frac{1}{3} h + \frac{\alpha}{3} \, h_{\zeta\zeta\zeta} + 3 \left(F h\right)_{\zeta}\right]_{\zeta} + h_{\widetilde{\Theta}\widetilde{\Theta}} = 0;
\end{align}
for the purpose of studying the temporal transverse instability, we will look for solutions of the above linear equation in the form
\begin{align}
h = e^{\lambda \widehat{\tau}} f(\zeta,\widetilde{\Theta}),
\end{align}
which gives a PDE eigenvalue problem with variable coefficients:
\begin{align}
\label{EP:ncKdV:rescaled}
\left[\left(2 \, \lambda - \frac{1}{3}\right) f - \frac{2}{3} \left(\widehat{\zeta} f_{\widehat{\zeta}} - \widehat{\Theta} f_{\widehat{\Theta}}\right) + \frac{\alpha}{6} \, f_{\widehat{\zeta}\widehat{\zeta}\widehat{\zeta}} + \left(v^{2} f\right)_{\widehat{\zeta}}\right]_{\widehat{\zeta}} + f_{\widehat{\Theta}\widehat{\Theta}} = 0,
\end{align}
subject to $|f| \rightarrow 0$ for $\zeta, \widetilde{\Theta} \rightarrow \pm \infty$ since we are looking for perturbations of finite energy (in $L^{2}$-norm); in \eqref{EP:ncKdV:rescaled} we used the same variables $\zeta = 2^{1/3} \widehat{\zeta}$ and $F = 2^{1/3} \widehat{F}/3$ as in \eqref{ODE:ncKdV:soliton:simplified} along with the rescaling $\widehat{\Theta} = 2^{1/6} \widetilde{\Theta}$ as well as took into account that $F = v^{2}$ with $v(\widehat{\zeta})$ governed by \eqref{eqn:Painleve}. Hence, despite that the base state \eqref{sln:cKdV:self-similar} is time-dependent, the corresponding linear evolution problem for a superimposed perturbation can be reduced to eigenvalue problem \eqref{EP:ncKdV:rescaled} in the plane of self-similar variables $(\widehat{\zeta},\widehat{\Theta})$, as opposed to other familiar stability problems on time-dependent domains \citep{Homsy:1973,Krechetnikov:2017b}. As evident form the far-field behavior (\ref{asymptotics:Painleve},\ref{asymptotics:H}), the eigenvalue problem \eqref{EP:ncKdV:rescaled} is singular with aperiodically oscillating and growing coefficients, which makes it challenging for accurate numerical approximation and hence deserves a separate study. The latter is beyond the scope of the present work as we will develop analytical insights into stability picture below in this section as well as in \S \ref{subsec:KP:ncKdV:analysis} with the help of the \citet{Kadomtsev:1970} type analysis.

At the point, however, we may note an important property of \eqref{EP:ncKdV:rescaled}, namely its structure indicates that there exists no solution of the form $f \sim e^{\i k \widehat{\Theta}}$, i.e. which would be periodic in the angular coordinate $\widehat{\Theta}$ and produce a regularly spaced ``spike'' structure. This observation holds regardless how we would scale the angular variable with respect to time and, of course, is contrary to standard intuition, but can be seen as a consequence of an effective `nonlinearity' built-in the linear stability problem through the base state-dependent term $v^{2}$ manifesting itself in the interaction of two effects: as the single-soliton travels outwards (1) the circular domain is stretching, which inevitably leads to insertion of new wavelengths via the Eckhaus mechanism \citep{Knobloch:2014,Knobloch:2015,Krechetnikov:2017}, and (2) the soliton amplitude decrease, which affects the most unstable wavelength if one adopts the plane (1D) stability picture (\S \ref{sec:introduction}). The competition between these two effects is responsible for an irregular along $\widehat{\Theta}$ structure and non-existence of a single most amplified wavenumber thus demonstrating the crucial differences between the transverse instability of plane and cylindrical solitons. As we saw in \S \ref{sec:deep-water}, this phenomenon, however, does not happen in the deep water case, in particular due to the different underlying dispersive relation.

In the limit when the transverse part of \eqref{eqn:perturbation:ncKdV} can be considered as a perturbation, in particular for long times, one can see that stability changes to instability with the sign of parameter $\alpha$ based on the following simple considerations. Taking the Fourier transform of \eqref{eqn:perturbation:ncKdV} in $\Theta$, we get
\begin{align}
\left[2 \, \widehat{\eta}_{\tau} + \frac{1}{\tau} \widehat{\eta} + \frac{\alpha}{3} \widehat{\eta}_{\xi\xi\xi} + 3 \left(H \widehat{\eta}\right)_{\xi}\right]_{\xi} - \frac{k^{2}}{\tau^{2}} \widehat{\eta} = 0,
\end{align}
which after integrating twice w.r.t. $\xi$ gives
\begin{align}
\label{eqn:linearized:ncKdV:g}
2 g_{\tau\xi} + \frac{1}{\tau} g_{\xi} + \frac{\alpha}{3} g_{\xi\xi\xi\xi} + 3 H g_{\xi\xi} - \frac{k^{2}}{\tau^{2}} g = C_{1} \xi + C_{2},
\end{align}
where $g_{\xi\xi} = \widehat{\eta}$ and $C_{1}=C_{2}=0$ as $g \rightarrow 0$ for $\xi \rightarrow +\infty$ for $\alpha>0$. To simplify equation \eqref{eqn:linearized:ncKdV:g} further we use the transformation $g(\tau,\xi) = \tau^{-\frac{1}{2}} \chi(\tau,\xi)$, which brings it to
\begin{align}
2 \chi_{\tau\xi} + \frac{\alpha}{3} \chi_{\xi\xi\xi\xi} + 3 H \chi_{\xi\xi} - \frac{k^{2}}{\tau^{2}} \chi = 0.
\end{align}
Considering the last two terms as perturbations for $\xi \rightarrow +\infty$, by splitting the solution $\chi = \chi^{0} + \chi^{1}$ the problem can be recast into
\begin{subequations}
\begin{align}
\label{EVP:ncKdV:0}
2 \chi_{\tau\xi}^{0} + \frac{\alpha}{3} \chi_{\xi\xi\xi\xi}^{0} &= 0, \\
\label{EVP:ncKdV:1}
2 \chi_{\tau\xi}^{1} + \frac{\alpha}{3} \chi_{\xi\xi\xi\xi}^{1} &= - 3 H \chi_{\xi\xi}^{0} + \frac{k^{2}}{\tau^{2}} \chi^{0},
\end{align}
\end{subequations}
where the ``smallness'' of $H$ for $\xi \rightarrow +\infty$ follows from \eqref{asymptotics:H}. Looking for an asymptotic solution of the first of these equations at $\xi \rightarrow +\infty$, i.e. $\chi^{0} \sim C_{0} e^{\mu \xi} e^{\lambda \tau}$, we find $\mu^{3} = 6 \lambda/\alpha$ and the real part $\Re{(\mu)}$ of $\mu$ must be negative as physically relevant solutions must decay at $\xi \rightarrow +\infty$. Hence, regardless whether $\lambda$ and $\mu$ are complex or real, if $\alpha$ changes sign, then the real part of $\lambda$ must change sign as well. Hence, the behavior is analogous to that of the npKdV \eqref{eqn:KP} and qualitatively similar to that in the GP equation (\S \ref{subsec:spectral-analysis:ncNLS}), i.e. instability appears at sufficiently high Weber numbers, though in the latter case they are measured in the carrier wavelength $2\pi/k_{0}$ compared to the layer depth $h$ in the case of ncKdV. However, as we will see in the next section, the short-time stability characteristics of ncKdV with application to the self-similar solution \eqref{sln:cKdV:self-similar} are very different from the considered here long-time limit conforming to our intuition developed in the near planar case of npKdV -- and this difference is due to the essential time-dependence of the base state \eqref{sln:cKdV:self-similar}.

It is easy to show that $\chi^{1}$ essentially follows the time-evolution of $\chi^{0}$ albeit with an algebraic function of $\tau$ multiplying the exponential $e^{\lambda \tau}$. For example, focusing on the last term on the rhs of \eqref{EVP:ncKdV:1} responsible for the input of the azimuthal perturbation, we may look for a particular solution of \eqref{EVP:ncKdV:1} in the form
\begin{align}
\chi^{1} = A(\tau) e^{\mu \xi}, \ \text{where} \ A(\tau) \sim e^{- \frac{\alpha \mu^3}{6} \tau} \int^{\tau}{\frac{k^{2}}{\widetilde{\tau}^{2}} e^{\left(\lambda+\frac{\alpha \mu^3}{6}\right) \widetilde{\tau}} \, \d \widetilde{\tau}} \sim k^{2} \frac{e^{\lambda \tau}}{\tau} \ \text{for} \ \tau \gg 1.
\end{align}
As consistent with the observation made earlier, there is no preferred wavenumber $n$ in the azimuthal direction. The contribution of the base state, i.e. the first term on the rhs of \eqref{EVP:ncKdV:1}, can be computed analogously, after the transformation \eqref{variables:self-similar:ncKdV} to self-similar variables $(\tau,\xi) \mapsto (\tau,\zeta)$.

\subsection{Kadomtsev-Petviashvili type analysis} \label{subsec:KP:ncKdV:analysis}

Finally, let us develop analysis of transverse instability of the self-similar solution \eqref{sln:cKdV:self-similar} to the ncKdV equation \eqref{eqn:cKP} rewritten, for the ease of notation and comparison with the classical analysis of npKdV \citep{Kadomtsev:1970,Alexander:1997}, in the form
\begin{align}
\label{eqn:ncKdV:form-2}
2 \, \eta_{\tau} + \frac{1}{\tau} \eta + 3 \, \eta \, \eta_{\xi} + \eta_{\xi\xi\xi} + \frac{\beta}{\tau^{2}} \, \partial_{\xi}^{-1} \eta_{\Theta\Theta} = 0,
\end{align}
after moving the surface tension factor $\frac{\alpha}{3} = \frac{1}{3} - We$ to the last term in \eqref{eqn:ncKdV:form-2} via the rescaling of \eqref{eqn:cKP} with
\begin{align}
\tau \rightarrow \gamma \tau, \ \xi \rightarrow \left(\frac{\alpha \, \gamma}{3}\right)^{1/3} \xi, \ \eta_{0} \rightarrow \left(\frac{\alpha}{3 \, \gamma^{2}}\right)^{1/3} \eta, \ \Theta \rightarrow \left(\frac{\alpha \, \gamma^{4}}{3 \, \beta^{3}}\right)^{1/3} \Theta,
\end{align}
without intruding new notations for the variables, but dropping index $0$ in $\eta_{0}$; note that in the above rescalings the factor $\beta > 0$ for $\alpha>0$; if, on the other hand, $\alpha<0$, then the factor $\beta < 0$. When the solution does not depend on the transverse coordinate $\Theta$, equation \eqref{eqn:ncKdV:form-2} admits the self-similar solution \eqref{sln:cKdV:self-similar}:
\begin{align}
\label{solution:cKdV:self-similar}
\eta \, \tau^{2/3} = F(\xi \, \tau^{-1/3}) \ \Rightarrow \ \eta = \tau^{-2/3} F(\zeta_{0}), \ \zeta_{0}=\frac{\xi}{\tau^{1/3}}.
\end{align}
This is the solution the transverse instability of which we will study by perturbing its amplitude and phase in analogy to the analysis of \citet{Kadomtsev:1970} (see also \citet{Kodama:2018,Ablowitz:1981} for interpretative accounts), who performed stability analysis of 1D plane $\mathrm{sech^{2}}$-soliton \eqref{eqn:self-similar:1D-KdV} with the help of the Krylov-Bogoliubov method \citep{Bogoliubov:1961}, translated here onto the stability analysis of a self-similar solution \eqref{solution:cKdV:self-similar}:
\begin{align}
\label{sln-perturbed:ncKdV}
\eta(t,T,\xi,\widetilde{\Theta}) = \tau^{-2/3} \left[1 + A(T,\widetilde{\Theta})\right] F\left(\frac{\xi+\varphi}{\tau^{1/3}}\right);
\end{align}
here we will assume $A = \mathcal{O}(\epsilon)$ and $\varphi = \mathcal{O}(\epsilon^{c})$ with time $T = \epsilon^{a} \tau$ and slow transverse coordinate $\widetilde{\Theta} = \epsilon^{b} \Theta$; the exponents $a$, $b$, and $c$ are to be determined with the requirement that one must have $b > 0$ for long-wave instability. The time derivative is calculated to become
\begin{multline}
\eta_{\tau} = - \frac{2}{3} \, \tau^{-5/3} (1+A) \, F(\zeta) + \tau^{-2/3} \, A_{\tau} \, F(\zeta) \\
+ \tau^{-2/3} (1+A) \, F^{\prime}(\zeta) \left[-\frac{1}{3} \frac{\zeta}{\tau} + \frac{\varphi_{\tau}}{\tau^{1/3}}\right],
\end{multline}
with $\zeta=(\xi+\varphi)/\tau^{1/3}$, while the first derivative w.r.t. $\Theta$ reads
\begin{align}
\eta_{\Theta} = \tau^{-2/3} \, A_{\Theta} F(\zeta) + \tau^{-2/3} (1+A) \, F^{\prime}(\zeta) \frac{\varphi_{\Theta}}{\tau^{1/3}},
\end{align}
and the second derivative w.r.t. $\Theta$
\begin{multline}
\eta_{\Theta\Theta} = \mathop{\tau^{-2/3} A_{\Theta\Theta}}_{\mathcal{O}(\epsilon^{1+2b+\frac{2}{3}a})}F(\zeta) + \mathop{2 \, \tau^{-2/3} \, A_{\Theta} \, \varphi_{\Theta} \, \tau^{-1/3}}_{\mathcal{O}(\epsilon^{a+1+c+2b})} F^{\prime}(\zeta) + \\
\mathop{\tau^{-2/3} (1+A) \, \varphi_{\Theta}^{2} \, \tau^{-2/3}}_{\mathcal{O}(\epsilon^{\frac{4}{3}a+2c+2b})} F^{\prime\prime}(\zeta) + \mathop{\tau^{-2/3} (1+A) \, \varphi_{\Theta\Theta} \, \tau^{-1/3}}_{\mathcal{O}(\epsilon^{a+c+2b})} F^{\prime}(\zeta),
\end{multline}
where under each term we show its order of magnitude once time $\tau$ and transverse direction $\Theta$ derivatives are understood in their modulational counterparts $T$ and $\widetilde{\Theta}$, and orders of $A$ and $\varphi$ are taken into account; the nonlinear terms in the above expression, under appropriate justification, must be omitted in the linear analysis. Since $\eta_{\xi} = \tau^{-1} (1+A) F^{\prime}(\zeta)$, $\eta_{\xi\xi\xi} = \tau^{-5/3} (1+A) F^{\prime\prime\prime}(\zeta)$, and $\partial^{-1} \xi = \tau^{1/3} \partial^{-1} \zeta$, at the leading order we get the equation for a self-similar soliton:
\begin{multline}
2 \left[-\frac{2}{3} \tau^{-5/3} F(\zeta) + \tau^{-2/3} F^{\prime}(\zeta)\left(-\frac{1}{3} \frac{\zeta}{\tau}\right)\right] + \tau^{-5/3} F(\zeta) + \\
3 \, \tau^{-2/3} F(\zeta) \, \tau^{-1} F^{\prime}(\zeta) + \tau^{-2/3} F^{\prime\prime\prime}(\zeta) \, \tau^{-1} = 0, \nonumber
\end{multline}
or dividing w.r.t. $\tau^{-5/3}$:
\begin{align}
\label{eqn:soliton:1D:ncKdV:self-similar}
-\frac{1}{3} F(\zeta) - \frac{2 \, \zeta}{3} F^{\prime}(\zeta) + 3 \, F(\zeta) \, F^{\prime}(\zeta) + F^{\prime\prime\prime}(\zeta) = 0.
\end{align}
Since in the cylindrical case there is no translational symmetry, we must expand \eqref{eqn:soliton:1D:ncKdV:self-similar} about $\zeta_{0}$ as the shift of $x$ changes the stability properties of the cylindrical soliton. Thus, taking into account that
\begin{subequations}
\begin{align}
\zeta F^{\prime}(\zeta) &= \left(F^{\prime}(\zeta_{0}) + F^{\prime\prime}(\zeta_{0}) \frac{\varphi}{\tau^{1/3}}\right)\left(\zeta_{0} + \frac{\varphi}{\tau^{1/3}}\right), \\
F(\zeta) \, F^{\prime}(\zeta) &= \left(F(\zeta_{0}) + F^{\prime}(\zeta_{0}) \frac{\varphi}{\tau^{1/3}}\right)\left(F^{\prime}(\zeta_{0}) + F^{\prime\prime}(\zeta_{0}) \frac{\varphi}{\tau^{1/3}}\right),
\end{align}
\end{subequations}
linearization of equation \eqref{eqn:soliton:1D:ncKdV:self-similar} results in (the first-order perturbation):
\begin{multline}
-\frac{1}{3} F^{\prime}(\zeta_{0}) \frac{\varphi}{\tau^{1/3}} - \frac{2}{3} \left[F^{\prime}(\zeta_{0}) + \zeta_{0} F^{\prime\prime}(\zeta_{0})\right] \frac{\varphi}{\tau^{1/3}} + \\
3 \left[F(\zeta_{0}) F^{\prime\prime}(\zeta_{0}) + F^{\prime 2}(\zeta_{0})\right] \frac{\varphi}{\tau^{1/3}} + F^{(iv)}(\zeta_{0}) \frac{\varphi}{\tau^{1/3}} = 0,
\end{multline}
which must be added (after multiplying by $\tau^{-5/3}$) to the linearization of \eqref{eqn:ncKdV:form-2}:
\begin{align}
&2 \left[-\frac{2}{3} \tau^{-1} A F(\zeta_{0}) + A_{\tau} F(\zeta_{0}) - \frac{1}{3} \frac{\zeta_{0}}{\tau} A F^{\prime}(\zeta_{0}) + F^{\prime}(\zeta_{0}) \frac{\varphi_{\tau}}{\tau^{1/3}}\right] + \tau^{-1} A F(\zeta_{0}) + \\
&3 \, \tau^{-1} 2 \, A F(\zeta_{0}) F^{\prime}(\zeta_{0}) + \tau^{-1} A F^{\prime\prime\prime}(\zeta_{0}) + \beta \frac{\tau^{1/3}}{\tau^{2}} \partial_{\zeta}\left[A_{\Theta\Theta} F(\zeta_{0}) + \varphi_{\Theta\Theta} \tau^{1/3} F^{\prime}(\zeta_{0})\right] = 0, \nonumber
\end{align}
multiplied by $\tau^{-2/3}$, altogether producing
\begin{align}
\varphi \, \tau^{-2} &\left\{-\frac{4}{3} F^{\prime} - \frac{2}{3} \left[F^{\prime} + \zeta_{0} F^{\prime\prime}\right] + F^{\prime} + 3 \left(F F^{\prime\prime} + F^{\prime 2}\right) + F^{(iv)}\right\} \nonumber \\
+ 2 &\left\{\underline{-\frac{2}{3} \tau^{-5/3} A F} + \tau^{-2/3} A_{\tau} F + \underline{\tau^{-2/3} A F^{\prime} \left(-\frac{1}{3}\frac{\zeta_{0}}{\tau}\right)} + \tau^{-2/3} F^{\prime} \frac{\varphi_{\tau}}{\tau^{1/3}}\right\} \label{linearization:ncKdV} \\
+ &\underline{\tau^{-5/3} A F + 6 \, A \, \tau^{-5/3} F F^{\prime} + \tau^{-5/3} A F^{\prime\prime\prime}} + \beta \tau^{-2} \partial_{\zeta}^{-1} \left[\tau^{-1/3} A_{\Theta\Theta} F + \tau^{-2/3} \varphi_{\Theta\Theta} F^{\prime}\right] = 0. \nonumber
\end{align}
Multiplying \eqref{eqn:soliton:1D:ncKdV:self-similar} evaluated at $\zeta=\zeta_{0}$ by $A \, \tau^{-5/3}$ eliminates the terms underlined in \eqref{linearization:ncKdV} simplifying the latter to
\begin{align}
\label{eqn:KP:intermediate}
&\varphi \, \tau^{-2} \left[-\frac{4}{3} F^{\prime} - \frac{2}{3} \left(F^{\prime} + \zeta_{0} F^{\prime\prime}\right) + F^{\prime} + 3 \left(F F^{\prime\prime} + F^{\prime 2}\right) + F^{(iv)}\right] + \\
&2 \mathop{\tau^{-2/3} \frac{\varphi_{\tau}}{\tau^{1/3}} F^{\prime}}_{\mathcal{O}(\epsilon^{2 a + c})} + 2 \mathop{\tau^{-2/3} A_{\tau} F}_{\mathcal{O}(\epsilon^{1 + \frac{5}{3} a})} + \mathop{3 \, A \, \tau^{-5/3} F F^{\prime}}_{\mathcal{O}(\epsilon^{1 + \frac{5}{3} a})} + \beta \tau^{-2} \partial_{\zeta}^{-1} \left[\mathop{\tau^{-1/3} A_{\Theta\Theta} F}_{\mathcal{O}(\epsilon^{1 + \frac{7}{3} a + 2 b})} + \mathop{\tau^{-2/3} \varphi_{\Theta\Theta} F^{\prime}}_{\mathcal{O}(\epsilon^{\frac{8}{3} a + c + 2 b})}\right], \nonumber
\end{align}
where the expression in the first brackets also vanish because equation \eqref{eqn:soliton:1D:ncKdV:self-similar} differentiated once and evaluated at $\zeta=\zeta_{0}$ yields the same expression. As a result, we are left with five terms in \eqref{eqn:KP:intermediate} having, in general, four different exponents in the respective orders of $\epsilon$: \circled{1} $2 a + c$, \circled{2} $1 + \frac{5}{3} a$, \circled{3} $1 + \frac{7}{3} a + 2 b$, \circled{4} $\frac{8}{3} a + c + 2 b$. Consideration of all possible matching combinations leaves reasonable only two options:
\begin{enumerate}
  \item $\circled{1}=\circled{2}$ yields $c = 1 - \frac{1}{3} a$, in which case $\circled{3}=\circled{4}$. In this case, there is a possibility of a slow developing long-wave instability.
  \item $\circled{2}=\circled{4}$ yielding $c = 1 - a - 2 b$, while $\circled{1}=\circled{3}$ produces $c = 1 + \frac{1}{3} a + 2 b$. Altogether, this leads to $b = - \frac{1}{3} a$ and therefore $\circled{1}=\circled{2}=\circled{3}=\circled{4}$, in which case the instability is fast $(a<0)$, but still long-wave $(b>0)$.
\end{enumerate}

Given that the most interesting and physically relevant case is the second one, i.e. if instability develops at short times then it will dominate the subsequent dynamics and the case (i) becomes irrelevant, let us proceed with its analysis:
\begin{align}
\label{eqn:perturbation:ncKdV:KP}
2 \, A_{\tau} F + 2 \frac{\varphi_{\tau}}{\tau^{1/3}} F^{\prime} + 3 \, A \, \tau^{-1} F \, F^{\prime} + \beta \, \tau^{-5/3} A_{\Theta\Theta} \int^{\zeta_{0}}{F \, \d \widehat{\zeta}} + \beta \, \tau^{-2} \varphi_{\Theta\Theta} F = 0.
\end{align}
The challenge of applying the \citet{Kadomtsev:1970} type analysis to \eqref{eqn:perturbation:ncKdV:KP} consists, in particular, in the lesser degree of localization of the soliton (\ref{sln:cKdV:self-similar},\ref{solution:cKdV:self-similar}) compared to the plane 1D case \eqref{soliton:1D:KdV} as we saw in \S \ref{subsec:soliton:ncKdV}. The goal, however, is still the same -- to decompose \eqref{eqn:perturbation:ncKdV:KP} in functionally independent parts, which would lead to an amplitude equation for the perturbation.

Differentiating \eqref{eqn:perturbation:ncKdV:KP} w.r.t. $\zeta_{0}$,
\begin{align}
\label{eqn:perturbation:ncKdV:KP:prime}
\left(2 \, A_{\tau} + \beta \, \tau^{-2} \varphi_{\Theta\Theta}\right) F^{\prime} + 2 \frac{\varphi_{\tau}}{\tau^{1/3}} F^{\prime\prime} + 3 \, A \, \tau^{-1} \left(F \, F^{\prime}\right)^{\prime} + \beta \, \tau^{-5/3} A_{\Theta\Theta} F = 0,
\end{align}
multiplying by $F^{\prime}$ and integrating w.r.t. $\zeta_{0}$, for example for $a>0$ from $\zeta_{0}$ to $\infty$ as dictated by the asymptotic behavior \eqref{asymptotics:Painleve}, in the limit $\zeta_{0}=-\infty$ we get at the leading order
\begin{align}
\label{eqn-1:KP-analysis:ncKdV}
2 \, A_{\tau} + \beta \, \tau^{-2} \varphi_{\Theta\Theta} = 0.
\end{align}
In arriving at \eqref{eqn-1:KP-analysis:ncKdV} we took into account that $\int_{\zeta_{0}}^{\infty}{F^{\prime} \left(F F^{\prime}\right)^{\prime} \, \d \zeta} = \frac{1}{2} \int_{\zeta_{0}}^{\infty}{F^{\prime 3} \, \d \zeta}$ and in the limit $\zeta_{0} \rightarrow -\infty$ the following integrals simplify $\int_{\zeta_{0}}^{\infty}{F^{\prime} F^{\prime \prime} \, \d \zeta} = \frac{1}{2} \left.F^{\prime 2}\right|_{\zeta_{0}} \sim \const$, $\int_{\zeta_{0}}^{\infty}{F F^{\prime} \, \d \zeta} = \frac{1}{2} \left.F^{2}\right|_{\zeta_{0}} = 0$, as well as
\begin{align}
\label{condition:KP:limiting-divergence}
\lim_{\zeta_{0} \rightarrow - \infty}{\frac{\int_{\zeta_{0}}^{\infty}{F^{\prime 3} \, \d \zeta}}{\int_{\zeta_{0}}^{\infty}{F^{\prime 2} \, \d \zeta}}} = 0,
\end{align}
since in the limit $\zeta_{0} \rightarrow -\infty$ the integral $\int_{\zeta_{0}}^{\infty}{F^{\prime 2} \, \d \zeta}$ diverges as $\sim \zeta_{0}$, while the integral $\int_{\zeta_{0}}^{\infty}{F^{\prime 3} \, \d \zeta}$ grows slower than $\zeta_{0}$ due to cancellation of integrals of a fast oscillating function for $\zeta_{0} \rightarrow -\infty$ -- the property also known as the Riemann-Lebesgue lemma in the case of Fourier analysis. Hence, compared to the approach of \citet{Kadomtsev:1970}, we used the different rate of divergence of the corresponding integrals \eqref{condition:KP:limiting-divergence}. Note that integration in the case $a<0$ would have to be from $-\infty$ to $\zeta_{0}$ with the limit taken as $\zeta_{0} \rightarrow + \infty$ due to the asymptotic behavior of the soliton reversed compared to \eqref{asymptotics:Painleve}.

Similarly, multiplying \eqref{eqn:perturbation:ncKdV:KP:prime} by $F^{\nu}$ with $\nu > 3$ and integrating w.r.t. $\zeta_{0}$, for example for $a>0$ from $\zeta_{0}$ to $\infty$, leads to
\begin{align}
\label{eqn-2:KP-analysis:ncKdV}
2 \frac{\varphi_{\tau}}{\tau^{1/3}} + 3 \, A \, \tau^{-1} \mathcal{I}_{32} + \beta \, \tau^{-5/3} A_{\Theta\Theta} \mathcal{I}_{12} = 0,
\end{align}
where $\mathcal{I}_{32} = \mathcal{I}_{3}/\mathcal{I}_{2} > 0$ and $\mathcal{I}_{12} = \mathcal{I}_{1}/\mathcal{I}_{2} < 0$ with the corresponding finite integrals $\mathcal{I}_{1-3}$ defined as follows
\begin{subequations}
\begin{align}
\mathcal{I}_{1} &= \int_{\zeta_{0}}^{\infty}{F^{\nu + 1} \, \d \zeta} \ \text{converges for} \ \nu>1, \\
\mathcal{I}_{2} &= \int_{\zeta_{0}}^{\infty}{F^{\nu} F^{\prime\prime} \, \d \zeta} \ \text{converges for} \ \nu>3, \\
\mathcal{I}_{3} &= \int_{\zeta_{0}}^{\infty}{F^{\nu} \left(F F^{\prime}\right)^{\prime} \, \d \zeta} = \left.F^{\nu+1} F^{\prime}\right|_{\zeta_{0}}^{\infty} - \nu \int_{\zeta_{0}}^{\infty}{F^{\nu} F^{\prime 2} \, \d \zeta} \ \text{converges for} \ \nu>2;
\end{align}
\end{subequations}
note that in the last integral $\lim_{\zeta_{0} \rightarrow -\infty}{\left.F^{\nu+1} F^{\prime}\right|_{\zeta_{0}}^{\infty}} = 0$. In the deduction of \eqref{eqn-2:KP-analysis:ncKdV} we also took into account that $\int_{-\infty}^{\infty}{F^{\nu} F^{\prime} \, \d \zeta} = 0$ as well as $F(\zeta_{0}) \sim (-\zeta_{0})^{-1/2}$ for $\zeta_{0} \rightarrow - \infty$ as per \eqref{asymptotics:Painleve}.

\begin{figure}
	\setlength{\labelsep}{-3.0mm}
	\centering
    \sidesubfloat[]{\includegraphics[width=2.5in]{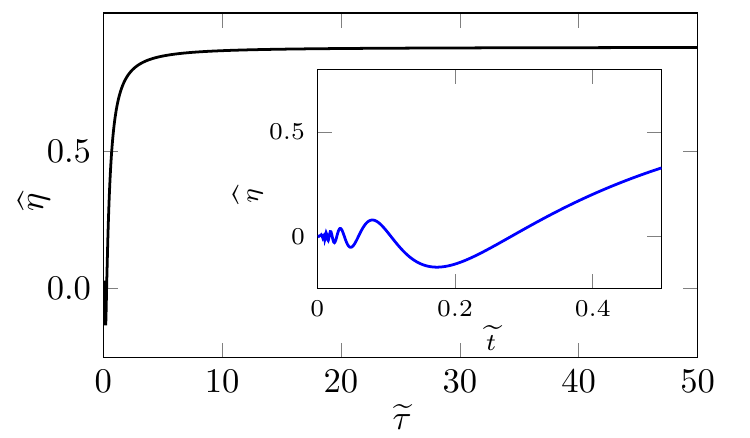}\label{fig:shorttime1}}
	\sidesubfloat[]{\includegraphics[width=2.5in]{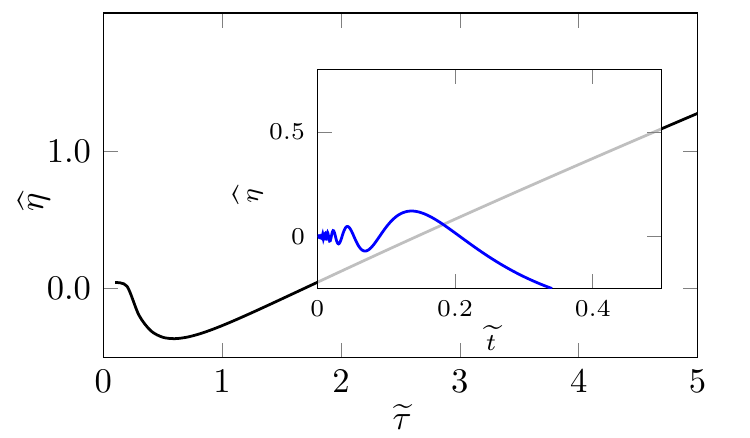}\label{fig:shorttime2}}
\caption{Behavior of two independent solutions (a,b) to \eqref{eqn:amplitude:ncKdV:perturbation:short-time} corresponding to the short-time asymptotics of \eqref{eqn:amplitude:ncKdV:perturbation}; insets show the oscillatory behavior near the origin.} \label{fig:shorttime}
\end{figure}
\begin{figure}
	\setlength{\labelsep}{-3.0mm}
	\centering
    \sidesubfloat[]{\includegraphics[width=2.5in]{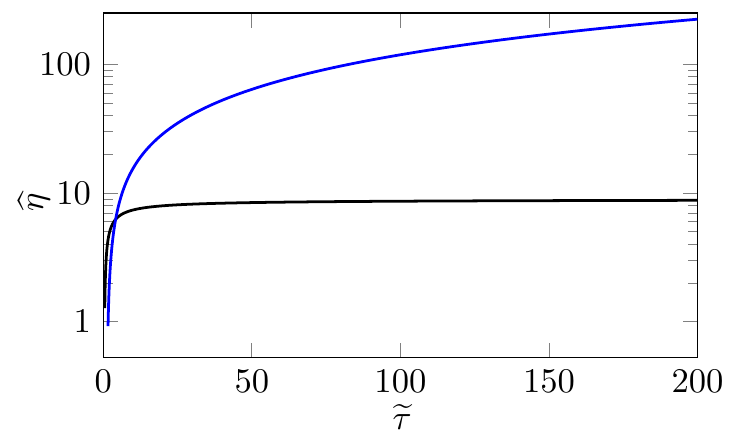}\label{fig:longtimep}}
	\sidesubfloat[]{\includegraphics[width=2.5in]{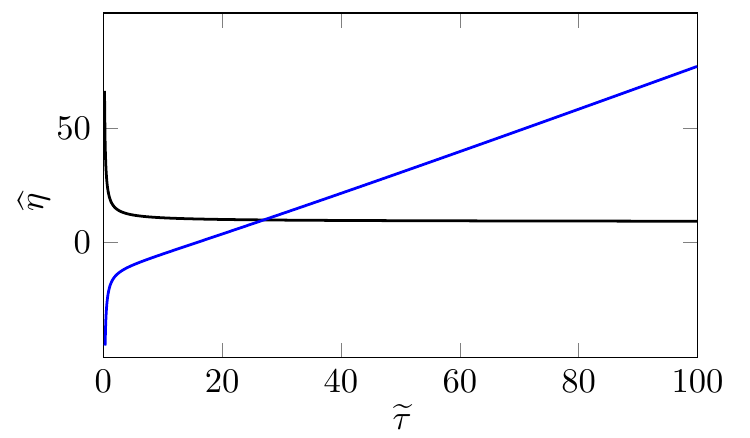}\label{fig:longtimem}}
\caption{Behavior of two independent solutions to \eqref{eqn:amplitude:ncKdV:perturbation:long-time} corresponding to (a) $\beta>0$ and (b) $\beta<0$.} \label{fig:longtime}
\end{figure}

As a result, the perturbation evolution is determined by the system (\ref{eqn-1:KP-analysis:ncKdV},\ref{eqn-2:KP-analysis:ncKdV}), which after the Fourier transform in the transverse direction becomes:
\begin{subequations}
\begin{align}
2 \, \widehat{A}_{\tau} - \beta \, k^{2} \, \tau^{-2} \widehat{\varphi} &= 0, \\
2 \frac{\widehat{\varphi}_{\tau}}{\tau^{1/3}} + 3 \, \widehat{A} \, \tau^{-1} \mathcal{I}_{32} - \beta \, k^{2} \, \tau^{-5/3} \widehat{A} \, \mathcal{I}_{12} &= 0,
\end{align}
\end{subequations}
and can be reduced to a single equation after elimination of $\widehat{\varphi}$ and substitution $\widehat{A} = \tau^{-1} \widehat{\eta}$:
\begin{align}
\label{eqn:amplitude:ncKdV:perturbation}
\widehat{\eta}_{\tau\tau} + \frac{\beta k^{2}}{4} \tau^{-10/3} \left(3 \, \mathcal{I}_{32} \, \tau^{3/2} - \beta k^{2} \mathcal{I}_{12}\right) \widehat{\eta} = 0.
\end{align}
The first observation to make about equation \eqref{eqn:amplitude:ncKdV:perturbation} is that the transverse wavenumber $k$ can be scaled out by $\tau \rightarrow k^{3} \widetilde{\tau}$ and hence no critical wavenumber exists, also in agreement with the conclusions of \S \ref{subsec:KdV:stability-preliminary}. While equation \eqref{eqn:amplitude:ncKdV:perturbation} corresponds to the short time instability, i.e. case (ii), within this asymptotic approximation we can consider the short- and long-time behavior of \eqref{eqn:amplitude:ncKdV:perturbation} in the proper multiple-scale sense. Clearly, for short times it is the second term in the brackets of \eqref{eqn:amplitude:ncKdV:perturbation}, which is dominant, thus leading to
\begin{align}
\label{eqn:amplitude:ncKdV:perturbation:short-time}
\widehat{\eta}_{\widetilde{\tau}\widetilde{\tau}} + \widetilde{\tau}^{-10/3} \widehat{\eta} = 0,
\end{align}
after the straightforward scaling out of the constant with the help of redefining the time variable and taking into account that $\mathcal{I}_{12}<0$. The solution of \eqref{eqn:amplitude:ncKdV:perturbation:short-time} is a linear combination of two independent modes
\begin{align}
\widehat{\eta}(\widetilde{\tau}) = C_{1} \widetilde{\tau}^{1/2} J_{3/4}{\left(\frac{3}{2} \widetilde{\tau}^{-2/3}\right)} +
C_{2} \widetilde{\tau}^{1/2} J_{-3/4}{\left(\frac{3}{2} \widetilde{\tau}^{-2/3}\right)},
\end{align}
which are shown in figure~\ref{fig:shorttime} -- the first is approaching a constant plateau, while the second one grows linearly in time. The long-time asymptotics of \eqref{eqn:amplitude:ncKdV:perturbation} is dictated by the first term in the brackets, which after scaling out the numerical coefficient produces
\begin{align}
\label{eqn:amplitude:ncKdV:perturbation:long-time}
\widehat{\eta}_{\widetilde{\tau}\widetilde{\tau}} \pm \widetilde{\tau}^{-8/3} \widehat{\eta} = 0,
\end{align}
where the plus sign corresponds to $\beta >0$ and negative to $\beta < 0$. The solutions of \eqref{eqn:amplitude:ncKdV:perturbation:long-time} read for $\beta>0$:
\begin{align}
\widehat{\eta}(\widetilde{\tau}) &= C_{1} \widetilde{\tau} \left[-\varrho \cos{\varrho} + \sin{\varrho}\right] = C_{2} \widetilde{\tau} \left[\cos{\varrho} + \varrho \sin{\varrho}\right],
\end{align}
where $\varrho = 3 \, \widetilde{\tau}^{-1/3}$, and are illustrated in figure~\ref{fig:longtimep}. In the case $\beta<0$, the solution becomes
\begin{align}
\widehat{\eta}(\widetilde{\tau}) &= C_{1} \widetilde{\tau} \left[\varrho \cosh{\varrho} - \sinh{\varrho}\right]  = C_{2} \widetilde{\tau} \left[\cosh{\varrho} - \varrho \sinh{\varrho}\right]
\end{align}
and is illustrated in figure~\ref{fig:longtimem}. In both cases, one of the solutions approaches a non-zero constant, while the other one grows linearly (the one in figure~\ref{fig:longtimep} is shown on logarithmic scale). Thus, taking into account the transformation $\widehat{A} = \tau^{-1} \widehat{\eta}$ connecting $\widehat{A}$ and $\widehat{\eta}$, we conclude that initial perturbations, measured relative to the unit amplitude of the self-similar solution as per \eqref{sln-perturbed:ncKdV}, are able to grow from infinitesimal values and approach some finite value, so that nonlinear effects start playing a role -- this behavior is atypical for linear stability problems, usually exhibiting either exponential growth or decay, and more characteristic to nonlinear behavior predicated earlier in \S \ref{subsec:KdV:stability-preliminary} based on the properties of equation \eqref{EP:ncKdV:rescaled}. Therefore, it is the short-time behavior governed by \eqref{eqn:amplitude:ncKdV:perturbation} which dictates the transverse stability properties of ncKdV, and makes the appearance of transverse instability possible. This situation is not unusual for stability problems involving time-dependent base states such as in the Rayleigh-Plateau instability of a growing cylindrical liquid blob \citep{Krechetnikov:2017b}.

\section{Conclusions}

With the goal to study stability of axisymmetric solitary waves, in the present work we deduced a proper envelope equation for solitary waves on deep water, which proves to include an  inverse-square potential and hence be of Gross-Pitaevskii type (\ref{ncNLS},\ref{ncNLS-ST:abstract}); in the shallow water limit we rederived a ncKdV equation \eqref{eqn:cKP} by including surface tension effects and under asymptotic assumptions different from what was known before. In the former case, our derivation is set apart from previous studies which postulated that the corresponding NLS for axisymmetric case has the Laplace operator unchanged -- our analysis (\S\S \ref{subsec:ncNLS},\ref{subsec:heuristic-analysis}) demonstrates that the covariance principle does not apply to envelope equations despite their ``universal'' character. Given the novelty of the deduced GP equation for deep water waves, we studied its general properties -- conservation laws (\S \ref{subsec:GP:conservation-laws}), Hamiltonian structure (\S \ref{subsec:Lagrange-Dirichlet}), finite-time-singularity (\S \ref{subsec:GP:conservation-laws}) -- as well as axisymmetric base states along with geometric and mechanistic interpretations of their varieties (\S \ref{subsec:ncNLS:BS}, Appendix~\ref{appx:mechanistic-interpretation}).

The stability analysis in the deep water case was performed with the help of both spectral (in the limit of long wavelengths, cf. \S \ref{subsec:spectral-analysis:ncNLS}) and Hamiltonian (for general wavelengths, cf. \S \ref{subsec:Lagrange-Dirichlet}) methods, which complement each other. The challenge of the spectral stability problem \eqref{EVP:cNLS} was its singular nature dictated by the particularities of the base states (\S \ref{subsec:ncNLS:BS}), which nevertheless enable analytical approaches. We revealed the crucial differences in stability characteristics between cylindrical and plane solitons: namely, there is a threshold in the Weber number $We_{c}=\frac{1}{2}$ above which instability appears in the deep water case as opposed to the nearly plane NLS \eqref{eqn:NLS-1D}, the 1D plane solitons of which are always unstable to transverse perturbations \citep{Zakharov:1974} regardless of the value of $We$. Thus, surface tension must be sufficiently strong to induce a transverse instability of a cylindrical soliton on the deep water\footnote{A qualitative interpretation one may offer is that in the case of a plane soliton surface tension breaks it similar to a Rayleigh-Plateau instability of a rectilinear liquid column, which takes place for any magnitude of surface tension as long as it is non-zero, while in the case of a cylindrical soliton the Rayleigh-Plateau instability competes with stabilizing effect of the transverse curvature in the plane of the soliton propagation as well as with the time-dependence of the base state.}.

In the shallow water case, we performed an analysis (\S \ref{subsec:KP:ncKdV:analysis}) in the spirit of \citet{Kadomtsev:1970} extending it not only to cylindrical geometry but also to self-similar solitons \eqref{sln:cKdV:self-similar}, with the resulting linear amplitude equation \eqref{eqn:amplitude:ncKdV:perturbation}, which governs perturbation evolution, being highly-nonautonomous and exhibiting transient growth of transverse perturbations regardless of the value of the Weber number in contrast to its plane counterpart, where there is a non-zero critical Weber number\footnote{Obviously, the effect of the solid bottom plays a stabilizing role in the case of plane solitons and thus requires strong enough surface tension to induce a transverse instability, i.e. to get into a Rayleigh-Plateau regime. On the other hand, in the case of a cylindrical soliton the effect of the base soliton time-dependence overpowers any other effects thus leading to transient growth of perturbations, which should trigger nonlinear effects before the subsequent linear dynamics would lead to a decay of the perturbation.}. For long times the stability picture is consistent with the intuition that the dynamics should approach that of npKdV (\S \ref{subsec:KdV:stability-preliminary}). Also, for general wavenumbers, from the reduction to a 2D eigenvalue problem \eqref{EP:ncKdV:rescaled} in the self-similar plane, we made an unexpected conclusion that the transverse perturbations must have an irregular structure in the azimuthal $\theta$-direction, cf. \S \ref{subsec:KdV:stability-preliminary}. Numerical study of \eqref{EP:ncKdV:rescaled}, however, represents a challenge for future efforts.

While here we explored only the basic properties of the GP and ncKdV solutions, one might expect that similar to the standard (near planar) versions of these equations, the behavior of their solutions is very rich \citep{Cai:2002}. Including higher-order terms \citep{Dysthe:1979} or generalization onto finite depth \citep{Hasimoto:1972} of the GP equation may offer further insights in the axisymmetric water waves, same as establishing a relation between GP and ncKdV similar to that between NLS and KdV \citep{Boyd:2001} as well as considering the near-critical values of the Weber number $We \rightarrow We_{c}=\frac{1}{3}$ in the ncKdV equation, which should bring up fifth-order derivatives \citep{Green:1983,Hunter:1988}. Also, in the derivation of ncKdV equation with surface tension (\S \ref{subsec:ncKdV}) we neglected the resonance between the linear (carrier) wave speed $c_{0}$ and the linear phase speed $\omega(k)/k \approx g^{1/2} (1 + We \, k^{2} h^{2})^{1/2} h$, which exhibits itself in the far-field \citep{Boyd:1988,Grimshaw:2003,Grimshaw:2005} and occurs because the graph $\omega(k)/k$ is not monotonic when $0 < We < \frac{1}{3}$; for $We>\frac{1}{3}$ the graph of $\omega(k)/k$ is monotonic and hence the derivation of ncKdV does not require corrections.

Both types of envelope equations -- on deep and shallow water -- could be amenable to the inverse scattering transform methods, which may serve as yet another method for studying transverse stability as it was done by \citet{Zakharov:1975} for the KdV solitons. While some solutions of the cKdV \eqref{eqn:cKdV} were constructed with inverse scattering transform, cf. \citet{Calogero:1978,Johnson:1979,Freeman:1980}, the feasibility of the inverse scattering for the NLS with an inverse-square potential has not been fully explored yet, cf. \citet{Murphy:2019} and references therein, even though NLS with potentials could be suitable to inverse scattering transform analysis and represents an active area of research, cf. \citet{Sasaki:2008,Fajun:2019}.

\section*{Acknowledgements}

his work was partially supported by the National Science Foundation (NSF) CAREER award under Grant No. 1054267 and the Natural Sciences and Engineering Research Council of Canada (NSERC) under Grant No. 04374. Declaration of interests: the author reports no conflict of interest.

\appendix

\section{Mechanical analogy} \label{appx:mechanistic-interpretation}

\begin{figure}
	\setlength{\labelsep}{-3.0mm}
	\centering
    \sidesubfloat[]{\includegraphics[width=1.7in]{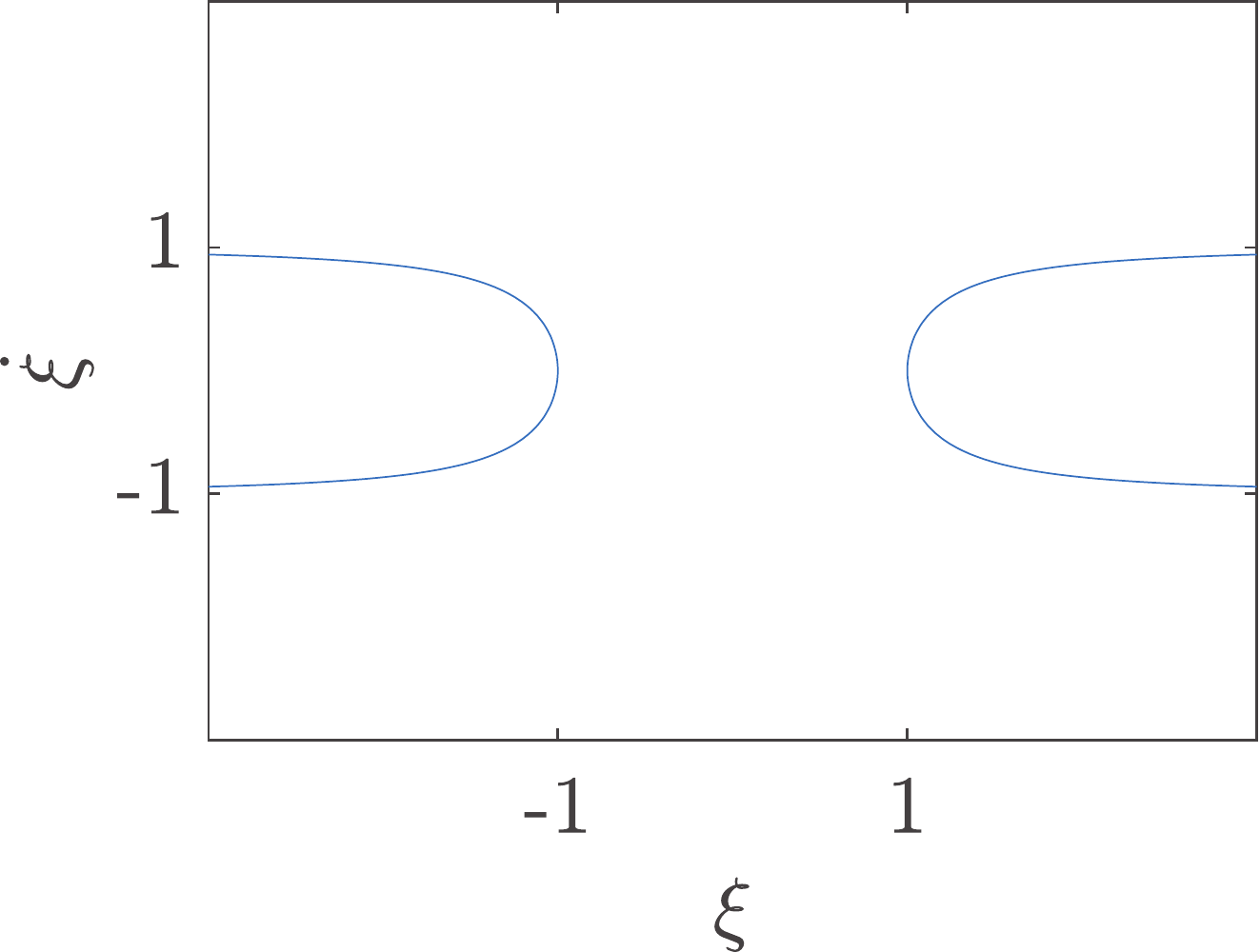}\label{fig:hamiltonianpp}} \
    \sidesubfloat[]{\includegraphics[width=1.7in]{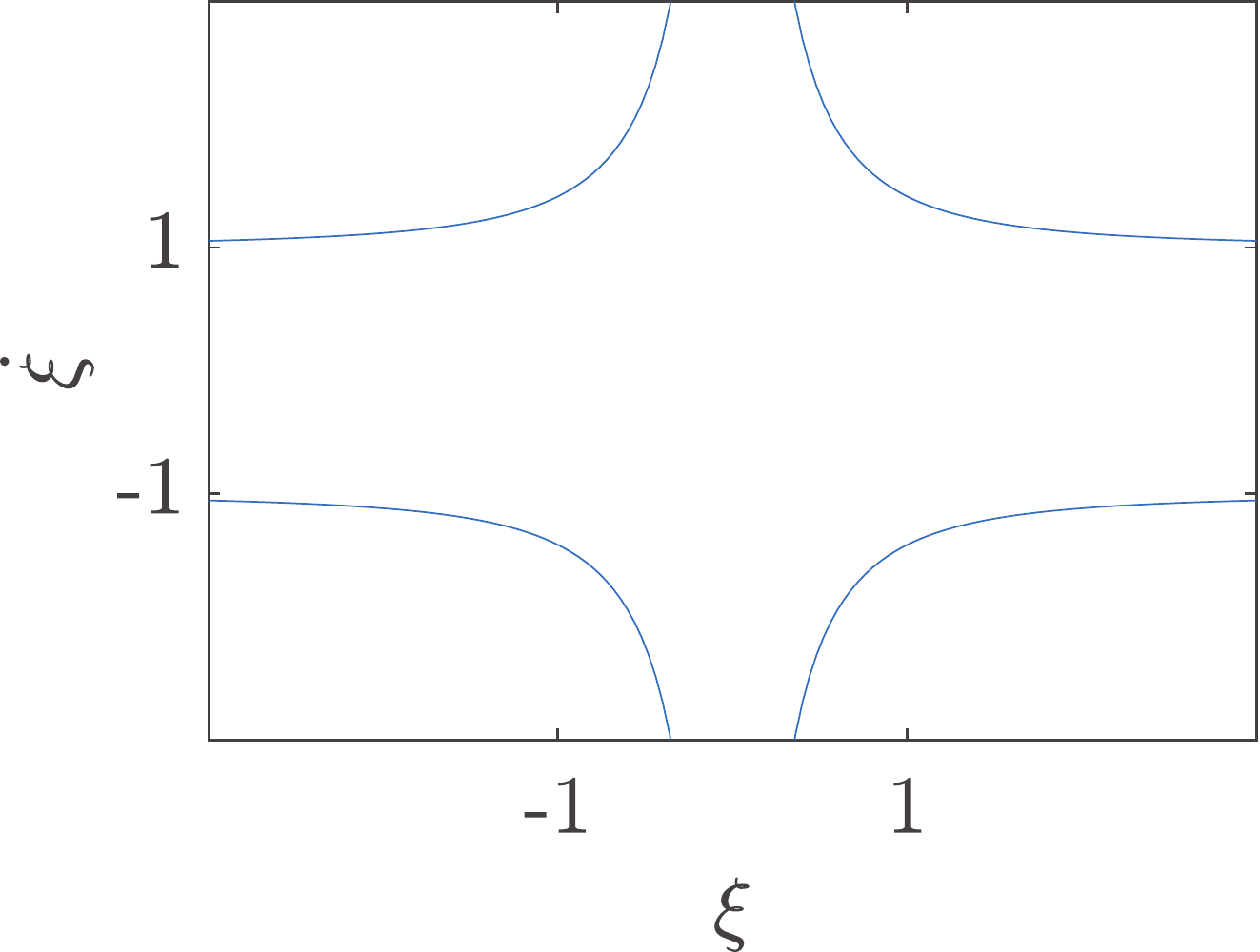}\label{fig:hamiltonianpm}} \
    \sidesubfloat[]{\includegraphics[width=1.7in]{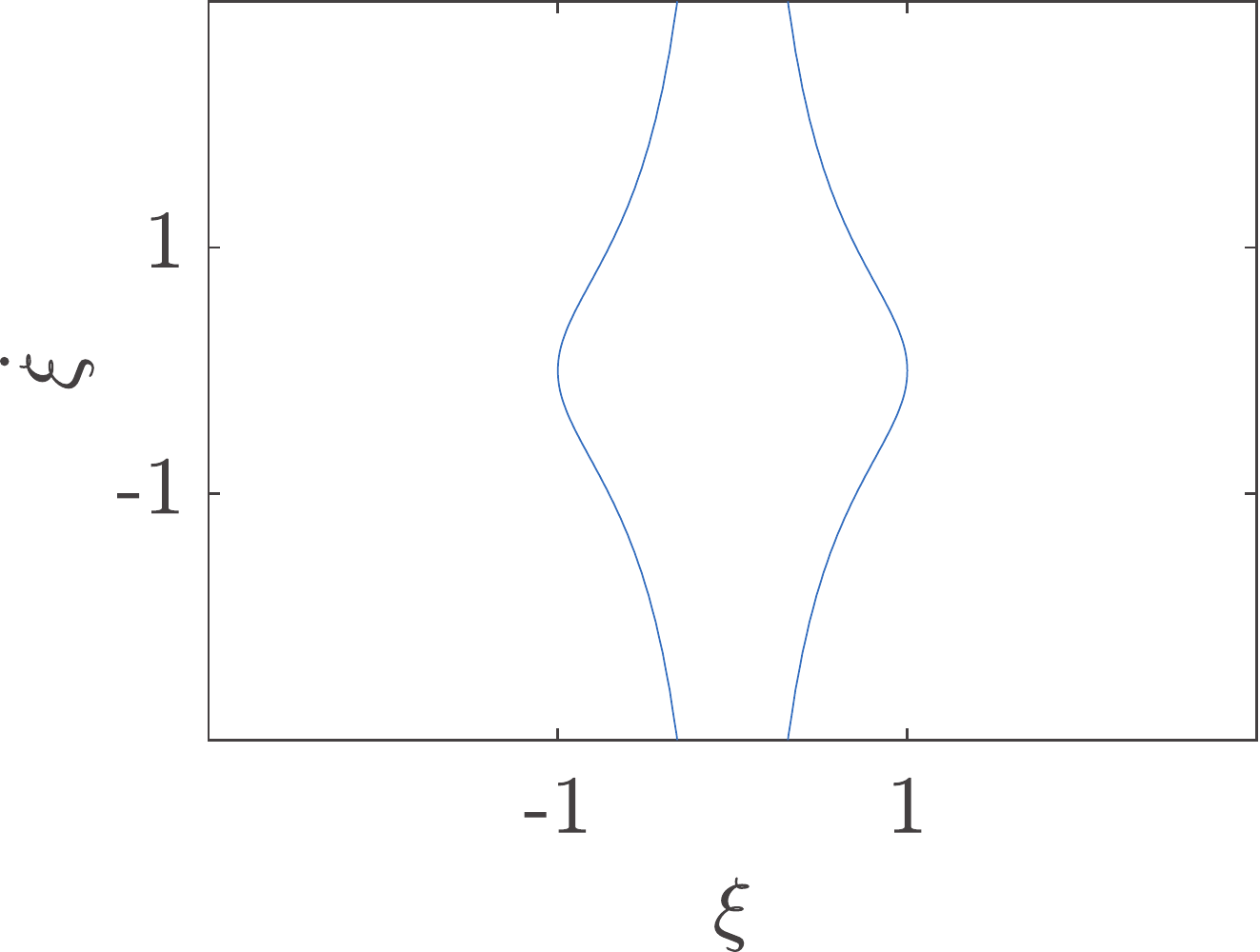}\label{fig:hamiltonianmp}}
\caption{Isolines of canonical Hamiltonians (a) $\dot{\xi}^{2} + \frac{1}{\xi^{2}} = 1$, (b) $\dot{\xi}^{2} - \frac{1}{\xi^{2}} = 1$, and (c) $- \dot{\xi}^{2} + \frac{1}{\xi^{2}} = 1$.} \label{fig:hamiltonian}
\end{figure}
\begin{figure}
	\setlength{\labelsep}{-3.0mm}
	\centering
    \sidesubfloat[]{\includegraphics[width=2.5in]{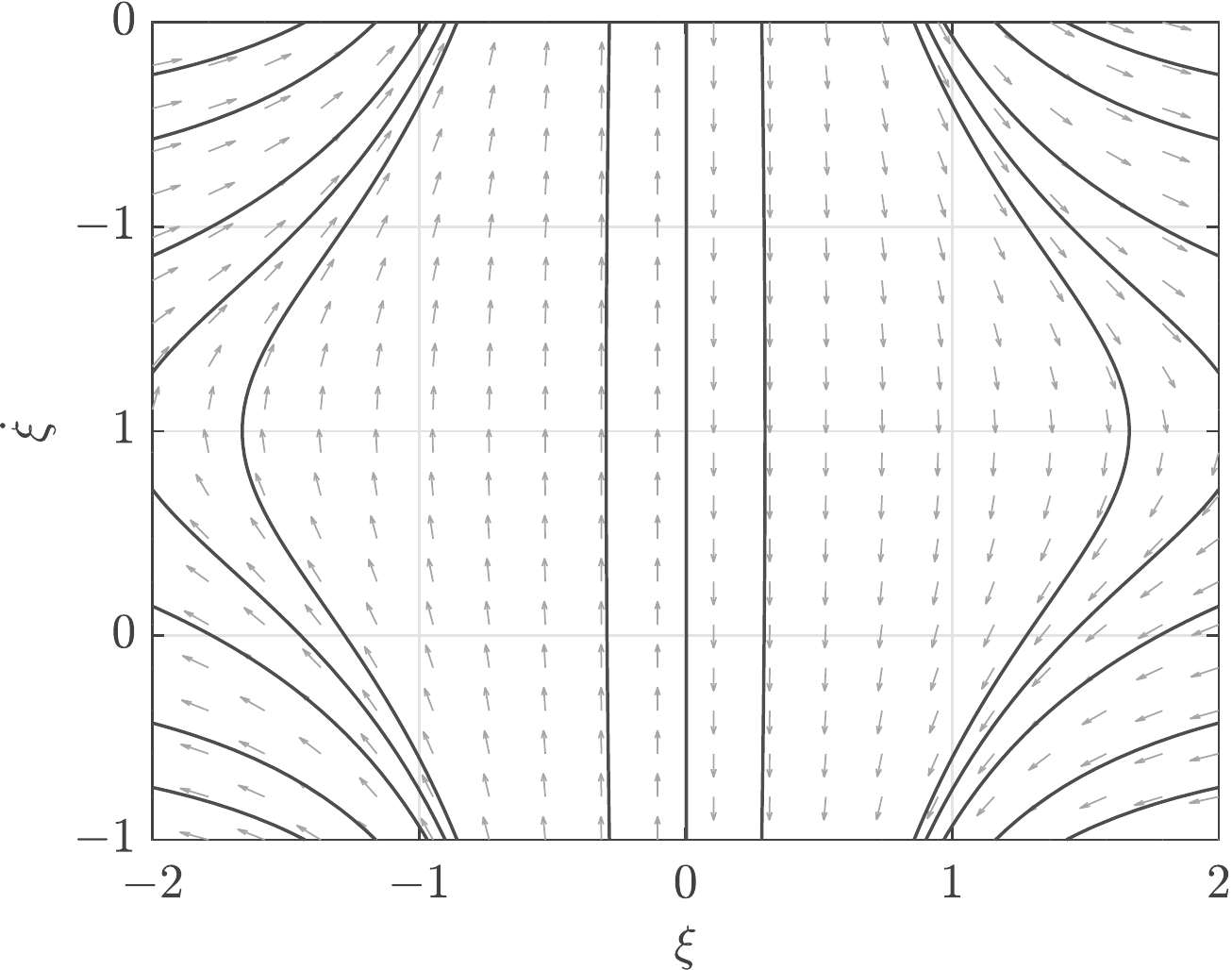}\label{fig:mechanicsm}} \quad
	\sidesubfloat[]{\includegraphics[width=2.5in]{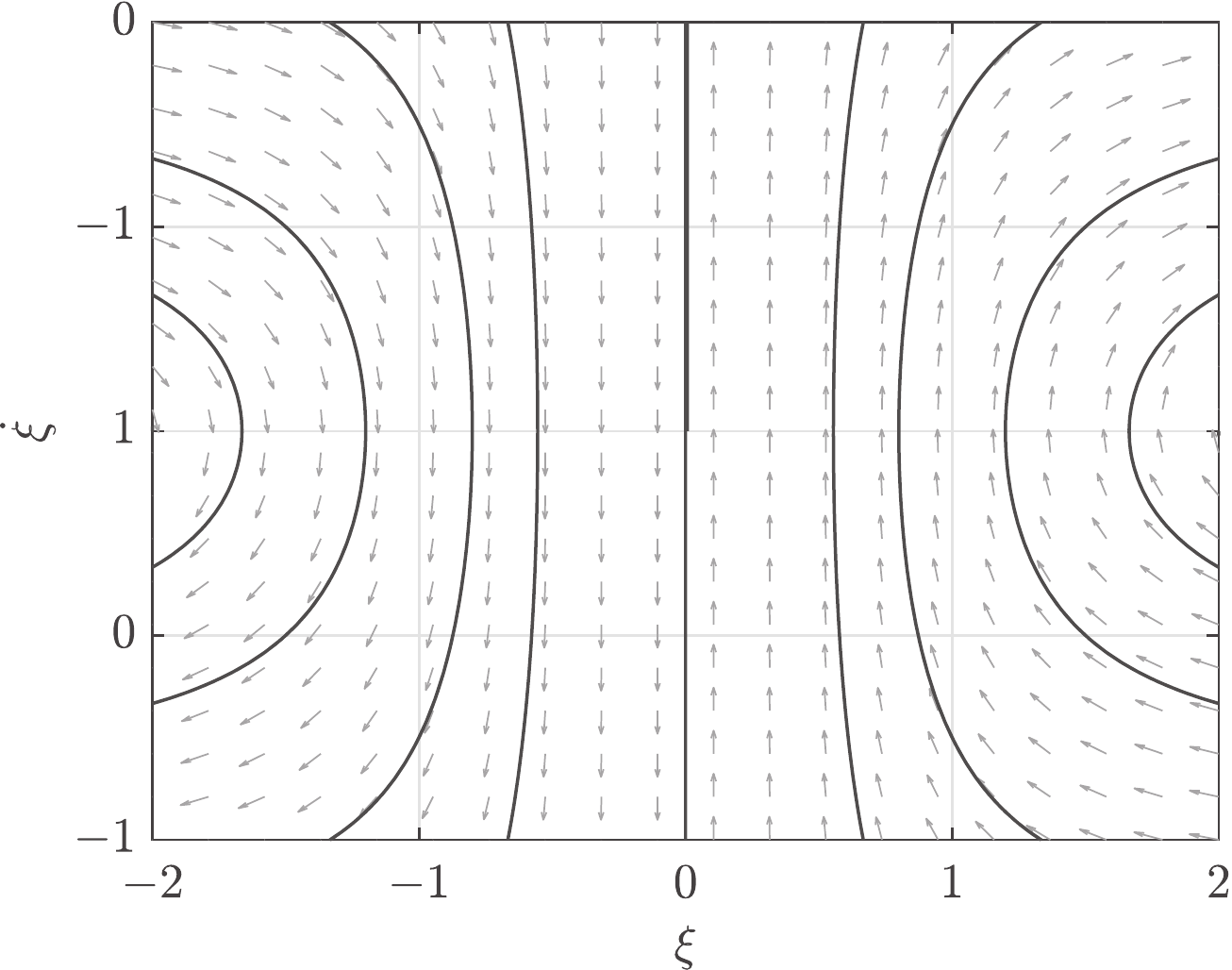}\label{fig:mechanicsp}}
\caption{Phase plots of (a) $\ddot{\xi} = - \frac{4}{\xi^{3}}$ and (b) $\ddot{\xi} = \frac{4}{\xi^{3}}$.} \label{fig:mechanics}
\end{figure}

Equation \eqref{ncNLS-ST:abstract} admits a mechanistic interpretation by looking for solutions in the form $\psi_{0} = \mathrm{H}^{1/2} \, e^{\i S(R)}$, which leads to the Hamiltonian
\begin{align}
\label{eqn:Hamiltonian:mechanics}
- \frac{\lambda_{\infty}}{\chi_{\infty}} p^{2} + \frac{\lambda_{\infty}^{\prime}}{\chi_{\infty}} \frac{1}{R^{2}} \equiv \mathrm{H},
\end{align}
where $p = S_{R}$ is the momentum, while the factors $\frac{\lambda_{\infty}}{\chi_{\infty}}$ and $\frac{\lambda_{\infty}^{\prime}}{\chi_{\infty}}$ may assume all possible sign combinations as per \eqref{coefficients:GP}. Of course, motion of a particle in a central field such as our inverse-square potential would also bring in conservation of angular momentum, which in turn leads to an ``effective'' $\sim R^{-2}$ potential \citep{Arnold:1989} as well; hence the potential in \eqref{eqn:Hamiltonian:mechanics} can be considered as a total one. Most importantly, if both central and effective potentials are $\sim R^{-2}$, then there is no local minimum of the total potential and hence no non-zero lower bound for the particle trajectory. In arriving at the Hamiltonian \eqref{eqn:Hamiltonian:mechanics} we omitted the terms $S_{RR} + \frac{1}{R} S_{R}$ as in the WKB method, which is to be verified a posteriori. Without loss, we may scale out the factors in \eqref{eqn:Hamiltonian:mechanics} and thus consider canonical Hamiltonians as in figure~\ref{fig:hamiltonian}. Clearly, all trajectories are unbounded. This can also be seen by introducing an ``effective'' time $t$ to arrive at the corresponding Hamitonian systems: $\ddot{\xi} = \pm \frac{4}{\xi^{3}}$, integration of which indeed gives diverging trajectories as per the phase portraits in figure~\ref{fig:mechanics}; these trajectories correspond to solutions of \eqref{ncNLS-ST:abstract} oscillatory at infinity. Hence $S_{R} \approx \left(-\frac{\chi_{\infty}}{\lambda_{\infty}} \mathrm{H}\right)^{1/2} \left(1 - \frac{\lambda_{\infty}^{\prime}}{\chi_{\infty}} \frac{1}{2 \mathrm{H} R^{2}} + \ldots\right)$ for large $R$ and thus neglecting the omitted terms $S_{RR} + \frac{1}{R} S_{R}$ in \eqref{eqn:Hamiltonian:mechanics} is justified.

\section{Variance} \label{appx:variance}

Finally, let us derive an equation for the variance $\mathcal{V}(\tau) = \int{R^{2} |\psi|^{2} \, \d \nu}$. Taking the first derivative, we find
\begin{multline}
\label{variance:derivative:first-prelim}
\i \frac{\d \mathcal{V}}{\d \tau} = \i \int{R^{2} \left(\overline{\psi} \psi_{\tau} + \psi \overline{\psi}_{\tau}\right) \, \d \nu} = - \lambda_{\infty} \int{R^{2} \left(\overline{\psi} \Delta_{R} \psi - \psi \Delta_{R} \overline{\psi}\right) \, \d \nu} \\
- \mu_{\infty} \int{\frac{1}{R^{2}} \left(\overline{\psi} \psi_{\theta\theta} - \psi \overline{\psi}_{\theta\theta}\right) \, \d \nu},
\end{multline}
which follows from multiplying \eqref{ncNLS-ST:abstract} by $\overline{\psi}$, i.e. equation \eqref{eqn:cNLS-times-psiconj}, and subtracting the complex conjugate of \eqref{ncNLS-ST:abstract} multiplied by $\psi$:
\begin{align}
\label{eqn:cNLSconj-times-psi}
- \i \, \psi \, \overline{\psi}_{\tau} + \lambda_{\infty} \psi \, \Delta_{R} \overline{\psi} + \frac{\lambda_{\infty}^{\prime}}{R^{2}} |\psi|^{2} + \frac{\mu_{\infty}}{R^{2}} \psi \overline{\psi}_{\theta\theta} - \chi_{\infty} \, |\psi|^{4} = 0,
\end{align}
results in
\begin{align}
\i \, \partial_{\tau} |\psi|^{2} = - \lambda_{\infty} \left(\overline{\psi} \Delta_{R} \psi - \psi \Delta_{R} \overline{\psi}\right) - \frac{\mu_{\infty}}{R^{2}} \left(\overline{\psi} \psi_{\theta\theta} - \psi \overline{\psi}_{\theta\theta}\right).
\end{align}
The first integral on the rhs of \eqref{variance:derivative:first-prelim} is calculated via integration by parts and use of BCs \eqref{BCs:cNLS}:
\begin{align}
\label{eqn:cNSL:variance-derivation:1}
&\int_{0}^{\infty}{R^{2} \left[\overline{\psi} \left(\psi_{RR} + \frac{1}{R}\psi_{R}\right) - \psi \left(\overline{\psi}_{RR} + \frac{1}{R}\overline{\psi}_{R}\right)\right] R \, \d R} \\
= &R^{3} \left[\psi_{R} \overline{\psi} - \overline{\psi}_{R} \psi\right]_{0}^{\infty} - \int_{0}^{\infty}{\left[\psi_{R} \left(R^{3} \overline{\psi}\right)_{R} - \overline{\psi}_{R} \left(R^{3} \psi\right)_{R}\right] \d R} + \int_{0}^{\infty}{R^{2} \left(\overline{\psi} \psi_{R} - \psi \overline{\psi}_{R}\right) \, \d R} \nonumber \\
= &\int_{0}^{\infty}{3 R^{2} \left(\overline{\psi}_{R} \psi - \psi_{R} \overline{\psi}\right) \, \d R} + \int_{0}^{\infty}{R^{2} \left(\overline{\psi} \psi_{R} - \psi \overline{\psi}_{R}\right) \, \d R} = 2 \int_{0}^{\infty}{R^{2} \left(\overline{\psi}_{R} \psi - \psi_{R} \overline{\psi}\right) \, \d R}. \nonumber
\end{align}
The second integral on the rhs of \eqref{variance:derivative:first-prelim} is calculated also via integration by parts and use of periodicity in $\theta$:
\begin{align}
\begin{split}
\int_{0}^{2\pi}{\left(\overline{\psi} \psi_{\theta\theta} - \psi \overline{\psi}_{\theta\theta}\right) \, \d \theta} = \left[\overline{\psi} \psi_{\theta} - \psi \overline{\psi}_{\theta}\right]_{0}^{2\pi} - \int_{0}^{2\pi}{\left(\overline{\psi}_{\theta} \psi_{\theta} - \psi_{\theta} \overline{\psi}_{\theta}\right) \, \d \theta} = 0.
\end{split}
\end{align}
Therefore, the first derivative of the variance \eqref{variance:derivative:first-prelim} becomes:
\begin{align}
\label{variance:derivative:first}
\i \frac{\d \mathcal{V}}{\d \tau} = - 2 \lambda_{\infty} \int{R \left(\overline{\psi}_{R} \psi - \psi_{R} \overline{\psi}\right) \, \d \nu}.
\end{align}
Differentiating $\mathcal{V}$ second time, we obtain
\begin{align}
\label{variance:derivative:second-prelim}
- \frac{\i}{2\lambda_{\infty}} \frac{\d^{2} \mathcal{V}}{\d \tau^{2}} &= \int{R \left(\overline{\psi}_{R} \psi_{\tau} - \psi_{R} \overline{\psi}_{\tau} + \overline{\psi}_{R\tau} \psi - \psi_{R\tau} \overline{\psi}\right) \, \d \nu} \nonumber \\
&= \int_{0}^{2\pi}{\d\theta}\int_{0}^{\infty}{R^{2} \left(\overline{\psi}_{R} \psi_{\tau} - \psi_{R} \overline{\psi}_{\tau}\right) \, \d R} + \int_{0}^{2\pi}{R^{2} \left[\overline{\psi}_{\tau} \psi - \psi_{\tau} \overline{\psi}\right]_{R=0}^{\infty} \, \d\theta} \nonumber \\
&- \int_{0}^{2\pi}{\d\theta}\int_{0}^{\infty}{\left[\overline{\psi}_{\tau} \left(R^{2} \psi\right)_{R}
- \psi_{\tau} \left(R^{2} \overline{\psi}\right)_{R}\right] \, \d R} \nonumber \\
&= 2 \int_{0}^{2\pi}{\d\theta} \int_{0}^{\infty}{R^{2} \left(\overline{\psi}_{R} \psi_{\tau} - \psi_{R} \overline{\psi}_{\tau}\right) \, \d R} - 2 \int_{0}^{2\pi}{\d\theta} \int_{0}^{\infty}{R \left(\overline{\psi}_{\tau} \psi - \psi_{\tau} \overline{\psi}\right) \, \d R} \nonumber \\
&= 2 \int{R \left(\overline{\psi}_{R} \psi_{\tau} - \psi_{R} \overline{\psi}_{\tau}\right) \, \d \nu} - 2 \int{\left(\overline{\psi}_{\tau} \psi - \psi_{\tau} \overline{\psi}\right) \, \d \nu}.
\end{align}
The integrand in the second integral on the rhs of \eqref{variance:derivative:second-prelim} is computed by adding \eqref{eqn:cNLS-times-psiconj} to \eqref{eqn:cNLSconj-times-psi}:
\begin{align}
\i \left(\overline{\psi}_{\tau} \psi - \psi_{\tau} \overline{\psi}\right) = \lambda_{\infty} \left(\psi \, \Delta_{R} \overline{\psi} + \overline{\psi} \, \Delta_{R} \psi\right) + \frac{2 \lambda_{\infty}^{\prime}}{R^{2}} |\psi|^{2} + \frac{\mu_{\infty}}{R^{2}} \left(\psi \overline{\psi}_{\theta\theta} + \overline{\psi} \psi_{\theta\theta}\right) - 2 \chi_{\infty} \, |\psi|^{4}. \nonumber
\end{align}
To compute the corresponding integral, we note that $\psi \, \Delta_{R} \overline{\psi} + \overline{\psi} \, \Delta_{R} \psi = 2 \left(\psi^{r} \, \Delta_{R} \psi^{r} + \psi^{i} \, \Delta_{R} \psi^{i}\right)$ and
\begin{align}
\label{eqn:cNSL:variance-derivation:2}
\begin{split}
\int_{0}^{\infty}{\phi \Delta_{R} \phi \, R \, \d R} &= \left.\phi_{R} \phi R\right|_{0}^{\infty} - \int_{0}^{\infty}{\phi_{R} \left(\phi \, R\right)_{R} \, \d R} + \int_{\Bbb{R}}{\phi \phi_{R} \, \d R} \\
&= \left.\phi_{R} \phi R\right|_{0}^{\infty} - \int_{0}^{\infty}{R \, \phi_{R}^{2} \, \d R},
\end{split}
\end{align}
so that applying the latter result to $\phi = \psi^{r}, \psi^{i}$ and using BCs \eqref{BCs:cNLS} we get
\begin{align}
\int_{0}^{\infty}{\left(\psi \, \Delta_{R} \overline{\psi} + \overline{\psi} \, \Delta_{R} \psi\right) \, R \, \d R} = - \int_{0}^{\infty}{|\psi_{R}|^{2} \, R \, \d R}.
\end{align}
Similarly, since $\psi \, \overline{\psi}_{\theta\theta} + \overline{\psi} \, \psi_{\theta\theta} = 2 \left(\psi^{r} \, \psi^{r}_{\theta\theta} + \psi^{i} \, \psi^{i}_{\theta\theta}\right)$ and
\begin{align}
\int_{0}^{2 \pi}{\frac{\phi \phi_{\theta\theta}}{R} \, \d \theta} = \left.\frac{\phi \phi_{\theta}}{R}\right|_{0}^{2\pi} - \int_{0}^{2\pi}{\frac{\phi_{\theta}^{2}}{R} \, \d \theta} = - \int_{0}^{2\pi}{\frac{\phi_{\theta}^{2}}{R} \, \d \theta},
\end{align}
altogether the second integral on the rhs of \eqref{variance:derivative:second-prelim} becomes
\begin{multline}
\i \int{\left(\overline{\psi}_{\tau} \psi - \psi_{\tau} \overline{\psi}\right) \, \d \nu} = - 2 \lambda_{\infty} \int{|\psi_{R}|^{2} \, \d \nu} + 2 \lambda_{\infty}^{\prime} \int{\frac{|\psi|^{2}}{R^{2}} \, \d \nu} \\
- 2 \mu_{\infty} \int{\frac{|\psi_{\theta}|^{2}}{R^{2}} \, \d \nu} - 2 \chi_{\infty} \int{|\psi|^{4} \, \d \nu}.
\end{multline}

As for the first integral on the rhs of \eqref{variance:derivative:second-prelim}, it is computed by multiplying \eqref{ncNLS-ST:abstract} with $\overline{\psi}_{R}$ and subtracting the complex conjugate of \eqref{ncNLS-ST:abstract} multiplied by $\psi_{R}$, which results in
\begin{multline}
\i \left(\overline{\psi}_{R} \psi_{\tau} - \psi_{R} \overline{\psi}_{\tau}\right) + \lambda_{\infty} \mathop{\left(\overline{\psi}_{R} \, \Delta_{R} \psi + \psi_{R} \, \Delta_{R} \overline{\psi}\right)}_{\text{does not contribute}} + \left[\frac{\lambda_{\infty}^{\prime}}{R^{2}} - \chi_{\infty} |\psi|^{2}\right] \left(\overline{\psi}_{R} \psi + \psi_{R} \, \overline{\psi}\right) \\
+ \frac{\mu_{\infty}}{R^{2}} \mathop{\left(\overline{\psi}_{R} \, \psi_{\theta\theta} + \psi_{R} \, \overline{\psi}_{\theta\theta}\right)}_{\text{does not contribute}} = 0,
\end{multline}
where $\overline{\psi}_{R} \psi + \psi_{R} \, \overline{\psi} = \frac{\d}{\d R}|\psi|^{2}$, $\overline{\psi}_{R} \, \Delta_{R} \psi + \psi_{R} \, \Delta_{R} \overline{\psi} = 2 \left(\psi^{r}_{R} \, \Delta_{R} \psi^{r} + \psi^{i}_{R} \, \Delta_{R} \psi^{i}\right)$ and $\overline{\psi}_{R} \, \psi_{\theta\theta} + \psi_{R} \, \overline{\psi}_{\theta\theta} = 2 \left(\psi^{r}_{R} \, \psi^{r}_{\theta\theta} + \psi^{i}_{R} \, \psi^{i}_{\theta\theta}\right)$; also, given that
\begin{multline}
\label{eqn:cNSL:variance-derivation:3}
\int_{0}^{\infty}{R^{2} \, \phi_{R} \Delta_{R} \phi \, \d R} = \int_{0}^{\infty}{R^{2} \, \phi_{R} \frac{1}{R} \frac{\d}{\d R} \left(R \phi_{R}\right) \, \d R} \\
= \frac{1}{2} \int_{0}^{\infty}{\frac{\d}{\d R} \left(R \phi_{R}\right)^{2} \, \d R} = \frac{1}{2} \left.\left(R \phi_{R}\right)^{2}\right|_{0}^{\infty} = 0,
\end{multline}
by applying this result to $\phi = \psi^{r},\psi^{i}$ with BCs \eqref{BCs:cNLS} we find
\begin{align}
\int_{0}^{\infty}{R^{2} \left(\overline{\psi}_{R} \, \Delta_{R} \psi + \psi_{R} \, \Delta_{R} \overline{\psi}\right) \, \d R} = 0.
\end{align}
In the same fashion,
\begin{multline}
\int_{0}^{2\pi}\int_{0}^{\infty}{\phi_{R} \phi_{\theta\theta} \, \d\theta \d R} = \int_{0}^{\infty}{\left[\left.\phi_{R} \phi_{\theta}\right|_{0}^{2\pi} - \int_{0}^{2\pi}{\phi_{\theta}\phi_{R\theta} \, \d \theta}\right] \, \d R} \\
= - \frac{1}{2} \int_{0}^{2\pi}{\int_{0}^{\infty}{\frac{\d}{\d R}\left(\phi_{\theta}^{2}\right) \, \d R} \, \d\theta} =
- \frac{1}{2} \int_{0}^{2\pi}{\left.\phi_{\theta}^{2}\right|_{R=0}^{\infty} \, \d\theta} = 0,
\end{multline}
so that the first integral on the rhs of \eqref{variance:derivative:second-prelim} becomes
\begin{multline}
\i \int_{0}^{\infty}\int_{0}^{2\pi}{R^{2} \left(\overline{\psi}_{R} \psi_{\tau} - \psi_{R} \overline{\psi}_{\tau}\right) \, \d R} = - \lambda_{\infty}^{\prime} \int_{0}^{2\pi}{\left.|\psi|^{2}\right|_{R=0}^{\infty} \, \d\theta} + \chi_{\infty} \int{R \, |\psi|^{2} \frac{\d}{\d R} |\psi|^{2} \, \d\nu} \\
= 2 \pi \lambda_{\infty}^{\prime} |\psi(\tau,0)|^{2} + \chi_{\infty} \int{R \, |\psi|^{2} \frac{\d}{\d R} |\psi|^{2} \, \d\nu},
\end{multline}
where in the first term on the rhs in the above expression we used the equality
\begin{align}
\int_{0}^{\infty}{R^{2} \left[\frac{1}{R^{2}}\right] \frac{\d}{\d R}|\psi|^{2} \, \d R} &= \int_{0}^{\infty}{\frac{\d}{\d R}|\psi|^{2} \, \d R} = \left.|\psi|^{2}\right|_{0}^{\infty},
\end{align}
and in the last term
\begin{multline}
\label{eqn:cNSL:variance-derivation:4}
\int_{0}^{\infty}{R^{2} |\psi|^{2} \frac{\d}{\d R}|\psi|^{2} \, \d R} = \frac{1}{2} \int_{0}^{\infty}{R^{2} \frac{\d}{\d R}\left(|\psi|^{2}\right)^{2} \, \d R} \\
= \frac{1}{2} \left[\left.R^{2} |\psi|^{4}\right|_{0}^{\infty} - 2 \int_{0}^{\infty}{R \, |\psi|^{4} \, \d R}\right].
\end{multline}

As a result, the second derivative of the variance \eqref{variance:derivative:second-prelim}
\begin{multline}
\label{variance:derivative:second}
\frac{1}{4\lambda_{\infty}} \frac{\d^{2} \mathcal{V}}{\d \tau^{2}} = \i \int_{0}^{2\pi}{\d\theta}\int_{0}^{\infty}{R^{2} \left(\overline{\psi}_{R} \psi_{\tau} - \psi_{R} \overline{\psi}_{\tau}\right) \, \d R} \\
- \i \int_{0}^{2\pi}{\d\theta}\int_{0}^{\infty}{R \left(\overline{\psi}_{\tau} \psi - \psi_{\tau} \overline{\psi}\right) \, \d R}
\end{multline}
becomes
\begin{multline}
\frac{1}{4\lambda_{\infty}} \frac{\d^{2} \mathcal{V}}{\d \tau^{2}} = 2 \pi \lambda_{\infty}^{\prime} |\psi(\tau,0)|^{2} - \chi_{\infty} \int{|\psi|^{4} \, \d\nu} + 2 \lambda_{\infty} \int{|\psi_{R}|^{2} \, \d\nu} \\
- 2 \lambda_{\infty}^{\prime} \int{\frac{|\psi|^{2}}{R^{2}} \, \d\nu} + 2 \mu_{\infty} \int{\frac{|\psi_{\theta}|^{2}}{R^{2}} \, \d\nu} + 2 \chi_{\infty} \int{|\psi|^{4} \, \d\nu}.
\end{multline}
Using the expression \eqref{H:ncNLS:original} for the Hamiltonian, equation \eqref{variance:derivative:second} can be rewritten in the form \eqref{variance:derivative:second:final}.

\section{Asymptotics of ground states at infinity} \label{appx:asymptotics:infinity}

To find next-order corrections, we introduce a phase $\varphi(x)$ such that $|\varphi(x)| \ll x$, so that plugging $y(x) = C \cos{\left[x + \varphi(x)\right]}$ into \eqref{ncNLS-ST:abstract:y:case-1} produces
\begin{multline}
- \cos{\left[x + \varphi(x)\right]} \, (1+\varphi^{\prime})^{2} - \sin{\left[x + \varphi(x)\right]} \varphi^{\prime\prime} + \cos{\left[x + \varphi(x)\right]} \\
\frac{d}{x^{2}} \cos{\left[x + \varphi(x)\right]} + \frac{C^{2}}{x} \cos^{3}{\left[x + \varphi(x)\right]} = 0.
\end{multline}
After cancelling out the leading-order solution $\cos{\left[x + \varphi(x)\right]}$, the next order terms are
\begin{align}
- 2 \varphi^{\prime}(x) + \frac{C^{2}}{x} \cos^{2}{\left[x + \varphi(x)\right]} = 0 \ \Rightarrow \ \frac{\d \varphi}{\d x} = C^{2} \frac{1 + \cos{2 \left[x + \varphi(x)\right]}}{4 x}.
\end{align}
Since we are studying the asymptotics at infinity, it is convenient to integrate the latter equation from $x$ to some large, but finite $x_{\infty}$, which yields
\begin{align}
\varphi(x_{\infty}) - \varphi(x) = \frac{C^{2}}{4}\left(\ln{x_{\infty}} - \ln{x}\right) + \frac{C^{2}}{4} \left[\Ci(2 x_{\infty}) -\Ci(2x)\right],
\end{align}
where the last term follows from
\begin{align}
&\int_{x}^{x_{\infty}}{\frac{\cos{2 \left[x + \varphi(x)\right]}}{x} \d x} = \int_{x}^{x_{\infty}}{\frac{\cos{2 \widetilde{x}}}{\widetilde{x} - \varphi(x)} \frac{\d \widetilde{x}}{1 + \varphi^{\prime}(x)}} \mathop{\Longrightarrow}_{x \gg 1} \int_{x}^{x_{\infty}}{\frac{\cos{2 \widetilde{x}}}{\widetilde{x}} \d \widetilde{x}} \\
= &\int_{2 x}^{2 x_{\infty}}{\frac{\cos{2 \widetilde{x}}}{2 \widetilde{x}} \d (2 \widetilde{x})} = \int_{2 x}^{\infty}{\frac{\cos{2 \widetilde{x}}}{2 \widetilde{x}} \d (2 \widetilde{x})} - \int_{2 x_{\infty}}^{\infty}{\frac{\cos{2 \widetilde{x}}}{2 \widetilde{x}} \d (2 \widetilde{x})} \equiv - \Ci{(2 x)} + \Ci{(2 x_{\infty})}. \nonumber
\end{align}
Letting $\varphi(x_{\infty}) = \frac{C^{2}}{4} \ln{x_{\infty}} + \frac{C^{2}}{4} \Ci{(2 x_{\infty})}$, we find the phase
\begin{align}
\varphi(x) = \frac{C^{2}}{4} \ln{x} + \frac{C^{2}}{4} \Ci{(2 x)} = \frac{C^{2}}{4} \ln{x} + \mathcal{O}\left(\frac{1}{x}\right),
\end{align}
where we took into account the asymptotics of the cosine integral for large argument $z$:
\begin{align}
\Ci{z} = \frac{1}{z} \sin{z} - \frac{1}{z^{2}} \cos{z} + \ldots.
\end{align}
Altogether, the solution $y(x)$ of \eqref{ncNLS-ST:abstract:y:case-1} for large $x$ reads
\begin{align}
y(x) = C \cos{\left[x + \frac{C^{2}}{4} \ln{x} + \mathcal{O}\left(\frac{1}{x}\right)\right]},
\end{align}
which is equation \eqref{asymptotics:BS:cNLS:infinity:case1}. To find the next-order correction we may add a new phase function $\psi(x)$ to \eqref{asymptotics:BS:cNLS:infinity:case1}, i.e. $y(x) = \cos{\left[x + \frac{C^{2}}{4} \ln{x} + \psi(x)\right]}$; straightforward calculations yield
\begin{align}
\psi(x) = \frac{C^{2}}{4} \Ci{(2 x)} \approx \frac{C^{2}}{4} \left[\frac{1}{2x} \sin{2x} - \frac{1}{4x^{2}} \cos{2x}\right] + \ldots.
\end{align}

\bibliographystyle{jfm}

\end{document}